\documentclass[fleqn,usenatbib]{mnras}

\usepackage{newtxtext,newtxmath}
\usepackage{ulem}
\usepackage[T1]{fontenc}

\DeclareRobustCommand{\VAN}[3]{#2}
\let\VANthebibliography\thebibliography
\def\thebibliography{\DeclareRobustCommand{\VAN}[3]{##3}\VANthebibliography}

\usepackage{graphicx}	% Including figure files
\usepackage{amsmath}	% Advanced maths commands

\title[Vertical shear instabilities in rotating stellar radiation zones]{Vertical shear instabilities in rotating stellar radiation zones:\\ effects of the full Coriolis acceleration and thermal diffusion}

\author[J. Park \& S. Mathis]{
J. Park,$^{1}$\thanks{E-mail: junho.park@coventry.ac.uk}
S. Mathis,$^{2}$
\\
$^{1}$Centre of Fluid and Complex Systems, Coventry University, Coventry CV1 5FB, UK\\
$^{2}$Université Paris-Saclay, Université Paris Cité, CEA, CNRS, AIM, Gif-sur-Yvette, F-91191, France
}

\date{Accepted XXX. Received YYY; in original form ZZZ}

\pubyear{\the\year{}}

\begin{document}
\label{firstpage}
\pagerange{\pageref{firstpage}--\pageref{lastpage}}
\maketitle

\begin{abstract}
Rotation deeply impacts the structure and the evolution of stars. 
To build coherent 1D or multi-D stellar structure and evolution models, we must systematically evaluate the turbulent transport of momentum and matter induced by hydrodynamical instabilities of radial and latitudinal differential rotation in stably stratified thermally diffusive stellar radiation zones. 
In this work, we investigate vertical shear instabilities in these regions. 
The full Coriolis acceleration with the complete rotation vector at a general latitude is taken into account. 
We formulate the problem by considering a canonical shear flow with a hyperbolic-tangent profile. 
We perform linear stability analysis on this base flow using both numerical and asymptotic Wentzel-Kramers-Brillouin-Jeffreys (WKBJ) methods. 
Two types of instabilities are identified and explored: inflectional instability, which occurs in the presence of an inflection point in shear flow, and inertial instability due to an imbalance between the centrifugal acceleration and pressure gradient.
Both instabilities are promoted as thermal diffusion becomes stronger or stratification becomes weaker. 
Effects of the full Coriolis acceleration are found to be more complex according to parametric investigations in wide ranges of colatitudes and rotation-to-shear and rotation-to-stratification ratios.
Also, new prescriptions for the vertical eddy viscosity are derived to model the turbulent transport triggered by each instability.
We foresee that the inflectional instability will be responsible for turbulent transport in the equatorial region of strongly-stratified radiative zones in slowly rotating stars while the inertial instability triggers turbulence in the polar regions of weakly-stratified radiative zones in fast-rotating stars.
%On the one hand, the inflectional instability is stabilized as $\Omega_{0}/N$ increases, but at the equator $\theta=90^{\circ}$, the instability is not affected by rotation and is the strongest. On the other hand, the inertial instability is triggered as the rotation $\Omega_{0}$ becomes non-zero. Its latitudinal dependence varies with the ratio $\Omega_{0}/S_{0}$ and thermal diffusivity $\kappa_{0}$. For a non-diffusive case with $\kappa_{0}=0$, cyclonic rotation with $\Omega_{0}/S_{0}>0$ promotes strongly the inertial instability in the mid-latitude regions, while the instability is maximal at the poles for both cyclonic and anti-cyclonic cases when the diffusivity $\kappa_{0}$ is high. 
\end{abstract}


%%%%%%%%%%%%%%%%%%%%%%%%%%%%%%%%%%%%%%%%%%%%%%%%%%

%%%%%%%%%%%%%%%%% BODY OF PAPER %%%%%%%%%%%%%%%%%%

\section{Introduction}
The rotation of stars deeply modifies their evolution \citep[e.g.][]{Maeder2009}. 
In the case of rapidly-rotating stars, such as early-type stars \citep[e.g.][]{Royeretal2007} and young late-type stars \citep[e.g.][]{GalletBouvier2015}, the centrifugal acceleration modifies their hydrostatic structure \citep[e.g.][]{EspinosaLaraRieutord2013,Rieutordetal2016}. 
Simultaneously, the Coriolis acceleration and buoyancy are governing the properties of large-scale flows \citep[e.g.][]{Garaud2002,Rieutord2006}, waves \citep[e.g.][]{DintransRieutord2000,Mathis2009,Mirouhetal2016}, hydrodynamical instabilities \citep[e.g.][]{Zahn1983,Zahn1992,Mathisetal2018}, and magneto-hydrodynamical processes \citep[e.g.][]{Spruit1999,Fulleretal2019,Jouveetal2020} that develop in their radiative regions. 
These regions are the seat of a strong transport of angular momentum occurring in all stars of all masses as revealed by space-based asteroseismology \citep[e.g.][]{Mosseretal2012,Deheuvelsetal2014,VanReethetal2016} and of a mild mixing that modify the stellar structure and chemical stratification with multiple consequences from the life time of stars to their interactions with their surrounding planetary and galactic environments. 
After almost three decades of implementation of a large diversity of physical parametrisations of transport and mixing mechanisms in one-dimensional stellar evolution codes \citep[e.g.][]{Talonetal1997,Hegeretal2000,MeynetMaeder2000,MaederMeynet2004,Hegeretal2005,TalonCharbonnel2005,Decressinetal2009,Marquesetal2013,Cantielloetal2014}, stellar evolution modelling is now entering a new area with the development of a new generation of bi-dimensional stellar structure and evolution models such as the numerical code ESTER \citep[][]{EspinosaLaraRieutord2013,Rieutordetal2016,Mombargetal2023,Mombargetal2024}. 
This code simulates in 2D the secular structural and chemical evolution of rotating stars and their large-scale internal zonal and meridional flows. 
Similarly to 1D stellar structure and evolution codes, it needs physical parametrisations of small spatial scale and short time scale processes such as waves, hydrodynamical instabilities and turbulence. 
To model the evolution of rotating stars, we need to take coherently the combined action of the buoyancy force and of the full Coriolis acceleration into account for any value of their ratio (from the order of $10^2$ in the bulk of the radiative core of main-sequence solar-type stars to $\sim 5-10$ in the bulk of the radiative envelope in rapidly-rotating main-sequence early-type stars). 
Walking on the path previously done for 1D codes, among all the necessary progresses, a first step is to examine the properties of the hydrodynamical instabilities of the vertical and horizontal shear of the differential rotation. 
Recent efforts have been devoted to improving the modelling of the turbulent transport triggered by the instabilities of the horizontal differential rotation in stellar radiation zones with buoyancy, the Coriolis acceleration and heat diffusion being considered \citep[e.g.][]{Park2020,Park2021}. 
However, strong vertical differential rotation also develops because of stellar structure's adjustments or the braking of the stellar surface by stellar winds \citep[e.g.][]{Zahn1992,MeynetMaeder2000,Decressinetal2009}. 
Up to now, state-of-the-art prescriptions for the turbulent transport it can trigger ignore the action of the Coriolis acceleration \citep[e.g.][]{Zahn1992,Maeder1995,MaederMeynet1996,TalonZahn1997,PratLignieres2014,KulenthirarajahGaraud2018} or examine it in a specific equatorial set up \citep{Chang2021}. 
Therefore, it becomes mandatory to study the hydrodynamical instabilities of vertical shear by taking into account the combination of buoyancy, the full Coriolis acceleration and strong heat diffusion at any latitude.

Instability of vertically-sheared flow has been a subject of interest in fundamental fluid dynamics to understand its development towards turbulence \citep[][]{Miles1961,Klaassen1985,Caulfield2021}. 
Two famous examples of vertical shear instability in stratified flows are the Kelvin-Helmholtz Instability (KHI) and Holmboe instability \citep[][]{Helmholtz1868,Kelvin1871,Holmboe1962}. 
KHI was originally studied for piecewise profiles of streamwise velocity $U(z)$ and density $\rho(z)$ but the instability can also be found in flows with smooth profiles of vertical shear $\mathrm{d}U/\mathrm{d}z$ and negative density gradient $\mathrm{d}\rho/\mathrm{d}z<0$, the latter corresponding to the case of stable density stratification. 
The Richardson's criterion states a necessary condition for instability in terms of the local Richardson number $\mathrm{Ri}(z)$ as
\begin{equation}
\mathrm{Ri}(z)=\frac{g(-\mathrm{d}\rho/\mathrm{d}z)}{\rho\left(\mathrm{d}U/\mathrm{d}z\right)^{2}}\leq\frac{1}{4},
\end{equation}
where $g$ is the gravity \citep[see also,][]{Drazin2015}. 
KHI is typically expected to be stabilized by strong stratification when $Ri>1/4$.
However, in the layer where density gradient is sharp, the Holmboe instability can appear and plays an important role in mixing of stratified shear flows \citep[][]{Ortiz2002,Eaves2019}. 

\citet{Wang2014} investigated more extensively instabilities of vertical shear flows in stratified and rotating fluids in which the rotation axis is along the vertical.
Depending on the rotation, stratification, shear flow profiles and other parameters such as wavelengths of perturbation, they identified centrifugal instability as well as KHI for such vertical shear flows. 
The centrifugal instability occurs when there is an imbalance between the pressure gradient and centrifugal force due to the rotation.
The instability is also called the inertial instability in the Cartesian coordinates \citep[][]{Holton} and is analogous to the Goldreich-Schubert-Fricke (GSF) instability in the limit of high thermal diffusivity \citep[][]{Goldreich1967,Fricke1968} and symmetric instability \citep[][]{Zeitlin2018}. 
For certain profiles of base flow, \citet{Wang2014} also identified another type of instability called the ageostrophic instability that occurs due to interactions between the shear and an inertia-gravity wave, or between two inertia-gravity waves. 
They demonstrated that different instabilities appear depending on the Rossby number $Ro$, a non-dimensional number that scales as the inverse of rotation (e.g. $Ro$ is small when the rotation is strong), and the profile of shear flow.
Despite their detailed investigation, their results can not be directly applied to understand instabilities in stellar radiation zones as the study has been conducted with relevance in geophysical contexts where thermal or density diffusion is weak, while strong thermal diffusion occurs in the radiation zones. 

In highly diffusive and stably stratified fluids, \citet{Lignieresetal1999} investigated vertical shear instability occurring due to an inflection point of the shear flow, which we call hereafter the inflectional instability. 
They identified that there is a self-similarity in the high thermal-diffusivity limit (or in the limit of low P\'eclet number $Pe$ that scales as the inverse of the thermal diffusivity) where the marginal stability curves fall onto a single curve $k_{x}RiPe\simeq0.117$ where $k_{x}$ is the horizontal wavenumber in the streamwise (longitudinal) direction $x$. 
The information on these critical $Ri$ or the normalized number $R=RiPe$ is essential when proposing eddy-viscosity models used in the modelling of angular momentum transport and chemical mixing along stellar evolution \citep[][]{Zahn1992,Prat2014}. 
At the same time, such a turbulence modelling needs to take into account both stratification and rotation to properly describe anisotropic transport in the radial and latitudinal directions in stellar radiation zones \citep[e.g.][]{Zahn1992,Mathisetal2018}. 
Recent progress in direct numerical simulations allows us to further investigate quantitativly the properties of this transport \citep[][]{Prat2013,Prat2014,Chang2021,Garaud2021}. 
However, these works do not take the Coriolis acceleration into account or only in the equatorial plane.

Therefore, while vertical shear instabilities in stably stratified and rotating fluids have been studied in the broad contexts of astrophysics as well as geophysics, there are relatively few studies on the instabilities when the rotation is not aligned with the local vertical axis as illustrated in Fig.~\ref{Fig_cartoon}. 
To explore vertical shear instabilities at a general colatitude $0^{\circ}\leq\theta\leq180^{\circ}$, we need to consider the full Coriolis acceleration with both vertical (radial) and horizontal (latitudinal) rotation components. 
\citet{Zeitlin2018} investigated how symmetric instability, an analogue to the inertial instability, can be modified by the inclusion of the full Coriolis acceleration.
Studies by \cite{Barker2019,Barker2020} and \cite{Dymott2023} investigated thoroughly the GSF instability with a linear shear profile at the equator or at a general colatitude, or with arbitrary shear profiles comprised of both vertical (radial) and horizontal (latitudinal) differential rotation.  
Their numerical simulations revealed various non-linear properties of the GSF instability such as turbulence and angular momentum transport, layering and zonal jet formation in a preferred direction between the local angular momentum gradient and the vector perpendicular to the rotation axis.
While these studies unveiled remarkable properties of the GSF and symmetric instabilities of vertical shears, a limitation lies in their linear shear profile, which does not possess an inflection point and thus does not exhibit inflectional instability.

It is noteworthy that the inflectional instability is essential in studying strong shear-induced mixing in stellar interior \citep[][]{Bruggen2001} and shear flow such a hyperbolic-tangent type flow has been used to verify the Richardson's criterion, which is in turn crucial to deduce effective eddy-viscosities \citep[e.g.][]{Zahn1992,Maeder1995,MaederMeynet1996,TalonZahn1997,Mathisetal2018}.
Our recent work \citep[][]{Park2020,Park2021} studied horizontal shear instabilities with the hyperbolic tangent flow in wide ranges of parameters to reveal how the strong thermal diffusion and full Coriolis acceleration modify the inflectional and inertial instabilities. 
It is revealed that the full Coriolis acceleration is crucial to model horizontal shear instabilities at a general latitude. 
Nonetheless, this non-traditional rotation effect is still not fully understood in the context of vertical shear flows in rotating radiation zones.

This situation motivates our current study on vertical shear instabilities in rotating stellar radiation zones  (Fig.~\ref{Fig_cartoon}), in which fluids are stably stratified, with a focus on the effects of both the strong thermal diffusion and full Coriolis acceleration. 
In Sect.~\ref{sec:Problem}, we formulate equations for linear stability analysis of a hyperbolic-tangent vertical shear flow in stratified and thermally diffusive fluids under the influence of the full Coriolis acceleration.
In Sect.~\ref{sec:LSA}, we present examples of numerical results on inflectional and inertial instabilities of vertical shear flows and study parametric dependence of these instabilities. 
In Sect.~\ref{sec:WKBJ}, we perform the WKBJ analysis and derive analytical and asymptotic expressions of the dispersion relation for the inertial instability in the limits of either vanishing or very high thermal diffusivity. 
In Sect.~\ref{sec:SI}, we explain the link between the inertial instability and symmetric instability by demonstrating the mathematical equivalence between the two instabilities. 
In Sect.~\ref{sec:Turbulent}, we explore turbulent dissipation induced by vertical shear instabilities, derive a new prescription of eddy-viscosity and conduct parametric investigations for either inertial or inflectional instabilities.
In Sect.~\ref{sec:Conclusion}, conclusion and discussion are presented.

\section{Problem formulation}
\label{sec:Problem}
\subsection{Governing equations and base state}
%
%                                                One column figure
%----------------------------------------------------------------- 
   \begin{figure}
   \centering
   \includegraphics[height=5.5cm]{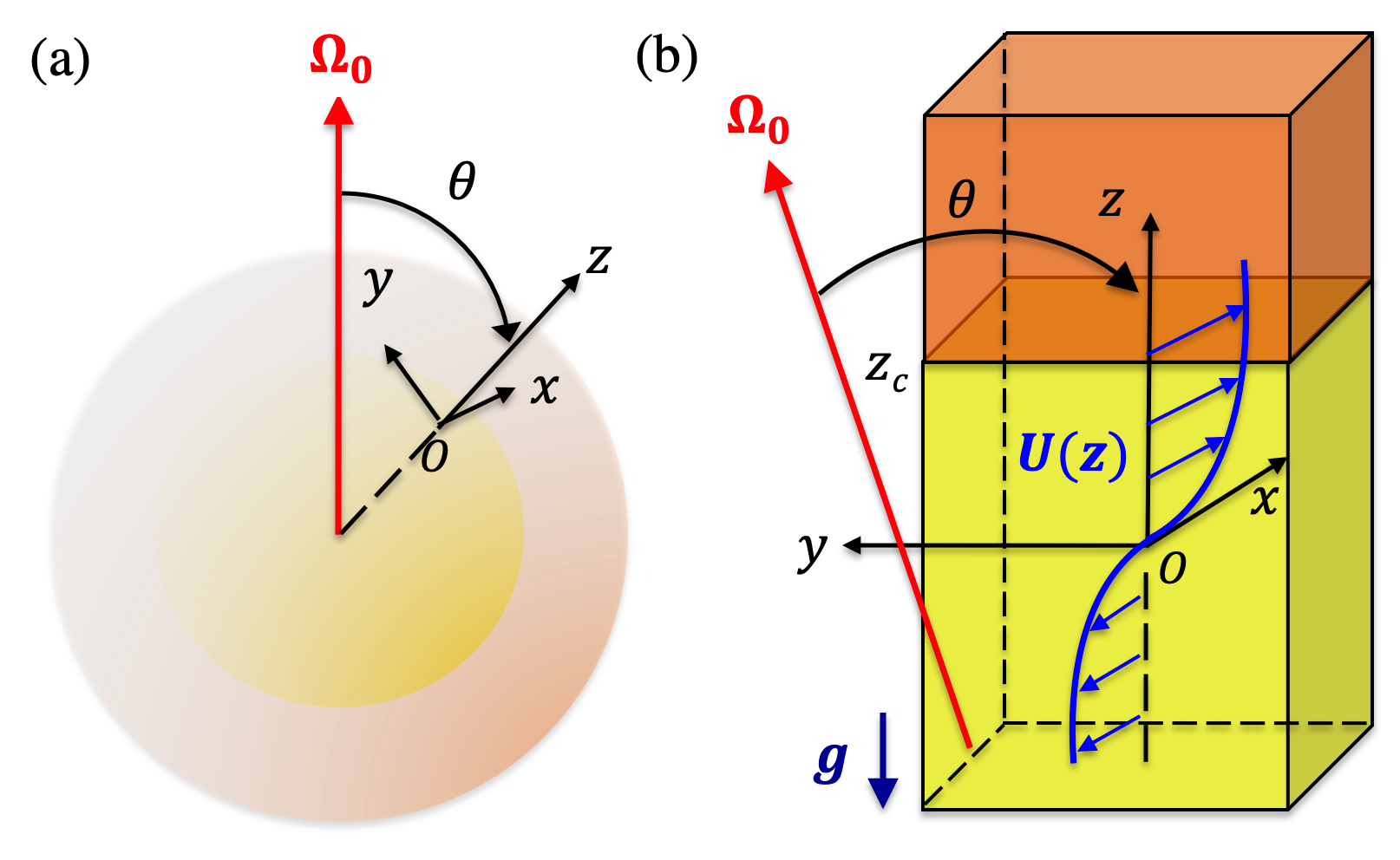}
      \caption{(a) Illustration of the radiation zone (yellow) and convective envelope (orange) of a low-mass star rotating with angular speed $\Omega_{0}$ at the pole and the local coordinates $(x,y,z)$ at a colatitude $\theta$. 
      (b) Vertical shear flow $U(z)$ on a local plane rotating with angular speed $\Omega_{0}$ in the rotating stellar radiation zone. $z_{c}$ is the transition altitude from the radiation zone to the convection zone.
      %Horizontal shear flow $U(y)$ on a local $f$-plane. The radiative and convective zones are colored as yellow and orange and $z_{c}$ is the transition altitude between the two zones, the configuration of low-mass stars. For the case of intermediate-mass and massive stars, the vertical structure is inverted. 
              }
         \label{Fig_cartoon}
   \end{figure}
%-----------------------------------------------------------------
We consider flow with shear along the vertical direction $z$ at a general colatitude $\theta$ in the radiation zone of rotating stars (see e.g., the illustration of the interior of low-mass main-sequence stars in Fig.~\ref{Fig_cartoon}; the position of radiation and convection zones can be reversed for main-sequence early-type stars).
The flow profile we consider here is relevant to shellular rotation where the angular rotation $\Omega(r)$ in stellar interior depends solely on the radial coordinate $r$ of a star. 
Such a shellular approximation has been used to treat the angular momentum transport into one dimension \citep[][]{Zahn1992}. 
To investigate instabilities of vertical shear flow, we consider the Navier-Stokes equations in a rotating frame under the Boussinesq approximation, in which unstable and turbulent motions have characteristic spatial scales smaller than the density and pressure height scales, and the heat advection-diffusion equation in the Cartesian coordinates $(x,y,z)$ (where $x,y,z$ denote the local longitudinal, latitudinal and vertical coordinates, respectively) as follows: 
\begin{equation}
\label{eq:total_continuity}
	\nabla\cdot\vec{u}=0,
\end{equation}
\begin{equation}
\label{eq:total_momentum}
	\frac{\partial\vec{u}}{\partial t}+\left(\vec{u}\cdot\nabla\right)\vec{u}+\vec{f}\times\vec{u}=-\frac{1}{\rho_{0}}\nabla p-\alpha_{0}\Theta\vec{g}+\nu_{0}\nabla^{2}\vec{u},
\end{equation}
\begin{equation}
\label{eq:total_diffusion}
	\frac{\partial \Theta}{\partial t}+\vec{u}\cdot\nabla \Theta=\kappa_{0}\nabla^{2}\Theta.
\end{equation}
We introduce $\vec{u}=\left(u,v,w\right)$ as the velocity, $p$ the pressure, $\Theta=T-T_{0}$ the temperature deviation from the reference temperature $T_{0}$, $\vec{f}=\left(0,f_{\rm{h}},f_{\rm{v}}\right)$ the Coriolis vector where $f_{\rm{h}}=2\Omega_{0}\sin\theta$ and $f_{\rm{v}}=2\Omega_{0}\cos\theta$ are its horizontal and vertical components, respectively, $\Omega_{0}$ the reference angular rotation at the northern pole, $\vec{g}=(0,0,-g)$ the gravity, $\rho_{0}$ the reference density, $\nu_{0}$ the reference viscosity, $\kappa_{0}$ the reference thermal diffusivity, $\alpha_{0}$ the thermal expansion coefficient, and $\nabla^{2}=\partial^{2}/\partial x^{2}+\partial^{2}/\partial y^{2}+\partial^{2}/\partial z^{2}$ the Laplacian operator. 

To perform linear stability analysis, we consider base velocity $\vec{U}=\left(U(z),0,0\right)$ with the vertically-sheared streamwise (longitudinal) velocity $U(z)$ in a canonical hyperbolic-tangent profile
\begin{equation}
\label{eq:base_shear}
	U(z)=U_{0}\tanh\left(\frac{z}{H_{0}}\right),
\end{equation}
where $U_{0}$ and $H_{0}$ are the reference velocity and length scales, respectively. 
The base flow remains steady in time if the balance between $U(z)$, the base pressure $P(y,z)$ and the base temperature $\bar{T}(y,z)$ is maintained as follows:
\begin{equation}
\label{eq:base_pressure}
\begin{aligned}
&f_{\rm{v}}U=-\frac{1}{\rho_{0}}\frac{\partial P}{\partial y},\\
&-f_{\rm{h}}U=-\frac{1}{\rho_{0}}\frac{\partial P}{\partial z}+\alpha_{0}g\bar{T}.
\end{aligned}
\end{equation}
By eliminating the pressure in (\ref{eq:base_pressure}), we can obtain the following thermal-wind balance between the vertical shear $U(z)$ and the base temperature $\bar{T}(y,z)$ 
\begin{equation}
\label{eq:tw_balance}
	f_{\rm{v}} \frac{\partial U}{\partial z}+\alpha_{0}g\frac{\partial\bar{T}}{\partial y}=0
\end{equation}
\citep[see also,][]{Wang2014}.
In addition, we assume that the base temperature is stably stratified in the vertical direction $z$ with a linear profile with the constant temperature gradient $\mathcal{T}_{\Theta}>0$. 
By considering this assumption and integrating the thermal-wind balance (\ref{eq:tw_balance}) in $y$, we find the base temperature as follows:
\begin{equation}
\label{eq:base_temperature}
	\bar{T}(y,z)=\mathcal{T}_{\Theta}z-\frac{f_{\rm{v}}}{\alpha_{0}g}\frac{\partial U}{\partial z}y.
\end{equation}

\subsection{Perturbation equations and linear stability analysis}
In addition to the steady base state, we consider the velocity perturbation $\tilde{\vec{u}}=\left(\tilde{u},\tilde{v},\tilde{w}\right)=\vec{u}-\vec{U}$, the pressure perturbation $\tilde{p}=p-P$ and temperature perturbation $\tilde{T}=\Theta-\bar{T}$. 
Applying the perturbed state into the equations (\ref{eq:total_continuity})-(\ref{eq:total_diffusion}), we derive the following equations for the perturbations:
\begin{equation}
\label{eq:ptb_continuity}
	\frac{\partial \tilde{u}}{\partial x}+\frac{\partial \tilde{v}}{\partial y}+\frac{\partial \tilde{w}}{\partial z}=0,
\end{equation}
\begin{equation}
\label{eq:ptb_x_mom}
	\frac{\partial \tilde{u}}{\partial t}+U\frac{\partial \tilde{u}}{\partial x}-f_{\rm{v}}\tilde{v}+\left(\frac{\partial U}{\partial z}+f_{\rm{h}}\right)\tilde{w}+\mathcal{N}_{\tilde{u}}=-\frac{1}{\rho_{0}}\frac{\partial \tilde{p}}{\partial x}+\nu_{0}\nabla^{2}\tilde{u},
\end{equation}
\begin{equation}
\label{eq:ptb_y_mom}
	\frac{\partial \tilde{v}}{\partial t}+U\frac{\partial \tilde{v}}{\partial x}+f_{\rm{v}}\tilde{u}+\mathcal{N}_{\tilde{v}}=-\frac{1}{\rho_{0}}\frac{\partial \tilde{p}}{\partial y}+\nu_{0}\nabla^{2}\tilde{v},
\end{equation}
\begin{equation}
\label{eq:ptb_z_mom}
	\frac{\partial \tilde{w}}{\partial t}+U\frac{\partial \tilde{w}}{\partial x}-f_{\rm{h}}\tilde{u}+\mathcal{N}_{\tilde{w}}=-\frac{1}{\rho_{0}}\frac{\partial \tilde{p}}{\partial z}+\alpha_{0}g\tilde{T}+\nu_{0}\nabla^{2}\tilde{w},
\end{equation}
\begin{equation}
\label{eq:ptb_diffusion}
	\frac{\partial \tilde{T}}{\partial t}+U\frac{\partial \tilde{T}}{\partial x}+\frac{\partial\bar{T}}{\partial y}\tilde{v}+\frac{\partial\bar{T}}{\partial z}\tilde{w}+\mathcal{N}_{\tilde{T}}=\kappa_{0}\nabla^{2}\tilde{T},
\end{equation}
where $\mathcal{N}_{\tilde{u}}$, $\mathcal{N}_{\tilde{v}}$, $\mathcal{N}_{\tilde{w}}$ and $\mathcal{N}_{\tilde{T}}$ are the nonlinear terms
\begin{equation}
\begin{aligned}
&\mathcal{N}_{\tilde{u}}=\tilde{u}\frac{\partial\tilde{u}}{\partial x}+\tilde{v}\frac{\partial\tilde{u}}{\partial y}+\tilde{w}\frac{\partial\tilde{u}}{\partial z},~~~
\mathcal{N}_{\tilde{v}}=\tilde{u}\frac{\partial\tilde{v}}{\partial x}+\tilde{v}\frac{\partial\tilde{v}}{\partial y}+\tilde{w}\frac{\partial\tilde{v}}{\partial z},\\
&\mathcal{N}_{\tilde{w}}=\tilde{u}\frac{\partial\tilde{w}}{\partial x}+\tilde{v}\frac{\partial\tilde{w}}{\partial y}+\tilde{w}\frac{\partial\tilde{w}}{\partial z},~~~
\mathcal{N}_{\tilde{T}}=\tilde{u}\frac{\partial\tilde{T}}{\partial x}+\tilde{v}\frac{\partial\tilde{T}}{\partial y}+\tilde{w}\frac{\partial\tilde{T}}{\partial z}.\\
\end{aligned}
\end{equation}
We note that the vertical temperature gradient 
\begin{equation}
\label{eq:base_temperature_vertical}
	\frac{\partial\bar{T}}{\partial z}=\mathcal{T}_{\Theta}-\frac{f_{\rm{v}}}{\alpha_{0}g}\frac{\partial^{2} U}{\partial z^{2}}y,
\end{equation}
is not homogeneous in the horizontal direction $y$ if the second derivative $\partial^{2}U/\partial z^{2}$ is not zero, which is the case for the hyperbolic-tangent profile.
If we keep the second term on the right-hand side of Eq.~(\ref{eq:base_temperature_vertical}), equations linearized from (\ref{eq:ptb_continuity})-(\ref{eq:ptb_diffusion}) cannot be expressed in a modal form due to the inhomogeneity in the $y$-direction. 
To avoid complexities of global stability analysis and simplify the problem, we assume that:
\begin{equation}
\label{eq:wide_jet_approximation}
	|y|\ll\left|\frac{\alpha_{0}g\mathcal{T}_{\Theta}}{f_{\rm{v}}}\right|/\max\left(\left|\frac{\partial^{2}U}{\partial z^{2}}\right|\right),
\end{equation}
which is called the wide jet approximation previously used for shear flows in geophysical contexts \citep[][]{Wang2014}.
The above assumption is valid if we limit our problem near $y=0$ or when the vertical Coriolis parameter $f_{\rm{v}}$ is small (e.g., near the equator) while the vertical gradient $\mathcal{T}_{\Theta}$ is large. 
The approximation (\ref{eq:wide_jet_approximation}) can also be rewritten in a non-dimensional form as
\begin{equation}
\label{eq:wide_jet_approximation_nondim}
\left|\frac{y}{H_{0}}\right|\ll\left|\frac{N^{2}H_{0}}{f_{\rm{v}}U_{0}}\right|=\left|RiR_{\mathrm{v}}\right|,
\end{equation}
where $N=\sqrt{\alpha_{0}g\mathcal{T}_{\Theta}}$ is the Brunt-V\"ais\"al\"a frequency, $Ri=N^{2}/S_{0}^{2}$ is the Richardson number with $S_{0}=U_{0}/H_{0}$ the reference shear obtained from the vertical velocity gradient $\partial U/\partial z$ evaluated at $z=0$, and $R_{\mathrm{v}}=S_{0}/f_{\mathrm{v}}$ is the vertical Rossby number based on the vertical Coriolis parameter $f_{\rm{v}}$.
If we consider this wide jet approximation, we can neglect the second term on the right-hand side of Eq.~(\ref{eq:base_temperature_vertical}) and the vertical temperature gradient $\partial\bar{T}/\partial z$ becomes constant as $\partial\bar{T}/\partial z\simeq \mathcal{T}_{\Theta}$.

As we consider the approximation (\ref{eq:wide_jet_approximation}) and assume that the perturbation is infinitesimal (i.e., $\mathcal{N}_{\tilde{u}}=\mathcal{N}_{\tilde{v}}=\mathcal{N}_{\tilde{w}}=\mathcal{N}_{\tilde{T}}\simeq0$), we can perform linear stability analysis using the normal mode
\begin{equation}
\label{eq:ptb_normal_mode}
	\left(\tilde{\vec{u}},\tilde{p},\tilde{T}\right)=\left(\hat{\vec{u}}(z),\rho_{0}\hat{p}(z),\mathcal{T}_{\Theta}\hat{T}(z)\right)\exp\left(\mathrm{i}k_{x} x+\mathrm{i}k_{y} y+\sigma t\right)+c.c.,
\end{equation}
where $c.c.$ denotes the complex conjugate, $\mathrm{i}^{2}=-1$, $\hat{\vec{u}}=\left(\hat{u},\hat{v},\hat{w}\right)$ is the velocity mode shape, $\hat{p}$ and $\hat{T}$ are the mode shapes for rescaled pressure and temperature perturbations, respectively, $k_{x}$ is the horizontal wavenumber in the longitudinal direction $x$, $k_{y}$ is the horizontal wavenumber in the latitudinal direction $y$, and $\sigma=\sigma_{r}+\mathrm{i}\sigma_{i}$ is the complex growth rate where the real part $\sigma_{r}$ denotes the temporal growth rate and the imaginary part $\sigma_{i}$ denotes the temporal frequency. 
In terms of the wavenumber $k_{y}$, the approximation (\ref{eq:wide_jet_approximation}) can be equivalently expressed as
\begin{equation}
\label{eq:wide_jet_approximation_ky}
	|k_{y}|\gg\max\left(\left|\frac{\partial^{2}U}{\partial z^{2}}\right|\right)/\left(\left|\frac{\alpha_{0}g\mathcal{T}_{\Theta}}{f_{\rm{v}}}\right|\right),
\end{equation}
which implies that the wavenumber $k_{y}$ should be sufficiently large enough (i.e., the wavelength $\lambda_{y}=2\pi/k_{y}$ should be small enough) to validate the wide jet approximation \citep[see also,][]{Wang2014}.
This also means that the latitudinal length scale of perturbations should be small compared to the length scale of the shear as the name `wide jet' signifies.
We can also define the cut-off minimum wavenumber $k_{y,\min}$ below which the wide jet approximation does not hold by rewriting (\ref{eq:wide_jet_approximation_ky}) in terms of nondimensional parameters as
\begin{equation}
\label{eq:wide_jet_approximation_ky_nondimensional}
\left|k_{y,\min}H_{0}\right|\approx \frac{1}{\left|RiR_{\rm{v}}\right|}.
\end{equation}

By applying the normal mode (\ref{eq:ptb_normal_mode}) to the perturbation equations (\ref{eq:ptb_continuity})-(\ref{eq:ptb_diffusion}) with zero nonlinear terms, we obtain the following set of linear stability equations
\begin{equation}
\label{eq:lse_continuity}
	\mathrm{i}k_{x}\hat{u}+\mathrm{i}k_{y}\hat{v}+\frac{\mathrm{d}\hat{w}}{\mathrm{d} z}=0,
\end{equation}
\begin{equation}
\label{eq:lse_x_mom}
	\left(\sigma+\mathrm{i}k_{x} U\right)\hat{u}-f_{\rm{v}}\hat{v}+\left(U'+f_{\rm{h}}\right)\hat{w}=-\mathrm{i}k_{x}\hat{p}+\nu_{0}\hat{\nabla}^{2}\hat{u},
\end{equation}
\begin{equation}
\label{eq:lse_y_mom}
	\left(\sigma+\mathrm{i}k_{x} U\right)\hat{v}+f_{\rm{v}}\hat{u}=-\mathrm{i}k_{y}\hat{p}+\nu_{0}\hat{\nabla}^{2}\hat{v},
\end{equation}
\begin{equation}
\label{eq:lse_z_mom}
	\left(\sigma+\mathrm{i}k_{x} U\right)\hat{w}-f_{\rm{h}}\hat{u}=-\frac{\mathrm{d}\hat{p}}{\mathrm{d} z}+N^{2}\hat{T}+\nu_{0}\hat{\nabla}^{2}\hat{w},
\end{equation}
\begin{equation}
\label{eq:lse_diffusion}
	\left(\sigma+\mathrm{i}k_{x} U\right)\hat{T}-\frac{f_{\rm{v}}U'}{N^{2}}\hat{v}+\hat{w}=\kappa_{0}\hat{\nabla}^{2}\hat{T},
\end{equation}
where prime $(')$ denotes the derivative with respect to $z$ and $\hat{\nabla}^{2}={\rm d}^{2}/{\rm d}z^{2}-k^{2}$ with $k^{2}=k_{x}^{2}+k_{y}^{2}$.
In this study, we will use the following nondimensional parameters such as the Richardson number $Ri$, the Reynolds number $Re$, the P\'eclet number $Pe$, the Prandtl number $Pr$, the vertical and horizontal Rossby numbers $R_{\rm{v}}$ and $R_{\rm{h}}$ defined as
\begin{equation}
\label{eq:Re_Pe_Pr_Ri_Ro}
\begin{aligned}
&Re=\frac{U_{0}H_{0}}{\nu_{0}},~~
Pe=\frac{U_{0}H_{0}}{\kappa_{0}},~~
Pr=\frac{\nu_{0}}{\kappa_{0}}=\frac{Pe}{Re},\\
&Ri=\frac{N^{2}}{S_{0}^{2}}=\frac{\alpha_{0}g\mathcal{T}_{\Theta}H_{0}^{2}}{U_{0}^{2}},~~
R_{\rm{v}}=\frac{S_{0}}{f_{\rm{v}}},~~
R_{\rm{h}}=\frac{S_{0}}{f_{\rm{h}}}.
\end{aligned}
\end{equation}
%\begin{equation}
%\label{eq:Rossby_number}
%\end{equation}

For numerical computation, the linear stability equations (\ref{eq:lse_continuity})-(\ref{eq:lse_diffusion}) can be simplified into a matrix form with three variables $\hat{u}$, $\hat{w}$ and $\hat{T}$ as
\begin{equation}
\label{eq:lse_matrix}
	\mathcal{A}
	\left(
	\begin{array}{c}
	\hat{u}\\
	\hat{w}\\
	\hat{T}
	\end{array}
	\right)=
	\sigma\mathcal{B}
	\left(
	\begin{array}{c}
	\hat{u}\\
	\hat{w}\\
	\hat{T}
	\end{array}
	\right),
\end{equation}
where $\mathcal{A}$ and $\mathcal{B}$ are the operator matrices in the form
\begin{eqnarray}
\label{eq:operator_matrices}
	\mathcal{A}&=&
	\left[
	\begin{array}{ccc}
	\mathcal{A}_{11} & \mathcal{A}_{12} & 0\\
	\mathcal{A}_{21} & \mathcal{A}_{22} & -N^{2}k_{y}k^{2}\\
	-k_{x}f_{\rm{v}}U'/(k_{y}N^{2}) & \mathcal{A}_{32} & \mathcal{A}_{33}
	\end{array}
	\right],\nonumber\\
	\mathcal{B}&=&
	\left[
	\begin{array}{ccc}
	k^{2} & -\mathrm{i}k_{x}\left({\mathrm{d}}/{\mathrm{d}z}\right) & 0\\
	0 & k_{y}\hat{\nabla}^{2} & 0\\
	0 & 0 & 1
	\end{array}
	\right],~~
\end{eqnarray}
where
\begin{equation}
\begin{aligned}
&\mathcal{A}_{11}=k^{2}\left(-\mathrm{i}k_{x}U+\nu_{0}\hat{\nabla}^{2}\right),\\
&\mathcal{A}_{12}=\left(\mathrm{i}k_{y}f_{\mathrm{v}}-\mathrm{i}k_{x}\nu_{0}\hat{\nabla}^{2}-k_{x}^{2}U\right)\frac{\mathrm{d}}{\mathrm{d}z}-k_{y}^{2}(U'+f_{\mathrm{h}}),\\
&\mathcal{A}_{21}=k^{2}\left(\mathrm{i}f_{\mathrm{v}}\frac{\mathrm{d}}{\mathrm{d}z}-k_{y}f_{\mathrm{h}}\right),\\
&\mathcal{A}_{22}=\mathrm{i}k_{x}k_{y}\left(U''+f_{\mathrm{h}}\frac{\mathrm{d}}{\mathrm{d}z}-U\hat{\nabla}^{2}\right)+k_{x}f_{\mathrm{v}}\frac{\mathrm{d}^{2}}{\mathrm{d}z^{2}}+k_{y}\nu_{0}\hat{\nabla}^{4},\\
&\mathcal{A}_{32}=\frac{\mathrm{i}f_{\mathrm{v}}U'}{N^{2}k_{y}}\frac{\mathrm{d}}{\mathrm{d}z}-1,\\
&\mathcal{A}_{33}=-\mathrm{i}k_{x}U+\kappa_{0}\hat{\nabla}^{2}.
\end{aligned}
\end{equation}
For the case with $k_{y}=0$, the eigenvalue problem (\ref{eq:lse_matrix}) becomes singular in $\mathcal{A}_{32}$ so we need to simplify differently the eigenvalue problem with different variables $\hat{v}$, $\hat{w}$, and $\hat{T}$ as
\begin{equation}
\label{eq:lse_matrix_CD}
	\mathcal{C}
	\left(
	\begin{array}{c}
	\hat{v}\\
	\hat{w}\\
	\hat{T}
	\end{array}
	\right)=
	\sigma\mathcal{D}
	\left(
	\begin{array}{c}
	\hat{v}\\
	\hat{w}\\
	\hat{T}
	\end{array}
	\right),
\end{equation}
where $\mathcal{C}$ and $\mathcal{D}$ are the operator matrices
\begin{equation}
\label{eq:operator_matrices_CD}
	\mathcal{C}=
	\left[
	\begin{array}{ccc}
	\mathcal{C}_{11} & \mathcal{C}_{12} & 0\\
	\mathcal{C}_{21} & \mathcal{C}_{22} & -N^{2}k_{x}^{2}\\
	{f_{\mathrm{v}}U'}/{N^{2}} & -1 & \mathcal{C}_{33}
	\end{array}
	\right],~~
	\mathcal{D}=
	\left[
	\begin{array}{ccc}
	1 & 0 & 0\\
	0 & \hat{\nabla}^{2} & 0\\
	0 & 0 & 1
	\end{array}
	\right],~~
\end{equation}
where
\begin{equation}
\begin{aligned}
&\mathcal{C}_{11}=-\mathrm{i}k_{x}U+\nu_{0}\hat{\nabla}^{2},\\
&\mathcal{C}_{12}=-\frac{\mathrm{i}f_{\mathrm{v}}}{k_{x}}\frac{\mathrm{d}}{\mathrm{d}z},\\
&\mathcal{C}_{21}=-\mathrm{i}k_{x}f_{\mathrm{v}}\frac{\mathrm{d}}{\mathrm{d}z},\\
&\mathcal{C}_{22}=\mathrm{i}k_{x}\left(U''-U\hat{\nabla}^{2}\right)+\nu_{0}\hat{\nabla}^{4},\\
&\mathcal{C}_{33}=-\mathrm{i}k_{x}U+\kappa_{0}\hat{\nabla}^{2}.
\end{aligned}
\end{equation}
For the case where the wavenumbers are zero as $k_{x}=k_{y}=0$, we obtain $\hat{w}=0$ from the continuity equation and the vanishing boundary condition as $|z|\rightarrow\infty$. 
Then we solve the eigenvalue problem for $\hat{u}$, $\hat{v}$, and $\hat{T}$ as follows:
\begin{equation}
\label{eq:lse_matrix_EF}
	\left[
	\begin{array}{ccc}
	\nu_{0}\hat{\nabla}^{2} & f_{\mathrm{v}} & 0\\
	-f_{\mathrm{v}} & \nu_{0}\hat{\nabla}^{2} & 0\\
	0 & {f_{\mathrm{v}}U'}/{N^{2}} & \kappa_{0}\hat{\nabla}^{2}
	\end{array}
	\right]
	\left(
	\begin{array}{c}
	\hat{u}\\
	\hat{v}\\
	\hat{T}
	\end{array}
	\right)=
	\sigma
	\left(
	\begin{array}{c}
	\hat{u}\\
	\hat{v}\\
	\hat{T}
	\end{array}
	\right).
\end{equation}
To solve the eigenvalue problems (\ref{eq:lse_matrix}), (\ref{eq:lse_matrix_CD}) or (\ref{eq:lse_matrix_EF}) numerically, we use in the $z$-direction the rational Chebyshev function that maps the spectral Chebyshev domain $z_{\mathrm{cheb}}\in(-1,1)$ onto the physical space $z\in (-\infty,\infty)$ by the mapping $z/{Z}_{\mathrm{map}}=z_{\mathrm{cheb}}/\sqrt{1-z_{\mathrm{cheb}}^{2}}$ where $Z_{\mathrm{map}}$ is the stretching factor in the mapping. 
In computing eigenmodes, we find both physical modes and spurious modes, the latter which are highly oscillatory. 
To distinguish physical modes from spurious modes, we apply a convergence criterion proposed by \citet{Fabre2004} who studied the behaviour of solutions such that the physical modes should not be highly oscillatory and decays to zero as $|z|\rightarrow\infty$. 
In this case, the coefficients of the Chebyshev functions at high orders are sufficiently low.
Also, in computation, we consider the number of collocation points in the $z$-direction between 100 and 200.
We found that the number is sufficiently large to confirm the convergence of physical modes in our parameter space of interest. 
To impose vanishing boundary conditions as $|z|\rightarrow\infty$, we suppress the first and last rows of each element of the operator matrices \citep[we refer to][for more details]{Antkowiak2005,Park2012}. 
%Numerical results of the horizontal shear instability are compared and validated with results of \citet{Deloncle2007,Arobone2012,PPM2020} in stratified and rotating fluids for the traditional case when $\tilde{f}=0$. 
\subsection{Simplified equations in the inviscid limit $\nu_{0}\rightarrow0$}
For the inviscid case ($\nu_{0}=0$), we can simplify the equations (\ref{eq:lse_continuity})-(\ref{eq:lse_diffusion}).
By eliminating the pressure and considering the continuity equation, we obtain the following differential equations for $\hat{w}$ and $\hat{T}$:
\begin{equation}
\label{eq:lse_2ndODE_1}
\begin{aligned}
&\frac{\mathrm{d}^{2}\hat{w}}{\mathrm{d}z^{2}}=\left(\frac{\mathrm{i}N^{2}k^{2}s}{s^{2}-f_{\mathrm{v}}^{2}}\right)\hat{T}+\left[\frac{k_{x}U'f_{\mathrm{v}}^{2}}{s(f_{\mathrm{v}}^{2}-s^{2})}+\frac{\mathrm{i}k_{y}f_{\mathrm{v}}(U'+2f_{\mathrm{h}})}{s^{2}-f_{\mathrm{v}}^{2}}\right]\frac{\mathrm{d}\hat{w}}{\mathrm{d}z}+\\
&\left[\frac{U''(k_{x}s+\mathrm{i}k_{y}f_{\rm{v}})}{s^{2}-f_{\rm{v}}^{2}}+\frac{k_{y}\left(U'+f_{\mathrm{h}}\right)\left(k_{y}f_{\rm{h}}s+\mathrm{i}k_{x}f_{\rm{v}}U'\right)}{s(f_{\rm{v}}^{2}-s^{2})}+\frac{k^{2}s^{2}}{s^{2}-f_{\rm{v}}^{2}}\right]\hat{w},
\end{aligned}
\end{equation}
\begin{equation}
\label{eq:lse_2ndODE_2}
\begin{aligned}
&\frac{\mathrm{d}^{2}\hat{T}}{\mathrm{d}z^{2}}=\frac{1}{\kappa_{0}}\left[\frac{f_{\mathrm{v}}U'\left(k_{x}f_{\mathrm{v}}-\mathrm{i}sk_{y}\right)}{N^{2}k^{2}s}\frac{\mathrm{d}\hat{w}}{\mathrm{d}z}+\left(1+\frac{\mathrm{i}k_{x}k_{y}f_{\rm{v}}U'(U'+f_{\rm{h}})}{N^{2}k^{2}s}\right)\hat{w}\right]\\
&+\left(k^{2}+\frac{\mathrm{i}s}{\kappa_{0}}\right)\hat{T},
\end{aligned}
\end{equation}
where $s=-\mathrm{i}\sigma+k_{x} U$ is the Doppler-shifted frequency \citep[][]{Park2013PoF}.
The above equations can further be simplified into a single fourth-order ordinary differential equation (ODE) in terms of $\hat{w}$ as follows:
\begin{eqnarray}
\label{eq:lse_4thODE}
	&&(s^{2}-f_{\mathrm v}^{2})\frac{{\mathrm d}^{2}\hat{w}}{{\mathrm d} z^{2}}+2f_{\mathrm{v}}\left[\frac{k_{x}f_{\mathrm{v}}U'}{s}-\mathrm{i}k_{y}\left(U'+f_{\mathrm{h}}\right)\right]\frac{{\mathrm d}\hat{w}}{{\mathrm d}z}\nonumber\\
	&&+\left[k^{2}(N^{2}-s^{2})-U''\left(k_{x}s+\mathrm{i}k_{ y}f_{\mathrm v}\right)\right.\nonumber\\
	&&+\left.k_{y}(U'+f_{\mathrm h})\left(\frac{2\mathrm{i}k_{x}f_{\mathrm v}U'}{s}+k_{y}f_{\mathrm h}\right)\right]\hat{w}\nonumber\\
	&&=\!\kappa_{0}\left(W_{4}\frac{{\mathrm d}^{4}\hat{w}}{{\rm d}z^{4}}+W_{3}\frac{{\rm d}^{3}\hat{w}}{{\rm d}z^{3}}+W_{2}\frac{{\rm d}^{2}\hat{w}}{{\rm d}z^{2}}+W_{1}\frac{{\rm d}\hat{w}}{{\rm d}z}+W_{0}\hat{w}\right)\!,\nonumber\\
\end{eqnarray}
where $\left\{W_{i}\right\}_{i\in\left[0,4\right]}$ are the operators as follows:
\begin{equation}
\begin{aligned}
&W_{4}=\mathrm{i}\left(\frac{f_{\rm v}^{2}}{s}-s\right),~~
W_{3}=-\mathrm{i}k_{x}U'\left(2+\frac{3f_{\rm v}^{2}}{s^{2}}\right)-\frac{k_{y}f_{\rm v}}{s}\left(U'+2f_{\rm h}\right),\\
&W_{2}=\mathrm{i}k^{2}s\left(2-\frac{f_{\rm v}^{2}}{s^{2}}\right)-\frac{3f_{\rm v}U''}{s^{2}}(\mathrm{i}k_{x}f_{\rm v}+k_{y}s)-\frac{\mathrm{i}k_{y}^{2}f_{\rm h}(U'+f_{\rm h})}{s}\\
&+\frac{k_{x}f_{\rm v}U'}{s^{2}}\left(5k_{y}f_{\rm h}+3k_{y}U'+\frac{6\mathrm{i}k_{x}f_{\rm v}U'}{s}\right),\\
&W_{1}=k^{2}\left[\mathrm{i}k_{x}U'\left(2+\frac{f_{\rm v}^{2}}{s^{2}}\right)+\frac{k_{y}f_{\rm v}(U'+2f_{\rm h})}{s}\right]\\
&+U'''\left[\mathrm{i}k_{x}\left(2-\frac{f_{\rm v}^{2}}{s^{2}}\right)-\frac{3k_{y}f_{\rm v}}{s}\right]+\frac{2\mathrm{i}k_{x}k_{y}^{2}f_{\rm h}U'(U'+f_{\rm h})}{s^{2}}\\
&+U''\left(\frac{6\mathrm{i}k_{x}^{2}f_{\rm v}^{2}U'}{s^{3}}+\frac{9k_{x}k_{y}f_{\rm v}U'}{s^{2}}+\frac{4k_{x}k_{y}f_{\rm v}f_{\rm h}}{s^{2}}-\frac{2\mathrm{i}k_{y}^{2}f_{\rm h}}{s}\right)\\
&-\frac{2k_{x}^{2}f_{\rm v}U'^{2}}{s^{3}}\left(4k_{y}f_{\rm h}+3k_{y}U'+\frac{3\mathrm{i}k_{x}f_{\rm v}U'}{s}\right),\\
&W_{0}=k^{2}\left[\frac{U''k_{y}f_{\rm v}}{s}-\mathrm{i}k^{2}s-(U'+f_{\rm h})\left(\frac{k_{x}k_{y}f_{\rm v}U'}{s^{2}}-\frac{\mathrm{i}k_{y}^{2}f_{\rm h}}{s}\right)\right]\\
&+U''''\left(\mathrm{i}k_{x}-\frac{k_{y}f_{\rm v}}{s}\right)+\frac{k_{y}U'''}{s^{2}}\left[{4k_{x}f_{\rm v}U'}+{f_{\rm h}(k_{x}f_{\rm v}-\mathrm{i}k_{y}s)}\right]\\
&+\frac{k_{x}k_{y}U''}{s^{2}}\left({3\mathrm{i}k_{y}f_{\rm h}U'}-\frac{6k_{x}f_{\rm v}f_{\rm h}U'}{s}+3f_{\rm v}U''+{\mathrm{i}k_{y}f_{\rm h}^{2}}\right.\\
&\left.-\frac{12k_{x}f_{\rm v}U'^{2}}{s}\right)+\frac{2k_{x}^{2}k_{y}U'^{2}(U'+f_{\rm h})}{s^{3}}\left(\frac{3k_{x}f_{\rm v}U'}{s}-\mathrm{i}k_{y}f_{\rm h}\right).
\end{aligned}
\end{equation}
If we also consider non-diffusive fluids (i.e., $\kappa_{0}=0$), the right-hand side of the ODE (\ref{eq:lse_4thODE}) is zero and the equation (\ref{eq:lse_4thODE}) becomes a second-order ODE. 
    
\section{Numerical results on vertical shear instabilities}
\label{sec:LSA}
%----------------------------------------------------------------- 
   \begin{figure*}
   \centering
   \includegraphics[height=5cm]{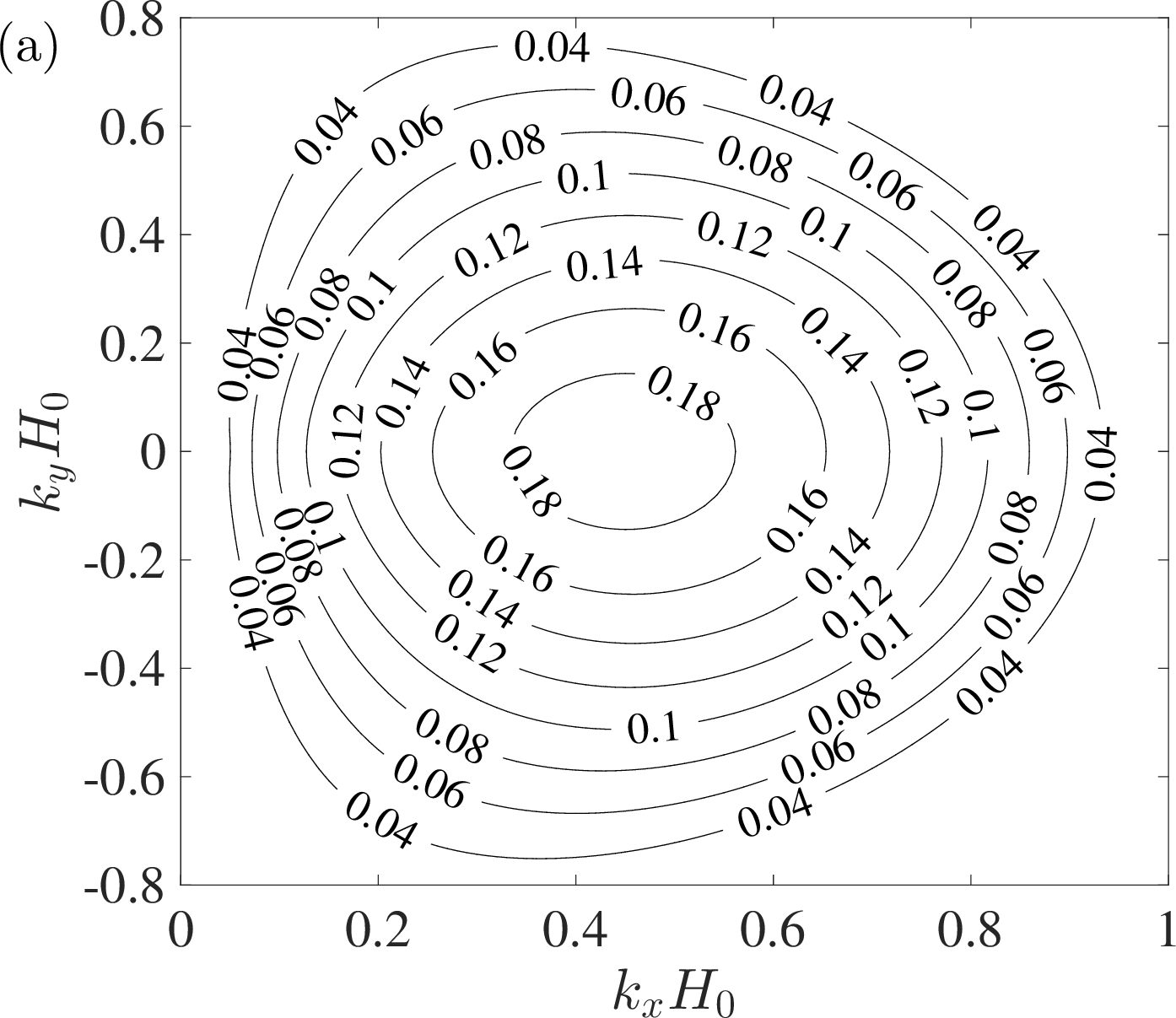}
      \includegraphics[height=5cm]{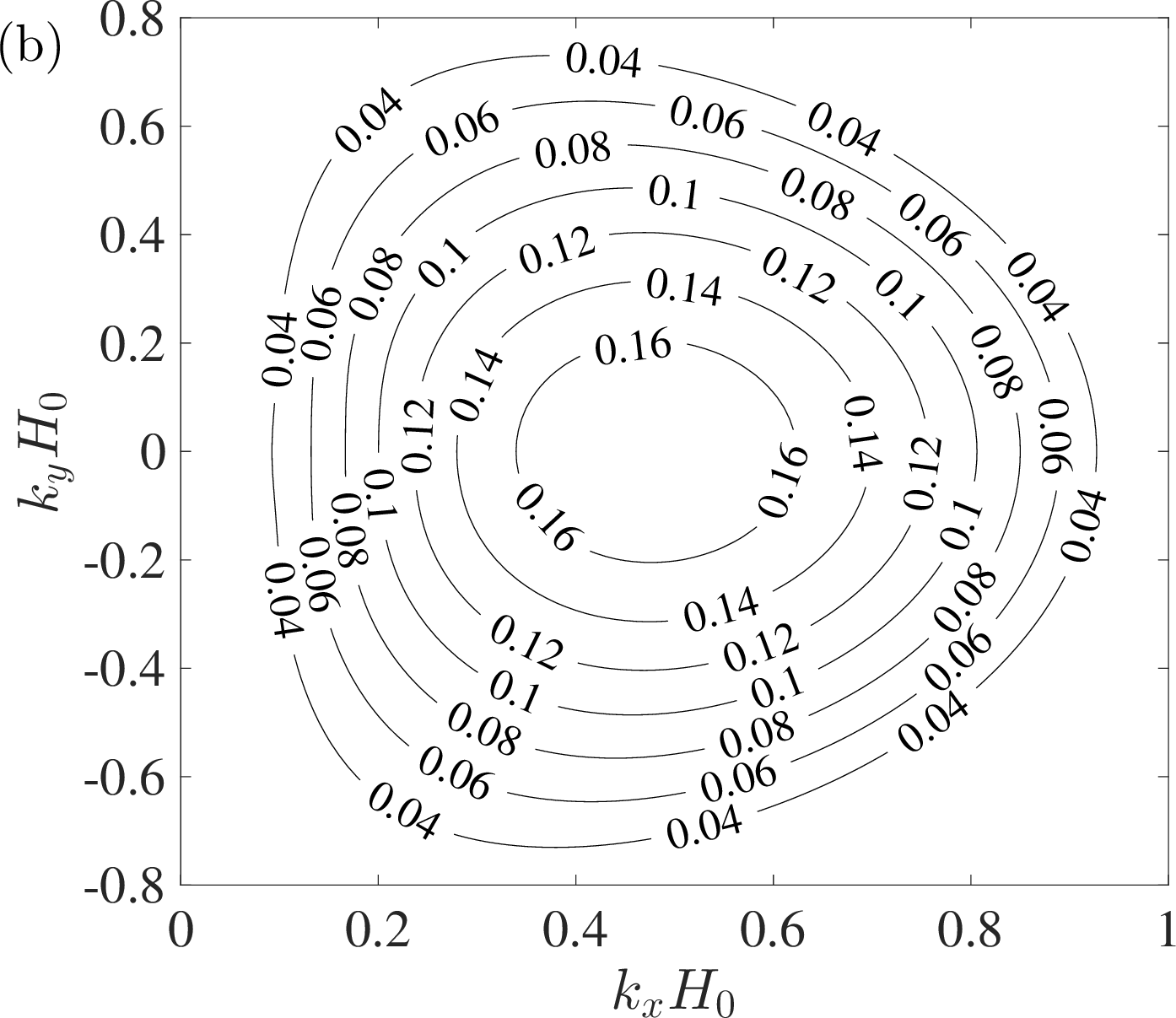}
         \includegraphics[height=5cm]{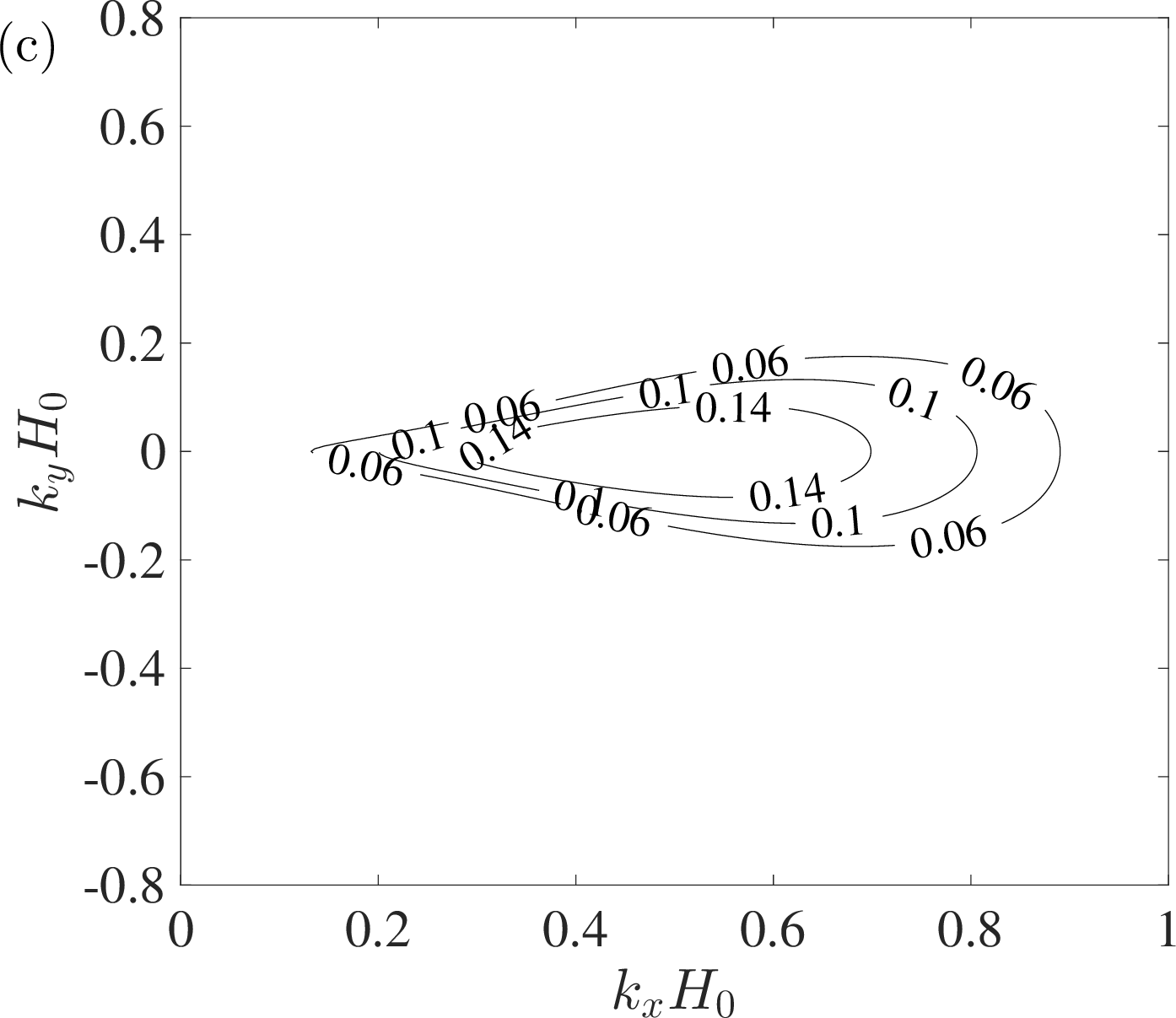}
      \includegraphics[height=5.05cm]{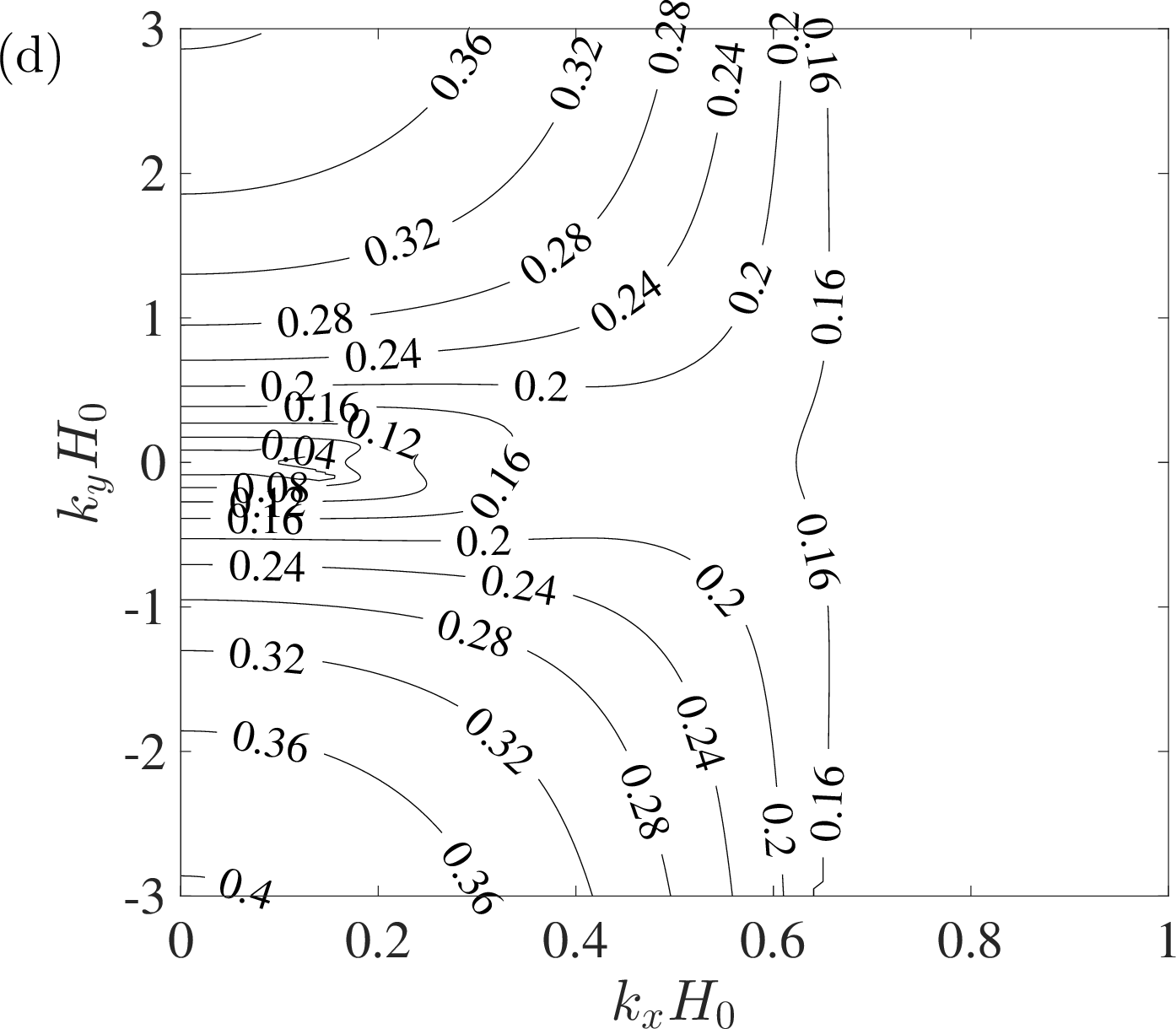}
         \includegraphics[height=5cm]{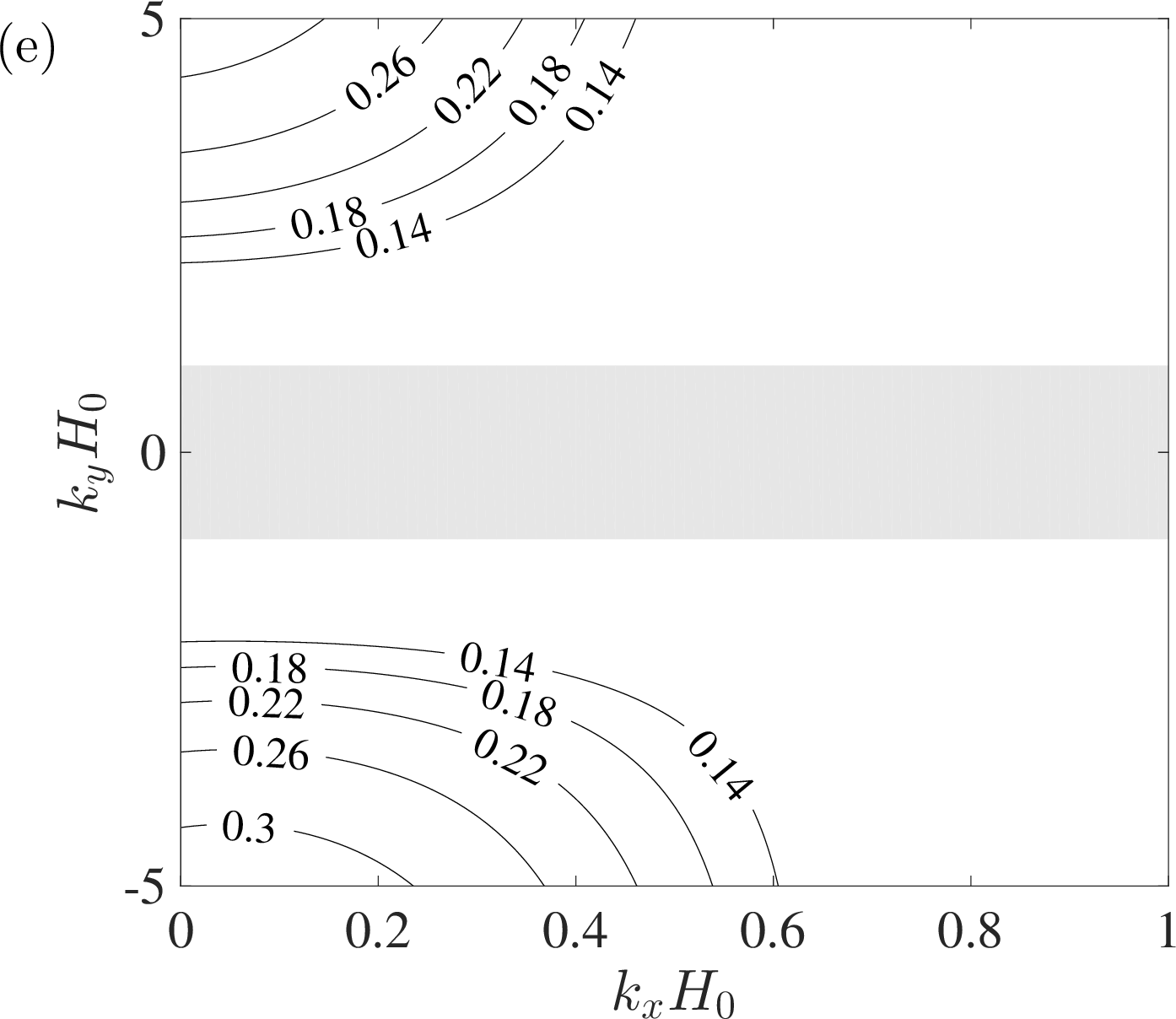}
      \includegraphics[height=5cm]{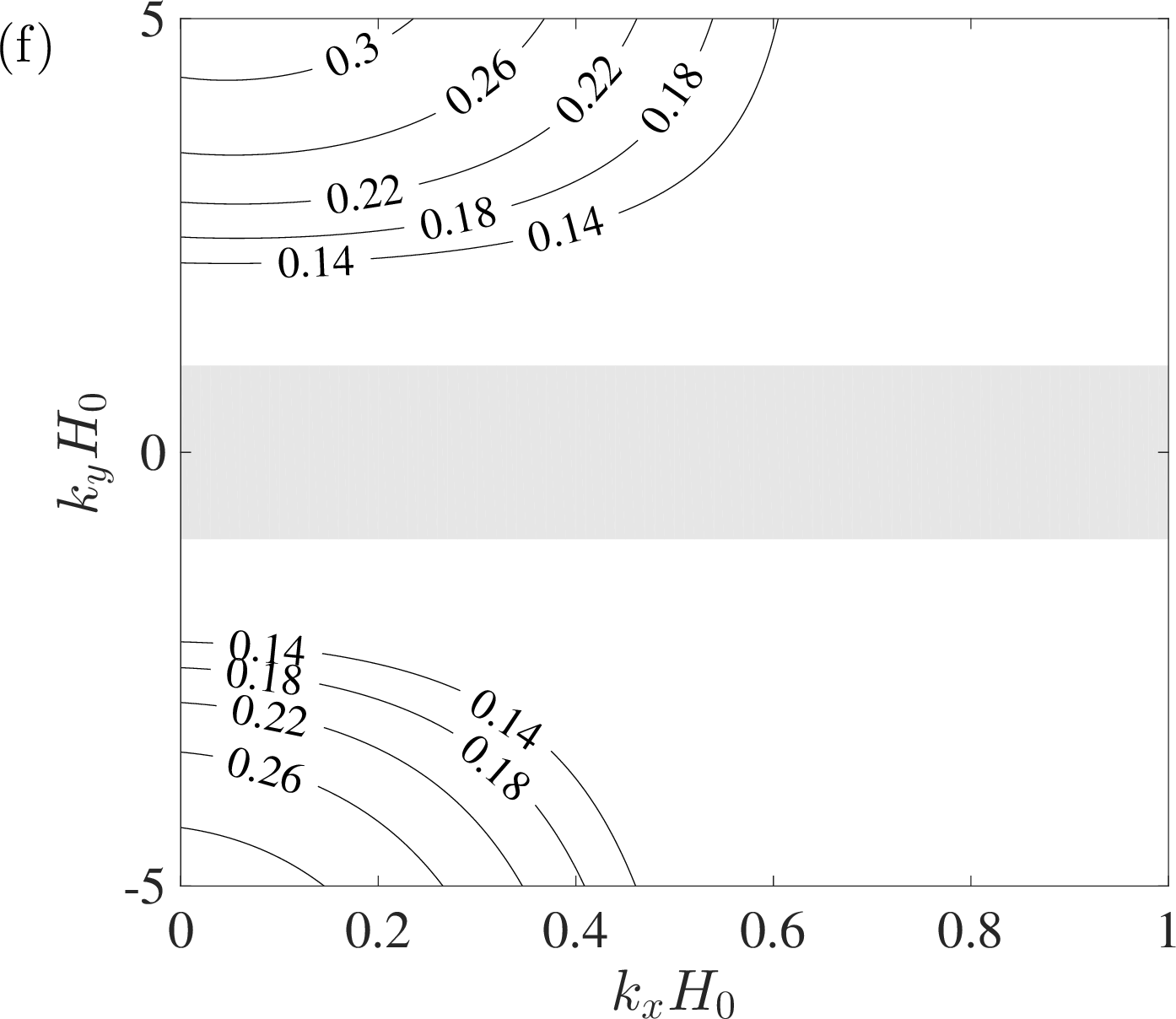}
      \caption{Contours of the growth rate $\sigma_{r}/S_{0}$ in the parameter space of wavenumbers $(k_{x},k_{y})$ for a highly diffusive and inviscid fluid with $Pe=0.01$ and $Re=\infty$ for (a) $(Ri,R_{\mathrm{h}},R_{\mathrm{v}})=(0.01,\infty,\infty)$ (a non-rotating case), (b) $(Ri,R_{\mathrm{h}},R_{\mathrm{v}})=(1,\infty,\infty)$ (a non-rotating case), (c) $(Ri,R_{\mathrm{h}},R_{\mathrm{v}})=(1,1,\infty)$ (a rotating case at the equator $\theta=90^{\circ}$ with $\Omega_{0}/S_{0}=0.5$), (d) $(Ri,R_{\mathrm{h}},R_{\mathrm{v}})=(1,-2,\infty)$ (a rotating case at the equator $\theta=90^{\circ}$ with $\Omega_{0}/S_{0}=-0.25$), (e) $(Ri,R_{\mathrm{h}},R_{\mathrm{v}})=(1,\infty,1)$ (a rotating case either at the northern pole $\theta=0^{\circ}$ with $\Omega_{0}/S_{0}=0.5$ or at the southern pole $\theta=180^{\circ}$ with $\Omega_{0}/S_{0}=-0.5$), and (f) $(Ri,R_{\mathrm{h}},R_{\mathrm{v}})=(1,\infty,-1)$ (a rotating case either at the northern pole $\theta=0^{\circ}$ with $\Omega_{0}/S_{0}=-0.5$ or at the southern pole $\theta=180^{\circ}$ with $\Omega_{0}/S_{0}=0.5$). 
      In panels e and f, the grey-shaded area denotes the range $\left|k_{y}H_{0}\right|\leq\left|RiR_{\mathrm{v}}\right|^{-1}$ of (\ref{eq:wide_jet_approximation_ky_nondimensional}) where the wide-jet approximation is not valid. 
      }
         \label{Fig_growth_contours}
   \end{figure*}
%-----------------------------------------------------------------
In this section, we present exemplary numerical results from linear stability analysis to understand effects of stratification, thermal diffusion and rotation on vertical shear instabilities.
The combined effects of the stratification and thermal diffusion on the inflectional instability were extensively studied by \citet{Lignieresetal1999}, thus we focus on the effects of rotation with the full Coriolis acceleration on vertical shear instabilities.  
Figure \ref{Fig_growth_contours} shows representative examples of contours of the growth rate $\sigma_{r}$ of the most unstable mode in the wavenumber space $(k_{x},k_{y})$ for various sets of parameters $(Ri,R_{\mathrm{h}},R_{\mathrm{v}})$ at $Pe=0.01$ and $Re=\infty$, the case where the fluid is inviscid and highly diffusive.
For all cases in Figure \ref{Fig_growth_contours}, the frequency $\sigma_{i}$ of the most unstable mode is found to be zero.
In panels a and b, we vary stratification by changing $Ri$ for the non-rotating case $\Omega_{0}=0$ (i.e., $R_{\mathrm{h}}=R_{\mathrm{v}}=\infty$), in which only the inflectional instability exists.  
The growth-rate contours in panel a for $Ri=0.01$ resemble those of the unstratified case at $Ri=0$ in \citet{Deloncle2007} and we find the maximum growth rate as $\sigma_{\max}/S_{0}\simeq0.1894$ at $(k_{x}H_{0},k_{y}H_{0})\simeq(0.45,0)$.
As the Richardson number increases to $Ri=1$ (panel b), the inflectional instability is weakened and the maximum growth rate is reduced as $\sigma_{\max}/S_{0}\simeq0.1767$ which is attained at $(k_{x}H_{0},k_{y}H_{0})\simeq(0.47,0)$. 
In panel c and d, we now consider rotating cases with $f_{\mathrm{h}}\neq0$ and $f_{\mathrm{v}}=0$ (i.e., $R_{\mathrm{v}}=\infty$), the cases at the equator of a star with differential rotation in the radial direction (i.e., shellular rotation).
Such a configuration with vertical (radial) shear and rotation perpendicular to the shear is equivalent to that with horizontal shear under the traditional $f$-plane approximation \citep[][]{Park2020}. 
In light of this consideration, if $f_{\rm{v}}=0$, we expect the hyperbolic-tangent base flow (\ref{eq:base_shear}) is inertially unstable if $-S_{0}<f_{\mathrm{h}}<0$ (i.e., $R_{\mathrm{v}}<-1$) according to the Rayleigh criterion in rotating fluids \citep[][]{Kloosterziel1991,Park2012,Park2020}.
We see in panel c that the cyclonic rotation case with $f_{\mathrm{h}}>0$ has no inertial instability, which occurs for large $k_{y}$, while the inflectional instability is suppressed significantly as we can see that the regime of instability shrinks in the wavenumber space $(k_{x},k_{y})$. 
We note, however, that this cyclonic case has the maximum growth rate $\sigma_{\max}/S_{0}\simeq0.1767$ at $(k_{x}H_{0},k_{y}H_{0})\simeq(0.47,0)$, which is the same as the non-rotating case in panel b.
This invariance of the growth rate for a two-dimensional inflectional instability (i.e., $k_{y}=0$) is due to the fact that the fourth-order ODE (\ref{eq:lse_4thODE}) becomes invariant with $f_{\mathrm{h}}$ if we consider $k_{y}=f_{\mathrm{v}}=0$ as follows:
\begin{equation}
\label{eq:lse_4thODE_reduced}
\begin{aligned}
&s\left\{s\frac{{\mathrm d}^{2}\hat{w}}{{\mathrm d} z^{2}}+\left[k_{x}^{2}\left(\frac{N^{2}}{s}-s\right)-k_{x}U''\right]\hat{w}\right\}+\mathrm{i}\kappa_{0}\left[s\frac{{\mathrm d}^{4}\hat{w}}{{\rm d}z^{4}}+2k_{x}U'\frac{{\rm d}^{3}\hat{w}}{{\rm d}z^{3}}\right.\\
&\left.-2k_{x}^{2}s\frac{{\rm d}^{2}\hat{w}}{{\rm d}z^{2}}-2k_{x}\left(k_{x}^{2}U'+U'''\right)\frac{{\rm d}\hat{w}}{{\rm d}z}+k_{x}\left(k_{x}^{3}s-U''''\right)\hat{w}\right]=0.
\end{aligned}
\end{equation}
We see that the equation is no longer dependent on $f_{\mathrm{h}}$.

For the anticyclonic case with $f_{\mathrm{h}}/S_{0}=-0.5$ (i.e., $R_{\mathrm{v}}=-2$), the flow is inertially unstable and thus both the inertial and inflectional instabilities co-exist.
From the growth rate contours in panel d, we found that the growth rate reaches its local maximum $\sigma/S_{0}\simeq0.1767$ at $(k_{x}H_{0},k_{y}H_{0})\simeq(0.47,0)$ corresponding to the inflectional instability while the global maximum growth rate $\sigma_{\max}/S_{0}\simeq0.5$ is attained as $|k_{y}|\rightarrow\infty$ at $k_{x}=0$ due to the inertial instability.
The growth rate contours in panel d are reminiscent of those for horizontal shear instabilities in rotating stellar radiation zones with the traditional $f$-plane approximation \citep[][]{Park2020}.
In panels e and f, we now consider rotating cases with $f_{\mathrm{v}}\neq0$ and $f_{\mathrm{h}}=0$ (i.e., $R_{\mathrm{h}}=\infty$), which correspond to configurations at the poles where the vertical shear varies along the direction of the Coriolis vector.
The situation is analogous to horizontal shear cases at the equator when the full Coriolis acceleration is taken into account \citep[][]{Park2021}. 
For vertical shear at the poles, the classical Rayleigh criterion based on the traditional $f$-plane approximation is no longer valid and thus we need to derive a new criterion for the inertial instability, which is presented in Sect.~\ref{sec:WKBJ}. 
In panels e and f where the vertical Rossby numbers are non-zero as $R_{\mathrm{v}}=\pm 1$, we delimit the zones where the wide jet approximation is not valid for small $|k_{y}|$ as $|k_{y}H_{0}|<|k_{y,\min}H_{0}|=1/|RiR_{\mathrm{v}}|$ according to (\ref{eq:wide_jet_approximation_ky_nondimensional}).
Unlike other cases in panels a--d, there is no inflectional instability in panels e and f in the parameter space of validity $|k_{y}|\geq|k_{y,\min}|$ and only the inertial instability is found for sufficiently large $k_{y}$.
For the positive vertical Rossby number case $R_{\mathrm{v}}=1$, the growth rate tends to be larger in the lower half ($k_{y}<0$) while the maximum growth rate $\sigma_{\max}/S_{0}\simeq 0.455$ is, however, attained in both limits: $k_{y}\rightarrow\pm\infty$ at $k_{x}=0$.
The growth rate at the negative vertical Rossby number $R_{\mathrm{v}}=-1$ is larger in the upper half ($k_{y}>0$) while the maximum growth rate $\sigma_{\max}/S_{0}\simeq 0.455$ is attained as $|k_{y}|\rightarrow\infty$ at $k_{x}=0$. 
Growth-rate contours in panels e and f with $R_{\mathrm{v}}=\pm1$ are mirrored by the line $k_{y}=0$ since the fourth-order ODE (\ref{eq:lse_4thODE}) is invariant with the substitution $(k_{y},f_{\mathrm{v}})\rightarrow(-k_{y},f_{\mathrm{v}})$ when $f_{\mathrm{h}}=0$.

\subsection{Inflectional instability}
%----------------------------------------------------------------- 
   \begin{figure*}
   \centering
   \includegraphics[height=3.9cm]{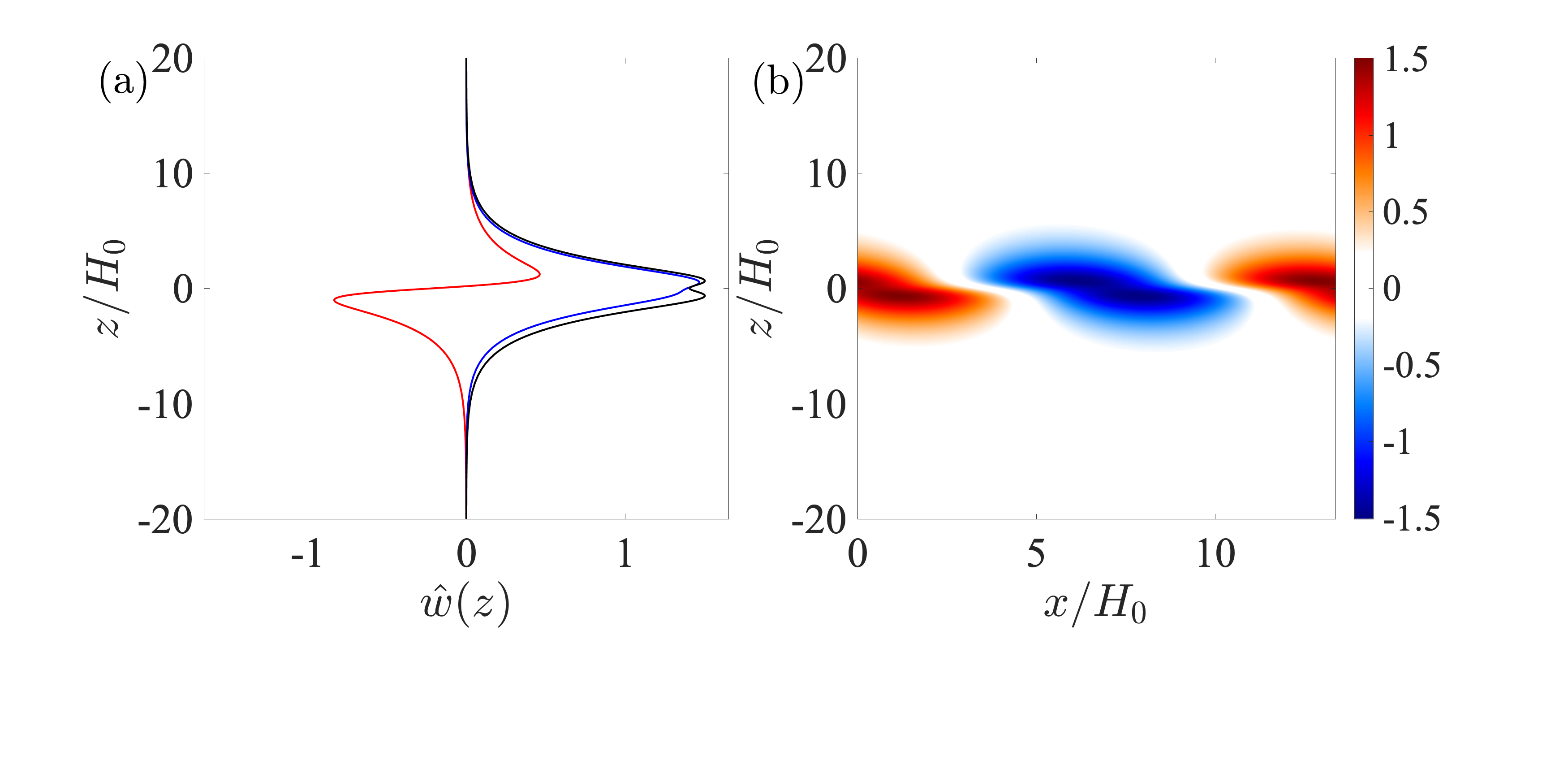}
      \includegraphics[height=3.9cm]{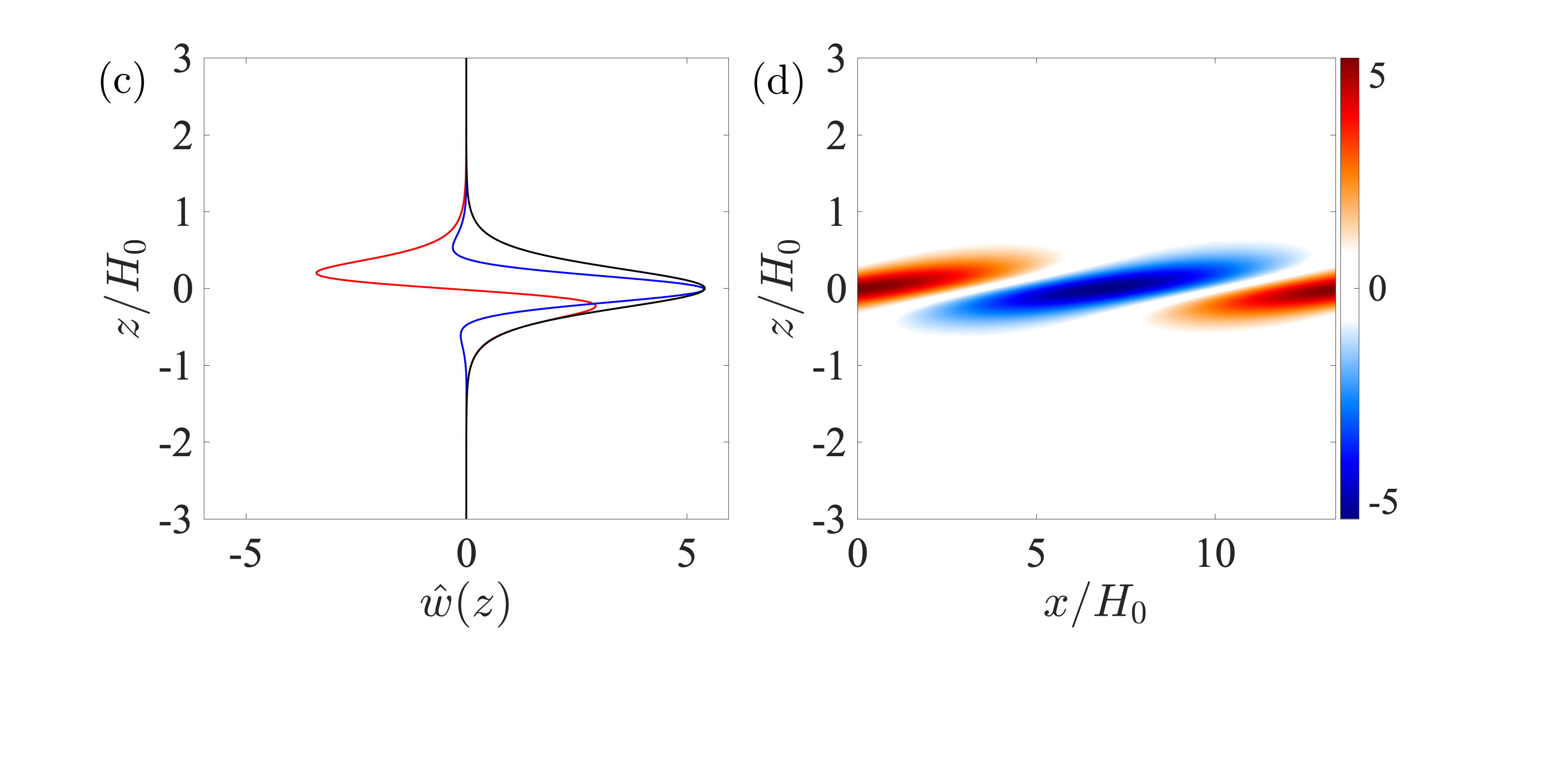}
              \caption{ (a,c) Eigenfunction $\hat{w}(z)$ and (b,d) perturbation $w(x,z)$ for $Pe=0.01$, $Re=\infty$, $Ri=1$, $(R_{\mathrm{h}},R_{\mathrm{v}})=(-2,\infty)$ (i.e. $\theta=90^{\circ}$ and $\Omega_{0}/S_{0}=-0.25$) for (a,b) the inflectional-instability mode at $(k_{x}H_{0},k_{z}H_{0})=(0.47,0)$ and (c,d) the inertial-instability mode at $(k_{x}H_{0},k_{z}H_{0})=(0.47,5)$. In (a) and (c), blue, red, and black lines denote the real, imaginary, and absolute parts of the eigenfunction $\hat{w}$, respectively.  
      }
         \label{Fig_modes}
   \end{figure*}
Panels a and b in Fig.~\ref{Fig_modes} show the eigenfunction $\hat{w}(z)$ and the corresponding perturbation $w(x,z)=\mathrm{Re}\left[\hat{w}(z)\exp(\mathrm{i}k_{x}x)\right]$ (where $\mathrm{Re}$ denotes the real part) for $Pe=0.01$, $Re=\infty$, $Ri=1$, $(R_{\mathrm{h}},R_{\mathrm{v}})=(-2,\infty)$ and $(k_{x}H_{0},k_{y}H_{0})=(0.47,0)$.
The perturbation at these wavenumbers corresponds to the inflectional instability mode that has a characteristic of being tilted against the base shear flow $U(z)$.
The tilted feature is responsible for the growth via the Orr mechanism and is similarly observed for the inflectional instability in other types of shear flows including horizontal shear flows \citep[][]{Park2020,Park2021}. 

\begin{figure*}
   \centering
   \includegraphics[height=4.6cm]{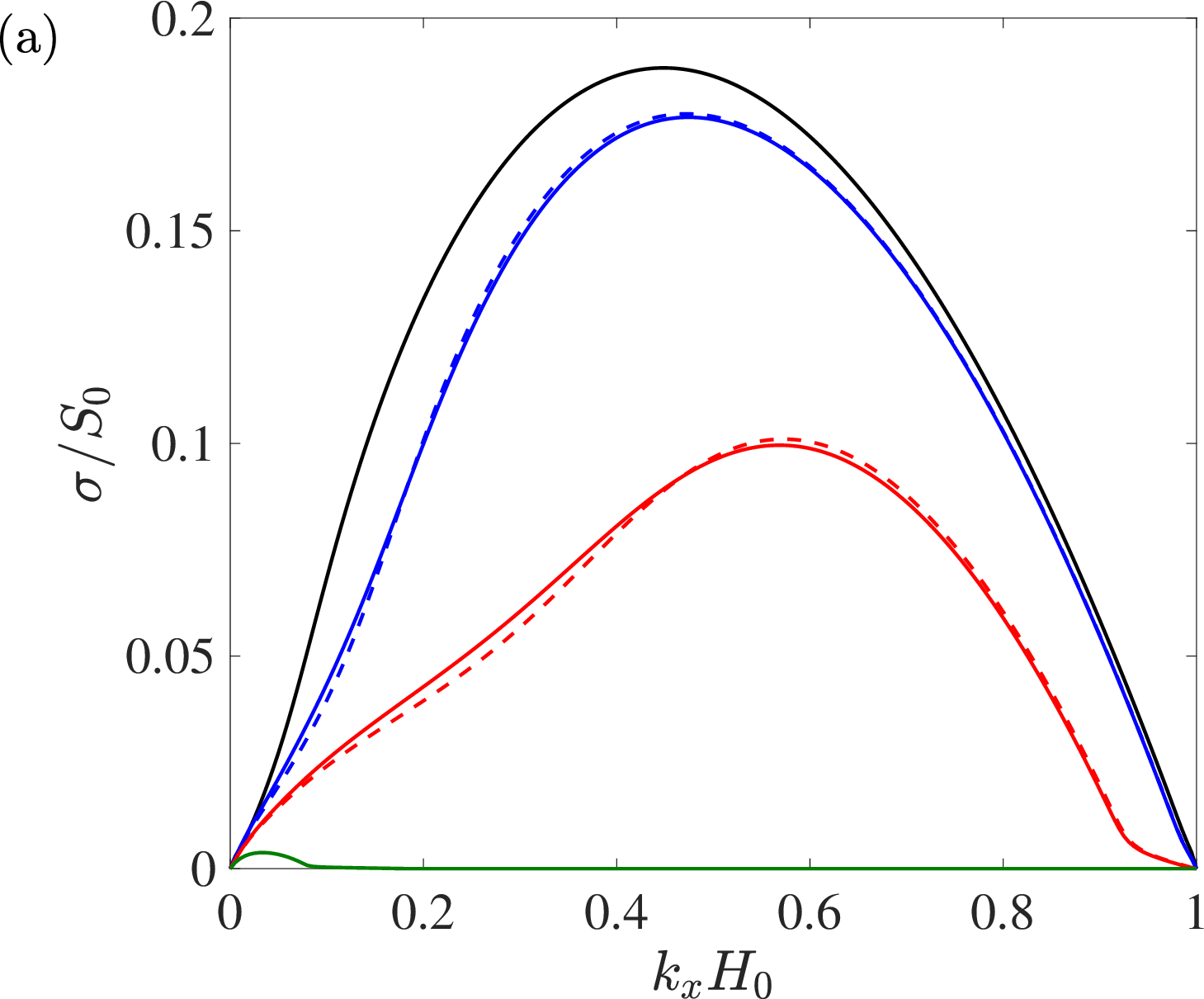}
    \includegraphics[height=4.6cm]{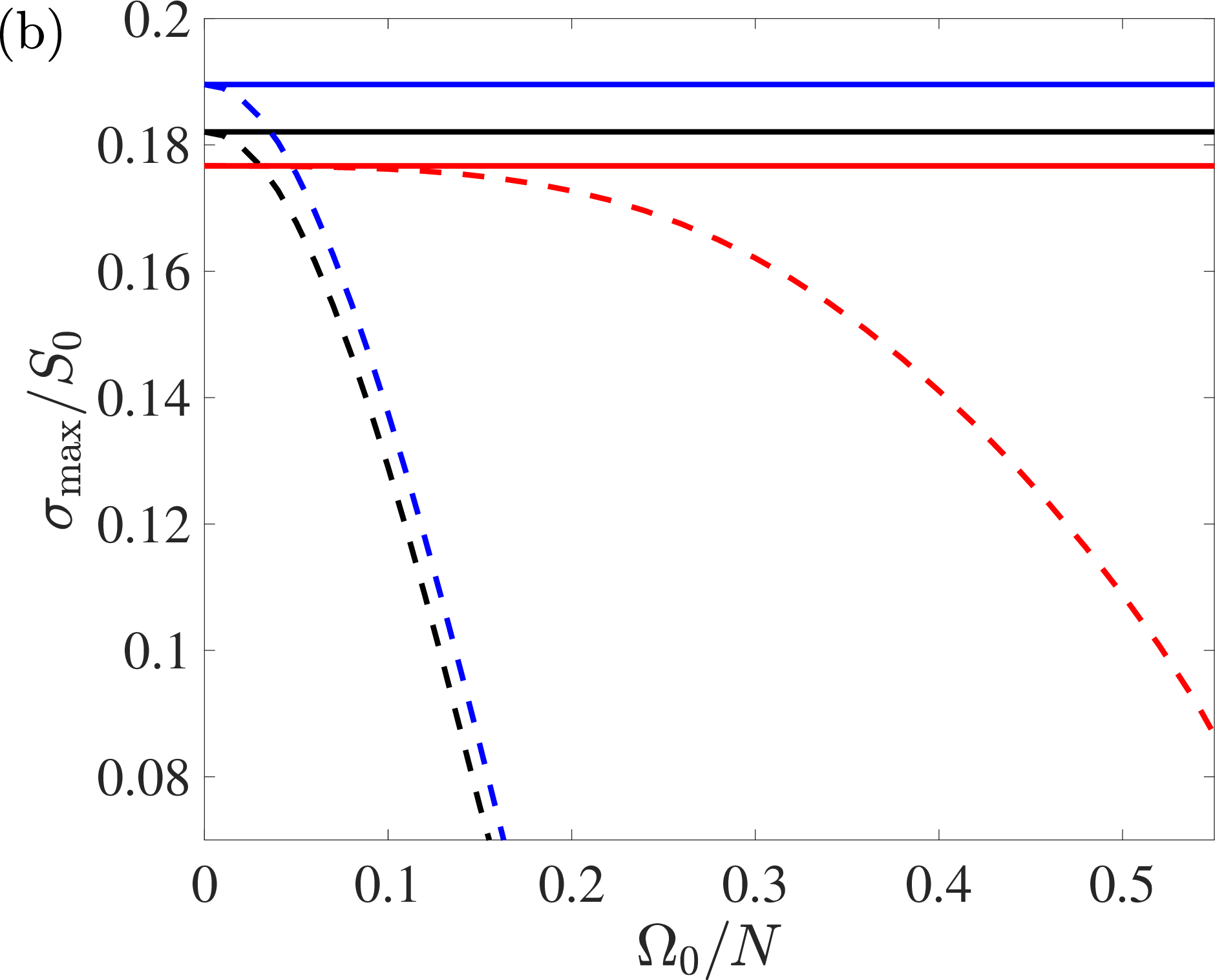}
    \includegraphics[height=4.6cm]{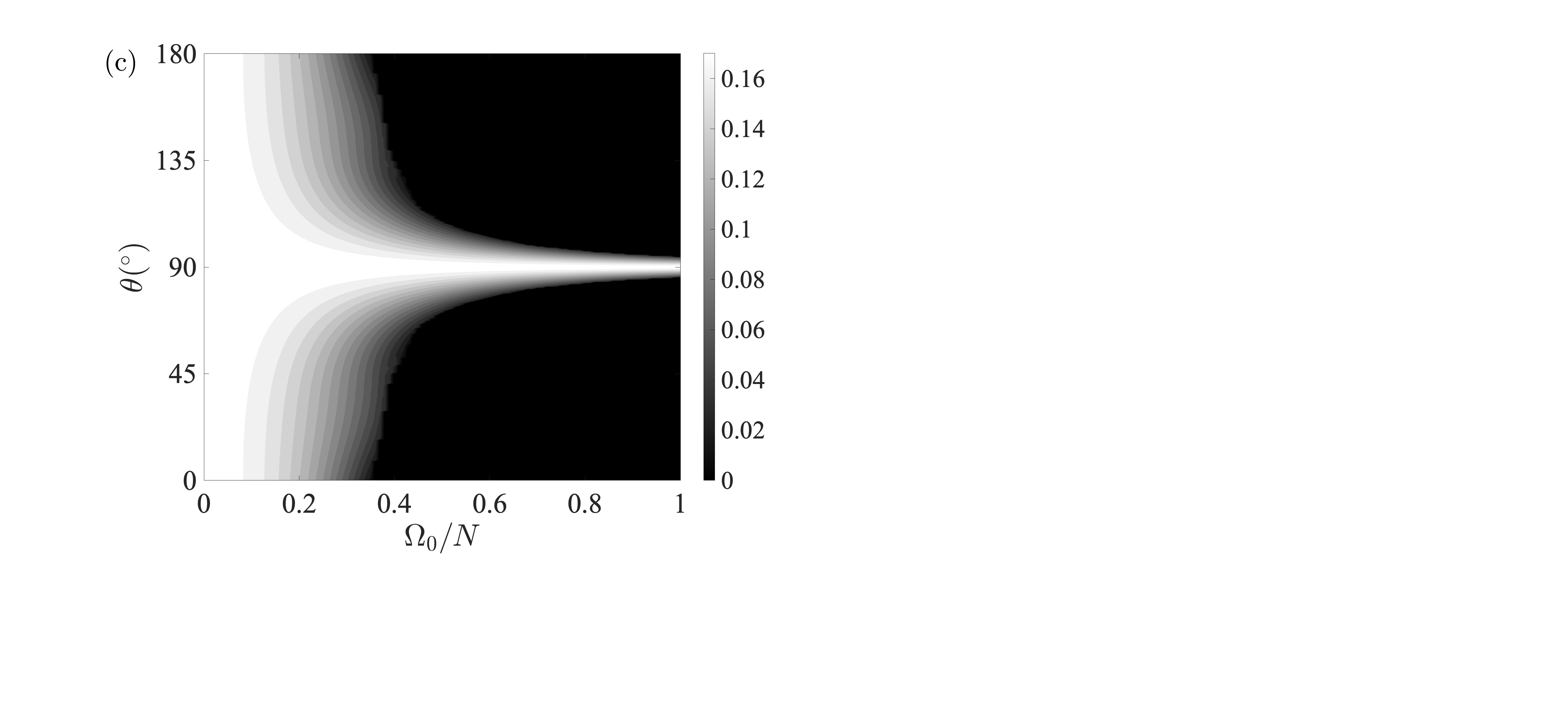}
              \caption{ (a) Growth rate $\sigma$ of the inflectional instability versus wavenumber $k_{x}$ at $k_{y}=0$, $\Omega_{0}=0$ and $Re=\infty$ for various sets of parameters $(Pe,Ri)$: (0.01,0.1) (black solid), (0.01,1) (blue solid), (0.01,10) (red solid), (0.1,0.1) (blue dashed), (0.1,1) (red dashed), and (1,1) (green solid). 
              (b) Maximum inviscid growth rate $\sigma_{\max}$ versus the ratio $\Omega_{0}/N$ at the equator (solid lines) and $\theta=100^{\circ}$ (dashed lines) for $(Ri,Pe)=(0.01,\infty)$ (black), $(Ri,Pe)=(0.01,0.01)$ (blue) and $(Ri,Pe)=(1,0.01)$ (red).   
              (c) Contours of the maximum growth rate $\sigma_{\max}/S_{0}$ in the parameter space ($\theta$, $\Omega_{0}/N$) at $Ri=1$, $Pe=0.01$ and $Pr=10^{-6}$.  
 %             (b) Growth rate $\sigma$ of the first mode of the inertial instability versus wavenumber $k_{y}$ for $k_{x}=0$ (solid lines) and $k_{x}H_{0}=0.47$ (dashed line) at $Re=\infty$, $Pe=0.01$, and $Ri=1$ for various sets of parameters $(\Omega_{0}/S_{0},\theta)$: $(-0.25,90^{\circ})$ (black), $(-0.25,45^{\circ})$ (blue), $(-0.5,0^{\circ})$ (red), $(-0.25,0^{\circ})$ (green), and $(0.25,135^{\circ})$ (purple). 
 %             Filled circles denote predictions from the WKBJ asymptotic dispersion relation (\ref{eq:WKBJ_ODE4_sigma}).
      }
         \label{Fig_growth_rates}
\end{figure*}
The inflectional instability is present regardless of rotation. % for different parameters of $Pe$, $Ri$ and $\theta$. 
For a non-rotating case $\Omega_{0}=0$ or a rotating case at the equator $\theta=90^{\circ}$ (i.e., $f_{\rm{v}}=0$ and $f_{\rm{h}}\neq0$), the growth rate of the inflectional instability is found to be maximal for two-dimensional modes at $k_{y}=0$ and a non-zero $k_{x}$ in the wavenumber range $0<k_{x}H_{0}<1$ as shown by panels a--c of Fig.~\ref{Fig_growth_contours}. 
In this case, it is found that three-dimensional inflectional instability modes with non-zero $k_{x}$ and $k_{y}$ have smaller growth rates than those of two-dimensional instability modes at $k_{y}=0$ . 
At $\Omega_{0}=k_{y}=0$, Fig.~\ref{Fig_growth_rates}a shows examples of the inviscid growth rate of the inflectional instability versus $k_{x}$ for different values of $Pe$ and $Ri$. 
On the one hand, we see that the growth-rate curves descend as $Ri$ increases at a fixed $Pe=0.01$, which implies that stratification suppresses the inflectional instability.
On the other hand, at a fixed $Ri=1$, the growth-rate curves ascend as $Pe$ decreases implying the promotion of the inflectional instability by the thermal diffusion. 
It is also noteworthy that the growth rates are quantitatively similar for the same $RiPe$ for low $Pe$ (e.g. the case with $(Ri,Pe)=(1,0.01)$ and the other with $(Ri,Pe)=(0.1,0.1)$ where both have $RiPe=0.01$).
These self-similar results in the low-P\'eclet number limit are expected and well described by \citet{Lignieresetal1999} for the non-rotating case.
It is not shown here but, as expected, the viscosity with $\nu_{0}>0$ stabilizes the inflectional instability. 

For rotating cases outside the equator (i.e. $\theta\neq90^{\circ}$ and $f_{\rm{v}}\neq0$), the wide-jet approximation (\ref{eq:wide_jet_approximation_ky_nondimensional}) needs to be taken into account. 
In this case, we found that three-dimensional modes with $k_{y}=k_{y,\min}$ and a non-zero $k_{x}$ in the range $0<k_{x}H_{0}<1$ have the maximum growth rate in the southern hemisphere $90^{\circ}<\theta<180^{\circ}$ while the modes at $k_{y}=-k_{y,\min}$ and for $0<k_{x}H_{0}<1$ have the maximum growth rate in the northern hemishpere $0^{\circ}<\theta<90^{\circ}$.
%As the rotation is taken into account, it is found that the inflectional instability is suppressed by the rotation when the latitude is off the equator, the situation in which the vertical rotation component is non-zero as $|f_{\mathrm{v}}|>0$. 
%In this case, we need to consider $|k_{y}|\geq|k_{y,\min}|$ due to the wide jet approximation and we found that the maximum growth rate $\sigma_{\max}$ over the range of $k_{x}$ is obtained at $k_{y}=k_{y,\min}$ if $\theta>90^{\circ}$ (i.e. in the southern hemisphere) or at $k_{y}=-k_{y,\min}$ if $\theta<90^{\circ}$ (i.e. in the northern hemisphere). 
We pick up the maximum of the growth rate $\sigma_{\max}$ over the range $0<k_{x}H_{0}<1$ at $|k_{y}|=|k_{y,\min}|$ and for a fixed rotation-to-stratification ratio $\Omega_{0}/N$, $Ri$ and $Pe$, and we display in panel b of Fig.~\ref{Fig_growth_rates} the maximum growth rate as a function of $\Omega_{0}/N$ for two cases: one at the equator $\theta=90^{\circ}$ and the other in the southern hemisphere $\theta=100^{\circ}$. 
At the equator, the inflectional instability is independent of the ratio $\Omega_{0}/N$ and thus its growth rate is invariant at a given $Ri$ and $Pe$. 
The inflectional instability is promoted slightly as the thermal diffusion becomes stronger (i.e., as $Pe$ decreases). 
In the southern hemisphere at $\theta=100^{\circ}$, for weakly stratified cases with $Ri=0.01$ at both $Pe=\infty$ and $Pe=0.01$, the instability is strongly suppressed as the ratio $\Omega_{0}/N$ increases. 
At $Pe=0.01$, the instability sustains for a high ratio of $\Omega_{0}/N$ as $Ri$ increases although the growth rate at $\Omega_{0}=0$ is smaller for a larger $Ri$. 
As the latitude is further away from the equator, the inflectional instability tends to be suppressed faster as the ratio $\Omega_{0}/N$ increases. 
This is clearly shown in Fig.~\ref{Fig_growth_rates}c by the contours of the maximum growth rate over the parameter space $(\theta,\Omega_{0}/N)$ at $Ri=1$, $Pe=0.01$ and now in a viscous case with $Pr=10^{-6}$ (i.e., $Re=Pe/Pr=10^{4}$). 
As $\Omega_{0}/N$ increases, the inflectional instability is suppressed as the colatitude $\theta$ is away from the equator while the growth rate is invariant with the ratio $\Omega_{0}/N$ at the equator $\theta=90^{\circ}$.
The growth rate contours are symmetric with respect to the equator.
This result implies that the inflectional instability is maximal near the equator and turbulent energy dissipation, momentum transport and matter mixing induced by the inflectional instability are therefore expected to be localised near the equator of fast-rotating stars and to vanish in polar regions. 
This latitudinal dependence is due to the action of the Coriolis acceleration. When its strength increases, gravito-inertial modes, which can become unstable because of the inflectional instability, become trapped around the equator \citep[e.g.][]{LeeSaio1997,DintransRieutord2000}. 
Therefore, it becomes mandatory to take into account the latitudinal dependences of hydrodynamical instabilities because of the Coriolis acceleration. 
More results on these turbulent processes are discussed in Sect.~\ref{sec:turbulent_viscosity}. 
  
%, the regime where the inflectional instability is suppressed and thus not observed. 
%$f_{\mathrm{v}}=0$ is, therefore, the case where the inflectional instability is the strongest. 
%At $f_{\mathrm{v}}=k_{y}=0$, it is noteworthy that the inflectional instability becomes independent of $f_{\mathrm{h}}$ according to the Eq.~(\ref{eq:lse_4thODE_reduced}).
%For $|k_{y}|>0$, the three-dimensional inflectional instability is suppressed in the inertially-stable regime (i.e., $f_{\mathrm{h}}>0$ or $f_{\mathrm{h}}<-S_{0}$ at $f_{\mathrm{v}}=0$) as shown in Fig.~\ref{Fig_growth_contours}c. 
%All of these suggest that the horizontal Coriolis parameter $f_{\rm{h}}$ can either suppress the inflectional instability if perturbations are three-dimensional (i.e., $k_{x}>0$ and $|k_{y}|>0$) or does not affect the instability if perturbations are two dimensional (i.e., $k_{x}>0$ and $k_{y}=0$).
%This implies that the inflectional instability is most unstable for cases without rotation. 
%$\sigma_{\max}$ computed at $k_{y}=k_{y,c}$.
%\textcolor{blue}{JP: I think this part is okay to remain here, as it connects Figures 3 and 4}.

\subsection{Inertial instability}
Panels c and d in Fig.~\ref{Fig_modes} display the eigenfunction $\hat{w}(z)$ and perturbation $w(x,z)$ at $(k_{x}H_{0},k_{z}H_{0})=(0.47,5)$, the case that corresponds to the inertial instability mode.
The eigenfunction has a wavelike structure around $|z|<1$ while it is evanescent and decreases exponentially as $|z|\rightarrow\infty$. 
This mode corresponds to the first mode that has the fewest zero-crossings; i.e., 1 and 2 zero-crossings in the imaginary and real parts of the eigenfunction $\hat{w}$, respectively. 
There are also higher-order modes with more zero-crossings along $z$ but we focus on the first mode with least zero-crossings since it is the most unstable mode. 
The existence of the inertial instability modes at the first and higher orders will also be discussed with quantization conditions derived by the WKBJ analysis in the next section. 

\begin{figure}
   \centering
      \includegraphics[width=6.5cm]{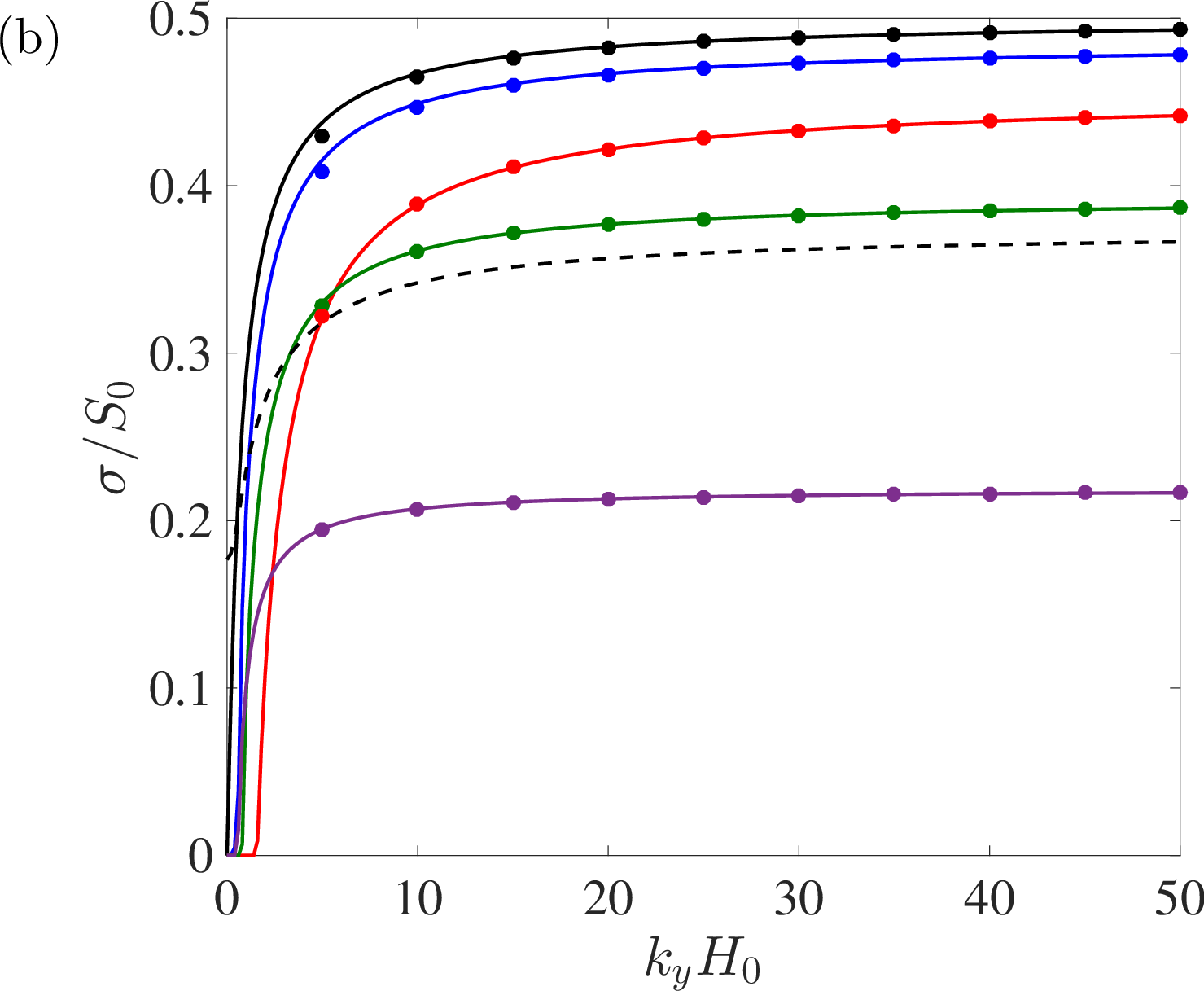}
              \caption{Growth rate $\sigma$ of the first mode of the inertial instability versus wavenumber $k_{y}$ for $k_{x}=0$ (solid lines) and $k_{x}H_{0}=0.47$ (dashed line) at $Re=\infty$, $Pe=0.01$, and $Ri=1$ for various sets of parameters $(\Omega_{0}/S_{0},\theta)$: $(-0.25,90^{\circ})$ (black), $(-0.25,45^{\circ})$ (blue), $(-0.5,0^{\circ})$ (red), $(-0.25,0^{\circ})$ (green), and $(0.25,135^{\circ})$ (purple). 
              Filled circles denote predictions from the WKBJ asymptotic dispersion relation (\ref{eq:WKBJ_ODE4_sigma}).
      }
         \label{Fig_growth_rate_inertial}
\end{figure}
Figure \ref{Fig_growth_rate_inertial} shows examples of the growth rate of the inertial instability at $k_{x}=0$, $Re=\infty$, $Pe=0.01$, and $Ri=1$ for various $\Omega_{0}$ and $\theta$. 
The anti-cyclonic rotation with $\Omega_{0}/S_{0}=-0.25$ at the equator $\theta=90^{\circ}$ (i.e., $f_{\mathrm{h}}/S_{0}=-0.5$ and $f_{\mathrm{v}}=0$, the case corresponding to the traditional $f$-plane approximation) drives the inertial instability. 
The growth rate of the inertial instability increases with $k_{y}$ and asymptotes as $k_{y}\rightarrow\infty$. 
Such an asymptotic behavior is also observed for non-zero $k_{x}$ cases (see e.g. black-dashed line for the case with $k_{x}H_{0}=0.47$) where the growth rate at $k_{y}=0$ is above zero due to the inflectional instability.
Nevertheless, the overall growth rate at $k_{x}H_{0}=0.47$ is smaller than the growth rate of the case with $k_{x}=0$. 
We note that the asymptotic behavior of the inertial instability as $k_{y}\rightarrow\infty$ allows us to use the WKBJ approximation, as previously applied to many other studies on the inertial instability of shear flows and centrifugal instability of vortices in stably stratified-rotating fluids \citep[][]{Park2013PoF,Park2020}. 
As the co-latitude $\theta$ decreases to zero (i.e., at the northern pole) while keeping the same rotation ratio $\Omega_{0}/S_{0}=-0.25$, we see that the inertial instability is slightly suppressed. 
Such suppression also occurs as the co-latitude $\theta$ increases to $\theta=180^{\circ}$ (i.e., at the southern pole) and it is found that the growth rates of at co-latitudes $\theta$ and $180^{\circ}-\theta$ are the same when the same rotation $\Omega_{0}$ is applied in the low-$Pe$ limit.
We also see in Fig.~\ref{Fig_growth_rate_inertial} how the growth rate curves vary as the magnitude $|\Omega_{0}|$ or the sign of $\Omega_{0}$ change.\\ 

From linear stability analysis in the inviscid limit $\nu_{0}=0$, the parametric dependence is numerically investigated for both the inflectional and inertial instabilities. 
In general, the inflectional instability has the maximum growth rate in the wavenumber range $0<k_{x}H_{0}<1$ at $k_{y}=0$ and tends to be suppressed by the stratification while strong thermal diffusion promotes the inflectional instability up to the level equivalent to the unstratified case. 
At the equator where $f_{\rm{v}}=0$, the three-dimensional inflectional instability with $k_{y}\neq0$ is stabilised by the horizontal component $f_{\rm{h}}$ of the rotation vector while the maximum growth rate attained for two-dimensional perturbation with $k_{y}=0$ is independent of $f_{\rm{h}}$. 
The inflectional instability is suppressed as the co-latitude $\theta$ is off the equator and as the ratio $\Omega_{0}/N$ increases.
For the inertial instability, it is clearly shown that its maximum growth rate is attained as $k_{y}\rightarrow\infty$ at $k_{x}=0$. 
We identified that the maximum growth rate of the inertial instability is higher for certain parameter sets of $\Omega_{0}$, $Pe$ and $Ri$ than the maximum growth rate of the inflectional instability, which is $\sigma_{\max}/S_{0}=0.1807$ obtained for the unstratified case. 
However, understanding the dependence on the rotation $\Omega_{0}$, the co-latitude $\theta$, and other parameters is more difficult for the inertial instability from numerical investigations. 
In the following section, we will apply the WKBJ approximation and derive analytic expressions of the dispersion relation to understand more explicitly the parametric dependence of the inertial instability. 

\section{Inertial instability and WKBJ analysis}
\label{sec:WKBJ}
Thanks to the nature of the inertial instability that occurs for large $k_{y}$, we can perform an asymptotic analysis in the limit $k_{y}\rightarrow\infty$ using the WKBJ approximation to derive analytical expressions of the dispersion relation and understand explicitly how the inertial instability varies with different parameters. 
The WKBJ method is useful to approximate solutions of ordinary differential equations (ODEs) such as Eq.~(\ref{eq:lse_4thODE}), which is obtained at $\nu_{0}=0$. 
The method is efficient to describe fast-oscillating wavelike solutions or evanescent solutions when a small parameter $\delta$ is considered. 
For the inertial instability, the parameter $\delta$ can be chosen by a proper scaling with large $k_{y}$ based on observation of the eigenmode which oscillates faster as $k_{y}$ increases.
We can obtain the WKBJ solutions from (\ref{eq:lse_4thODE}) for two cases: a non-diffusive case with $\kappa_{0}=0$ and a highly-diffusive case as $\kappa_{0}\rightarrow\infty$. 
We only consider for simplicity the zero streamwise wavenumber (i.e., $k_{x}=0$) at which the growth rate of the inertial instability is maximal. 
In the next subsections, the WKBJ solutions for inertial instability modes, asymptotic dispersion relations, and parametric dependence will be discussed case by case.  
 
\subsection{WKBJ analysis for non-diffusive cases with $\kappa_{0}=0$}
For the zero wavenumber $k_{x}=0$ and adiabatic configuration where there is no viscous and thermal diffusion ($\nu_{0}=\kappa_{0}=0$), the fourth-order ODE (\ref{eq:lse_4thODE}) can be simplified into the second-order ODE as
\begin{equation}
\label{eq:WKBJ_2ODE_k0}
\begin{aligned}
&\frac{\mathrm{d}^{2}\hat{w}}{\mathrm{d}z^{2}}+\frac{2\mathrm{i}k_{y}f_{\mathrm{v}}(U'+f_{\mathrm{h}})}{\sigma^{2}+f^{2}_{\mathrm{v}}}\frac{\mathrm{d}\hat{w}}{\mathrm{d}z}\\
&-\left[k_{y}^{2}\frac{\sigma^{2}+N^{2}+f_{\mathrm{h}}(U'+f_{\mathrm{h}})}{\sigma^{2}+f^{2}_{\mathrm{v}}}-\frac{\mathrm{i}k_{y}f_{\mathrm{v}}U''}{\sigma^{2}+f^{2}_{\mathrm{v}}}\right]\hat{w}=0.
\end{aligned}
\end{equation}
To the above ODE, we apply the WKBJ approximation
\begin{equation}
\label{eq:WKBJ}
\hat{w}\sim\exp\left[\frac{1}{\delta}\sum_{l=0}\delta^{l}K_{l}(z)\right],
\end{equation}
\citep[see also,][]{Park2020,Park2021}.
At the leading order, we find that the small parameter $\delta$ scales as $\delta=k_{y}^{-1}$ and the leading-order function $K_{0}(z)$ satisfies the following relation:
\begin{equation}
K'_{0}=-\frac{\mathrm{i}f_{\mathrm{v}}(U'+f_{\mathrm{h}})}{\sigma^{2}+f^{2}_{\mathrm{v}}}\pm\sqrt{\frac{\sigma^{2}+N^{2}+f_{\mathrm{h}}(U'+f_{\mathrm{h}})}{\sigma^{2}+f^{2}_{\mathrm{v}}}-\frac{f_{\mathrm{v}}^{2}\left(U'+f_{\mathrm{h}}\right)^{2}}{(\sigma^{2}+f_{\mathrm{v}}^{2})^{2}}}.
\end{equation}
We recall that $'$ denotes the derivative with respect to $z$. 
To express the WKBJ solutions more concisely, we define the function $\Delta$
\begin{equation}
\Delta=\frac{\sigma^{2}+N^{2}+f_{\mathrm{h}}(U'+f_{\mathrm{h}})}{\sigma^{2}+f^{2}_{\mathrm{v}}}-\frac{f_{\mathrm{v}}^{2}\left(U'+f_{\mathrm{h}}\right)^{2}}{(\sigma^{2}+f_{\mathrm{v}}^{2})^{2}}.
\end{equation}
On the one hand, when $\Delta(z)>0$, we expand $\hat{w}$ as
\begin{eqnarray}
\label{eq:WKBJ_2nd_evanescent}
\lefteqn{\hat{w}=\exp\left[-\frac{\mathrm{i}k_{y}f_{\mathrm{v}}(U+f_{\mathrm{h}}z)}{\sigma^{2}+f^{2}_{\mathrm{v}}}\right]\times}\nonumber\\
&&\left[A_{+}\exp\left(k_{y}\int_{z}\sqrt{\Delta}\mathrm{d}z\right)+A_{-}\exp\left(-k_{y}\int_{z}\sqrt{\Delta}\mathrm{d}z\right)\right]+O(1),\nonumber\\
\end{eqnarray}
where $A_{+}$ and $A_{-}$ are the constant amplitudes to be determined by the boundary conditions. 
The prefactor $\mathcal{P}$ defined as 
\begin{equation}
\mathcal{P}=\exp\left[-\frac{\mathrm{i}k_{y}f_{\mathrm{v}}(U+f_{\mathrm{h}}z)}{(\sigma^{2}+f^{2}_{\mathrm{v}})}\right],
\end{equation}
in the above equation (\ref{eq:WKBJ_2nd_evanescent}) implies that the mode shape $\hat{w}$ is not completely evanescent but has a wavelike behaviour if $f_{\mathrm{v}}\neq0$.
The next-order term of $O(1)$ will be neglected in the analysis.
On the other hand, if $\Delta(z)<0$, the mode shape $\hat{w}$ is wavelike at the leading order as
\begin{eqnarray}
\label{eq:WKBJ_2nd_wavelike}
\lefteqn{\hat{w}=\exp\left[-\frac{\mathrm{i}k_{y}f_{\mathrm{v}}(U+f_{\mathrm{h}}z)}{\sigma^{2}+f^{2}_{\mathrm{v}}}\right]\times}\nonumber\\
&&\left[B_{+}\exp\left(\mathrm{i}k_{y}\int_{z}\sqrt{-\Delta}\mathrm{d}z\right)+B_{-}\exp\left(-\mathrm{i}k_{y}\int_{z}\sqrt{-\Delta}\mathrm{d}z\right)\right],\nonumber\\
\end{eqnarray}
where $B_{+}$ and $B_{-}$ are constants. 

\begin{figure}
   \centering
   \includegraphics[width=5cm]{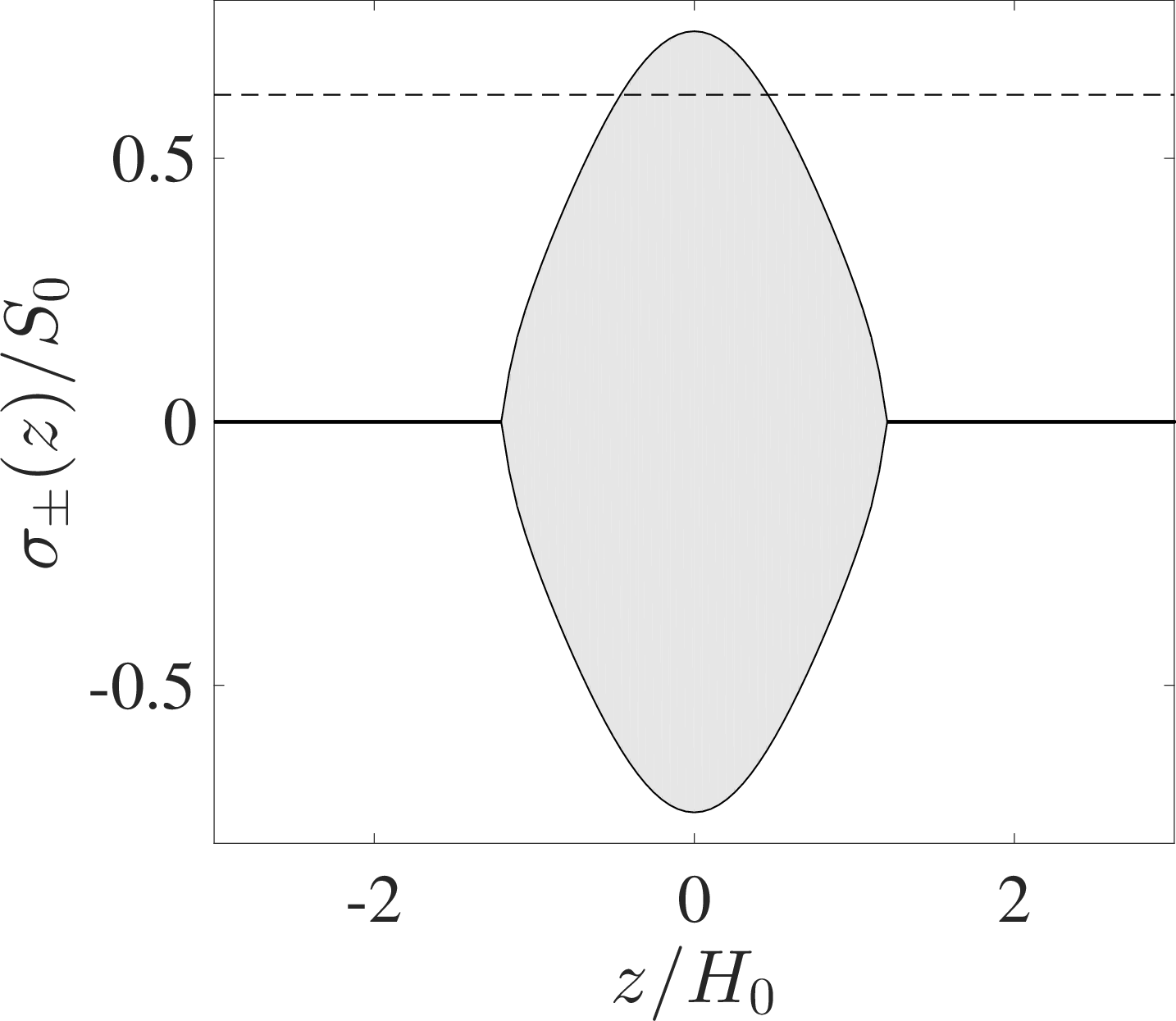}
              \caption{Epicyclic growth rates $\sigma_{\pm}$ (solid lines) for $R_{\mathrm{v}}=1$, $R_{\mathrm{h}}=\infty$, and $Ri=0.1$. 
              White and grey areas denote the regions where $\Delta(z)$ is positive and negative (i.e. the WKBJ solution is evanescent-like and wavelike), respectively. 
              The dashed line indicates the growth rate $\sigma/S_{0}=0.6206$ at $k_{x}=0$ and $k_{y}H_{0}=5$. 
      }
         \label{Fig_epicyclic}
\end{figure}
The two solutions (\ref{eq:WKBJ_2nd_evanescent}) and (\ref{eq:WKBJ_2nd_wavelike}) are distinguishable by the sign of $\Delta$.
Around the turning points $z_{t}$ where $\Delta(z_{t})=0$, the solutions (\ref{eq:WKBJ_2nd_evanescent}) and (\ref{eq:WKBJ_2nd_wavelike}) are no longer valid and we need to derive local solutions around the turning points to connect these two WKBJ solutions. 
We facilitate this turning point analysis by substituting $\hat{w}$ with a new variable $\hat{W}=\hat{w}/\mathcal{P}$ and implementing $\hat{W}$ into (\ref{eq:WKBJ_2ODE_k0}) to obtain the following second-order ODE at the leading order:
\begin{equation}
\label{eq:WKBJ_2ODE_k0_W}
\frac{\mathrm{d}^{2}\hat{W}}{\mathrm{d}z^{2}}-k_{y}^{2}\Delta\hat{W}=0.
\end{equation}
From (\ref{eq:WKBJ_2ODE_k0_W}), we clearly see that $\hat{W}$ can be expressed as an evanescent solution when $\Delta>0$:
\begin{equation}
\label{eq:WKBJ_2nd_evanescent_W}
\hat{W}=A_{+}\exp\left(k_{y}\int_{z}\sqrt{\Delta}\mathrm{d}z\right)+A_{-}\exp\left(-k_{y}\int_{z}\sqrt{\Delta}\mathrm{d}z\right),
\end{equation}
or a wavelike solution when $\Delta<0$:
\begin{equation}
\label{eq:WKBJ_2nd_wavelike_W}
\hat{W}=B_{+}\exp\left(\mathrm{i}k_{y}\int_{z}\sqrt{-\Delta}\mathrm{d}z\right)+B_{-}\exp\left(-\mathrm{i}k_{y}\int_{z}\sqrt{-\Delta}\mathrm{d}z\right).
\end{equation}
To figure out where turning points $z_{t}$ at which $\Delta(z=z_{t})=0$ are located at given parameters, it is practical to introduce the epicyclic growth rates $\sigma_{\pm}(z)$ defined as
\begin{equation}
\label{eq:epicyclic_growth_Rate}
\sigma_{\pm}(z)=\pm\sqrt{\frac{-F+\sqrt{F^{2}+4f_{\mathrm{v}}^{2}\left[U'\left(U'+f_{\rm{h}}\right)-N^{2}\right]}}{2}}
\end{equation}
where $F(z)=f_{\mathrm{v}}^{2}+N^{2}+f_{\mathrm{h}}(U'(z)+f_{\mathrm{h}})$ \citep[see also,][]{Park2012,Park2017}.
The epicyclic growth rates $\sigma_{\pm}$ are obtained after replacing $\sigma$ by $\sigma_{\pm}$ in $\Delta$ to locate where $\Delta=0$ at a given $\sigma$.
For instance, in Fig.~\ref{Fig_epicyclic}, the growth rate $\sigma/S_{0}=0.6206$ (horizontal dashed line), which is obtained for the parameters $R_{\mathrm{v}}=1$, $R_{\mathrm{h}}=\infty$, $Ri=0.1$, $\nu_{0}=\kappa_{0}=k_{x}=0$, and $k_{y}H_{0}=5$, crosses the epicylic growth rate $\sigma_{+}(z)$ at $z/H_{0}=\pm0.453$ and these are the corresponding turning points $z_{t\pm}$.   
The bounded grey region where $\sigma_{-}<\sigma<\sigma_{+}$ denotes the regime where $\Delta(z)<0$, while the regions $\sigma>\sigma_{+}(z)$ and $\sigma<\sigma_{-}(z)$ denote the regimes where $\Delta(z)>0$. 
%Crossings of the dashed line, which indicates the value of the growth rate, and the epicyclic growth rates denote the turning points. 
Therefore, for the growth rate $\sigma/S_{0}=0.6206$, the WKBJ solution of $\hat{W}$ is evanescent when $|z/H_{0}|>0.453$ and is wavelike when $|z/H_{0}|<0.453$.
This configuration in which the wavelike solution is bounded by evanescent solutions can construct an eigenfunction that satisfies the exponentially-decaying solution as $|z|\rightarrow\infty$ \citep[][]{Ledizes2005}. 
This implies that the eigenfunction can be constructed if the growth rate lies in the range $\min(\sigma_{-})<\sigma<\max(\sigma_{+})$.
One of the advantages in using the epicyclic growth rates is that we can easily figure out the growth rate ranges in which the eigenfunction can exist for various sets of parameters of $\left(R_{\mathrm{v}},R_{\mathrm{h}}\right)$ prior to actual stability computation. 

For the case with the growth rate lying in the range $\min(\sigma_{-})<\sigma<\max(\sigma_{+})$, we construct the eigenfunction as follows.
We have the two turning points $z_{t\pm}$ where $z_{t+}>z_{t-}$. 
For $z>z_{t+}$, we consider a solution that decays exponentially as $z\rightarrow\infty$:
\begin{equation}
\label{eq:WKBJ_2nd_ztp}
\hat{W}=A_{+}\exp\left(-k_{y}\int_{z_{t+}}^{z}\sqrt{\Delta(l)}\mathrm{d}l\right).
\end{equation}
Around the turning point $z=z_{t+}$ where the WKBJ approximation (\ref{eq:WKBJ_2nd_ztp}) is no longer valid, we can find a local solution using the following equation obtained by expanding the ODE (\ref{eq:WKBJ_2ODE_k0_W}) around a new coordinate $z_{\epsilon}=(z-z_{t+})/\epsilon$ with a small parameter $\epsilon$:
\begin{equation}
\label{eq:Airy_ztp}
\frac{\mathrm{d}^{2}\hat{W}}{\mathrm{d}z_{\epsilon}^{2}}-\epsilon^{3}k_{y}^{2}\Delta^{'}_{t+}z_{\epsilon}\hat{W}=0,
\end{equation}
where $\Delta^{'}_{t+}=\Delta^{'}(z_{t+})$ is the vertical gradient of $\Delta$ evaluated at $z=z_{t+}$. 
By taking $\epsilon=\left(k_{y}^{2}\Delta^{'}_{t+}\right)^{-1/3}$, the local equation (\ref{eq:Airy_ztp}) becomes the Airy equation and $\hat{W}$ can be expressed as the sum of the Airy functions:
\begin{equation}
\label{eq:Airy_ztp_solution}
\hat{W}=a_{1}\mathrm{Ai}(z_{\epsilon})+b_{1}\mathrm{Bi}(z_{\epsilon}),
\end{equation}
where $a_{1}$ and $b_{1}$ are constants \citep[][]{Abramowitz}.
Considering the asymptotic behaviors of the Airy functions as $z_{\epsilon}\rightarrow\pm\infty$ and matching the local solution (\ref{eq:Airy_ztp_solution}) in the limits $z_{\epsilon}\rightarrow\pm\infty$ with the WKBJ solution (\ref{eq:WKBJ_2nd_ztp}) in the limit $z\rightarrow z_{t+}$, we impose $b_{1}=0$ and find the wavelike WKBJ solution in the region $z_{t-}<z<z_{t+}$ as
\begin{equation}
\label{eq:WKBJ_2nd_ztpztm}
\hat{W}=B_{+}\exp\left(\mathrm{i}k_{y}\int^{z_{t+}}_{z}\sqrt{-\Delta(l)}\mathrm{d}l\right)+B_{-}\exp\left(-\mathrm{i}k_{y}\int^{z_{t+}}_{z}\sqrt{-\Delta(l)}\mathrm{d}l\right),
\end{equation}
where the constants $B_{\pm}$ satisfy
\begin{equation}
B_{+}=\exp\left(-\mathrm{i}\frac{\pi}{4}\right)A_{+}~~\mathrm{and}~~B_{-}=\exp\left(\mathrm{i}\frac{\pi}{4}\right)A_{+}
\end{equation}
\citep[see also,][]{Olver}.
We re-write the solution (\ref{eq:WKBJ_2nd_ztpztm}) as
\begin{equation}
\label{eq:WKBJ_2nd_ztpztm_2nd}
\hat{W}=C_{+}\exp\left(\mathrm{i}k_{y}\int_{z_{t-}}^{z}\sqrt{-\Delta(l)}\mathrm{d}l\right)+C_{-}\exp\left(-\mathrm{i}k_{y}\int_{z_{t-}}^{z}\sqrt{-\Delta(l)}\mathrm{d}l\right),
\end{equation}
where $C_{+}$ and $C_{-}$ are constants that are phase-shifted from $B_{+}$ and $B_{-}$ with satisfying the following relation: 
\begin{equation}
\frac{B_{-}}{C_{+}}=\frac{C_{-}}{B_{+}}=\exp\left(\mathrm{i}k_{y}\int_{z_{t-}}^{z_{t+}}\sqrt{-\Delta(l)}\mathrm{d}l\right).
\end{equation}
A similar turning point analysis can be performed around the turning point $z_{t-}$ to connect the WKBJ solution (\ref{eq:WKBJ_2nd_ztpztm_2nd}) in the region $z_{t-}<z<z_{t+}$ and the WKBJ solution in the region $z<z_{t-}$ decaying exponentially as $z\rightarrow-\infty$:
\begin{equation}
\label{eq:WKBJ_2nd_ztm}
\hat{W}=A_{-}\exp\left(-k_{y}\int^{z_{t-}}_{z}\sqrt{\Delta(l)}\mathrm{d}l\right).
\end{equation}
The connection between the two solutions (\ref{eq:WKBJ_2nd_ztm}) and (\ref{eq:WKBJ_2nd_ztpztm_2nd}) via the local turning point analysis around $z=z_{t-}$ leads to the following quantization condition: 
\begin{equation}
\label{eq:WKBJ_2nd_quantization}
k_{y}\int_{z_{t-}}^{z_{t+}}\sqrt{-\Delta(z)}\mathrm{d}z=\left(m-\frac{1}{2}\right)\pi,
\end{equation}
where $m$ denotes the mode number as a positive integer (i.e. $m=1,2,\cdots$).
The quantization condition (\ref{eq:WKBJ_2nd_quantization}) implies that there are modes with different number of oscillations between the turning points $z_{t\pm}$ depending on the mode number $m$.
Eq.~(\ref{eq:WKBJ_2nd_quantization}) can be used directly to compute the growth rate $\sigma$ in an implicit way \citep[][]{Billant2009,Park2013JFM}. 
However, we can further expand (\ref{eq:WKBJ_2nd_quantization}) to express $\sigma$ more explicitly in terms of other parameters.
We note that the right-hand-side term of (\ref{eq:WKBJ_2nd_quantization}) is finite and independent of $k_{y}$ and thus the left-hand-side term should remain finite as $k_{y}$ increases.
This implies that the integral on the left-hand side is of order $O(k_{y}^{-1})$ and the turning points $z_{t\pm}$ should approach each other to allow the integral becomes zero as $k_{y}\rightarrow\infty$ . 
As the turning points are symmetric to $z=0$, the two turning points behave as $z_{t+}\rightarrow0$ and $z_{t-}\rightarrow0$ as $k_{y}$ increases. 
Considering this behavior and following previous studies by \citet{Billant2005,Park2013PoF,Park2021}, we use the Taylor expansion of the growth rate $\sigma$:
\begin{equation}
\label{eq:WKBJ_2nd_Taylor}
\sigma=\sigma_{0}-\frac{\sigma_{1}}{k_{y}}+O\left(\frac{1}{k_{y}^{2}}\right),
\end{equation}
and apply (\ref{eq:WKBJ_2nd_Taylor}) to the quantization condition (\ref{eq:WKBJ_2nd_quantization}) after expanding it around $z=0$.
This leads to the following leading-order growth rate $\sigma_{0}$ and the first-order term $\sigma_{1}$ as
\begin{equation}
\label{eq:WKBJ_2nd_Taylor_sigma0}
\sigma_{0}=\sqrt{\frac{-F_{0}+\sqrt{F_{0}^{2}+4f^{2}_{\rm{v}}\left[S_{0}\left(S_{0}+f_{\rm{h}}\right)-N^{2}\right]}}{2}},
\end{equation}
where $F_{0}=f_{\mathrm{v}}^{2}+N^{2}+f_{\mathrm{h}}(S_{0}+f_{\mathrm{h}})$ and
\begin{equation}
\sigma_{1}=\left(m-\frac{1}{2}\right)\frac{(\sigma_{0}^{2}+f^{2}_{\rm{v}})\sqrt{2S_{0}\left[f_{\rm{v}}^{2}\left(f_{\rm{h}}+2S_{0}\right)-f_{\rm{h}}\sigma_{0}^{2}\right]}}{\sqrt{2}\sigma_{0}H_{0}\left[2\sigma_{0}^{2}+N^{2}+f^{2}_{\rm{v}}+f_{\rm{h}}(S_{0}+f_{\rm{h}})\right]}.
\end{equation}
The term $\sigma_{1}$ is positive, which implies that $\sigma=\sigma_{0}$ obtained as $k_{y}\rightarrow\infty$ is the maximum growth rate. 

\begin{figure}
   \centering
      \includegraphics[width=6cm]{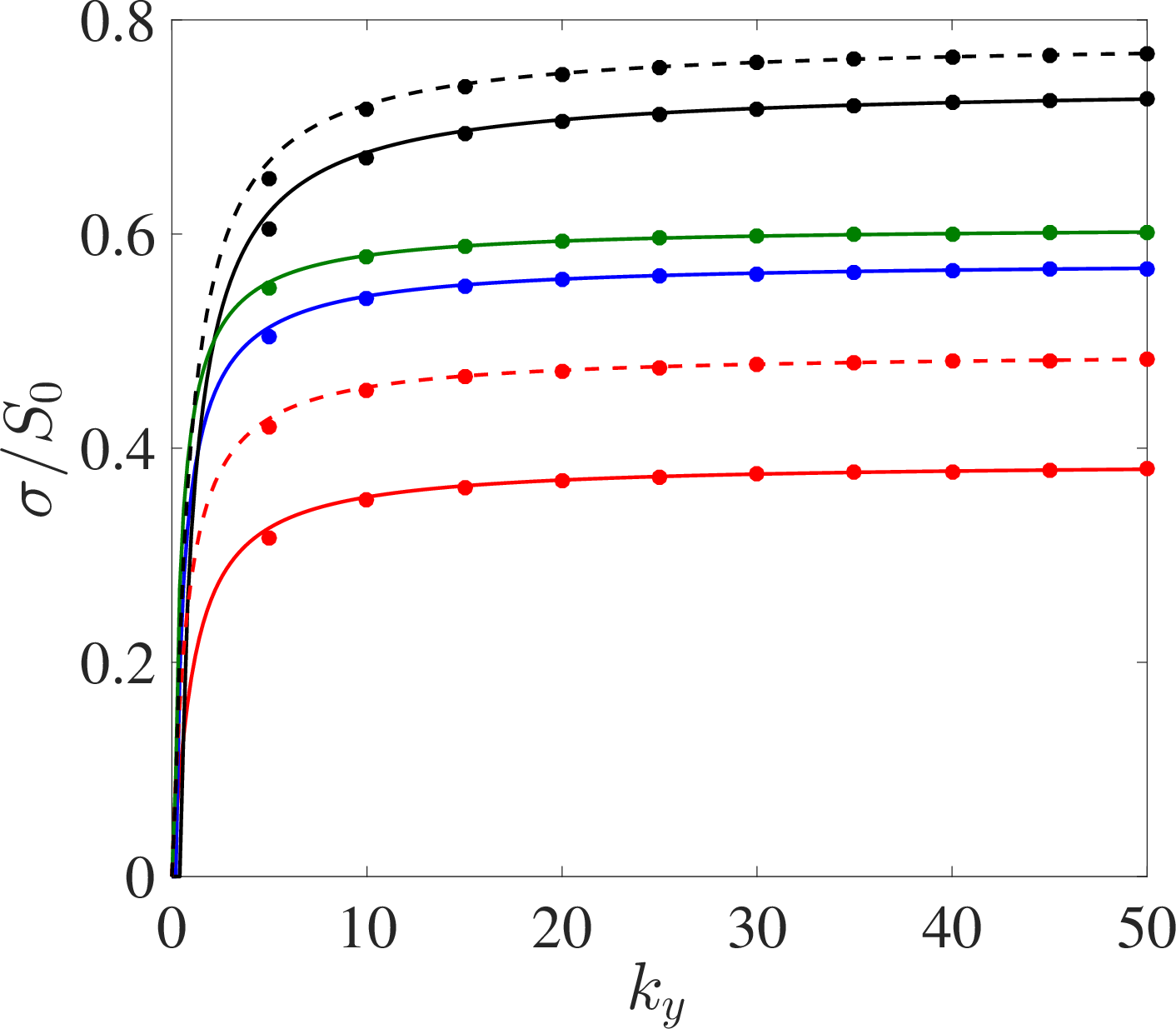}
              \caption{Growth rate $\sigma$ versus wavenumber $k_{y}$ at $k_{x}=\nu_{0}=\kappa_{0}=0$ for various sets of parameters $(Ri,\Omega_{0}/S_{0},\theta)$; from top to bottom, $(0.01,0.5,0^{\circ})$ (black dashed), $(0.1,0.5,0^{\circ})$ (black solid), $(0.1,0.5,45^{\circ})$ (green solid), $(0.1,0.25,180^{\circ})$ (blue solid), $(0.01,-0.25,90^{\circ})$ (red dashed), and $(0.1,-0.25,90^{\circ})$ (red dashed).
              Filled circles denote predictions from the WKBJ asymptotic dispersion relation (\ref{eq:WKBJ_2nd_Taylor}).
      }
         \label{Fig_growth_WKBJ}
\end{figure}
Figure \ref{Fig_growth_WKBJ} shows examples of the growth rates $\sigma$ as a function of the wavenumber $k_{y}$ for various parameters $Ri$, $\Omega_{0}$ and $\theta$ at $k_{x}=\nu_{0}=\kappa_{0}=0$. 
Numerical results (lines) have a very good agreement with the WKBJ predictions from (\ref{eq:WKBJ_2nd_Taylor}) (filed circles).
For all parameters, we see that the growth rate increases with $k_{y}$ and asymptotes to the predicted maximum growth rate $\sigma_{0}$ as $k_{y}\rightarrow\infty$. 
At fixed $\Omega_{0}$ and $\theta$, the growth-rate curves descend as $Ri$ increases. 
This indicates the stabilizing role of the stratification as similarly reported in previous studies on horizontal-shear inertial instability \citep[][]{Arobone2012,Park2021}. 
%For fixed $Ri$ and $\Omega_{0}$, the growth rate decreases as the co-latitude $\theta$ increases from zero, reaches the minimum at the equator $\theta=90^{\circ}$ and increases as $\theta\rightarrow180^{\circ}$.

\begin{figure}
   \centering
      \includegraphics[width=6cm]{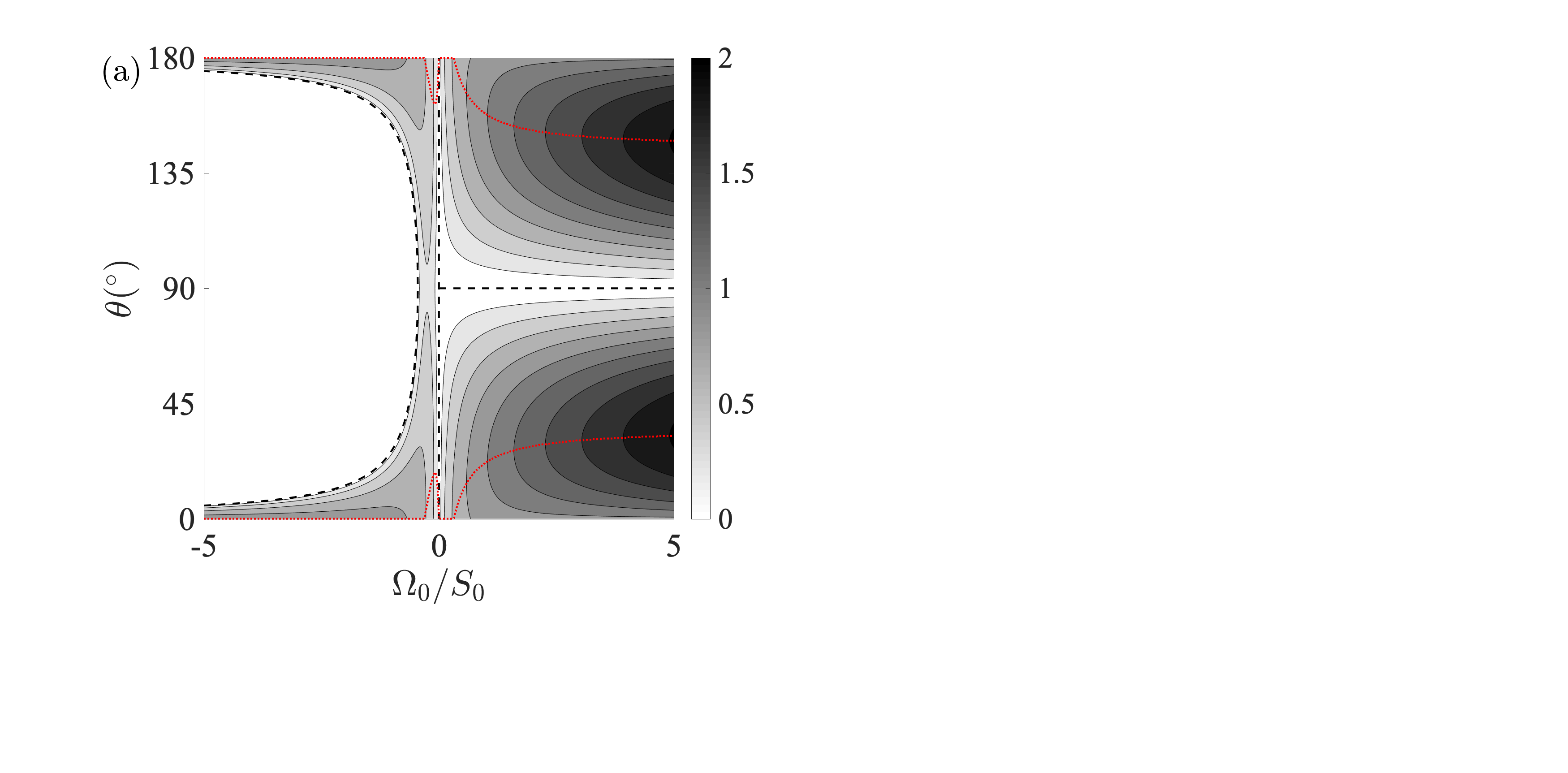}
            \includegraphics[width=6cm]{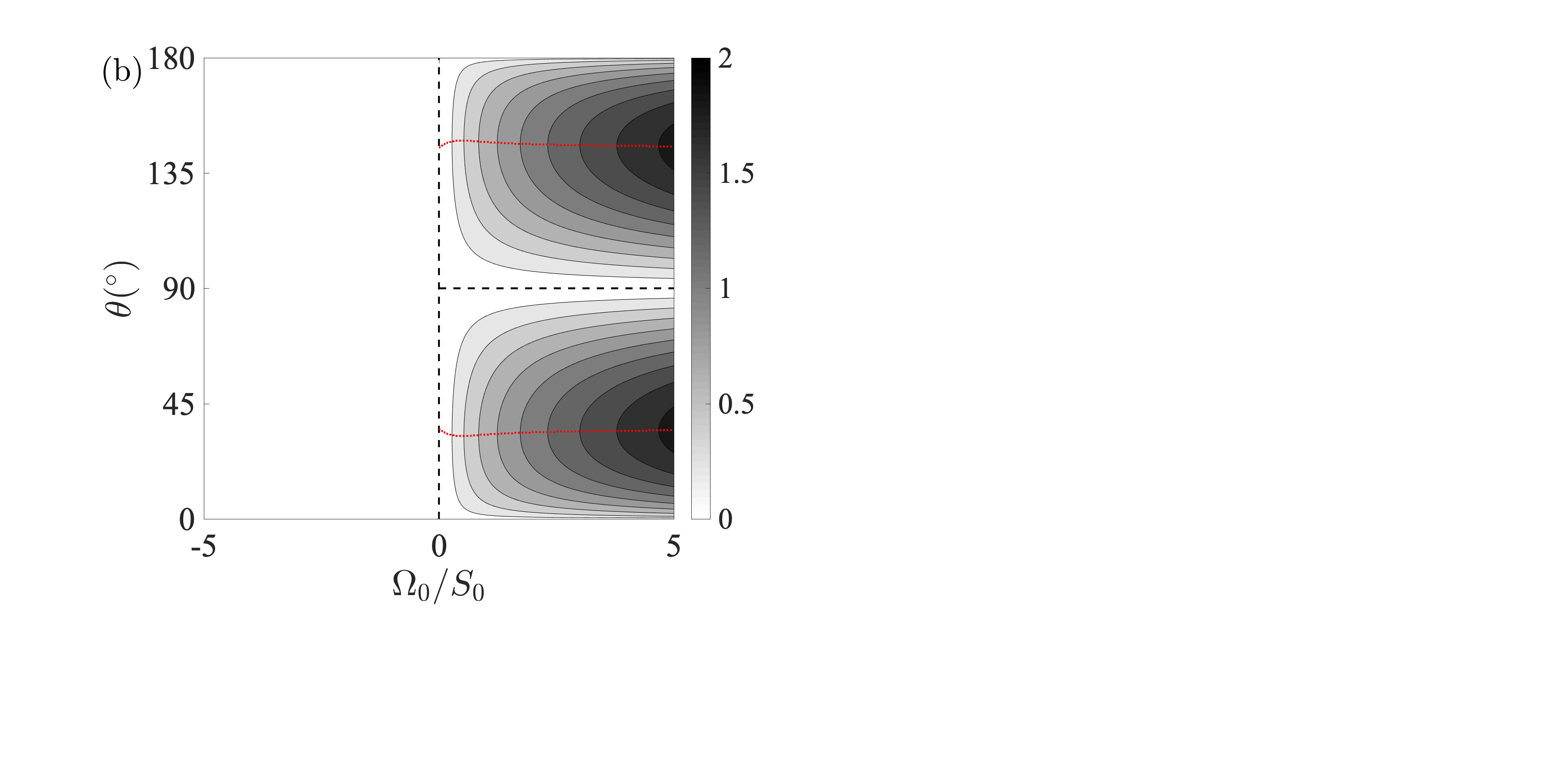}
              \caption{The maximum growth rate $\sigma_{0}/S_{0}$ (\ref{eq:WKBJ_2nd_Taylor_sigma0}) in the parameter space $(\Omega_{0},\theta)$ for (a) $Ri=0.1$ and (b) $Ri=1$. 
              Black dashed lines indicate the stability thresholds where the growth rate $\sigma_{0}$ becomes zero. 
              Red dotted lines denote the co-latitude $\theta$ at which the local maximum growth rate is. }
         \label{Fig_sigma0_kp0}
\end{figure}
Results like growth-rate curves as a function of the wavenumber $k_{y}$ in Fig.~\ref{Fig_growth_WKBJ} can explain the effects of various parameters but in a limited way. 
To better understand parametric dependence of the growth rate, we display in Fig.~\ref{Fig_sigma0_kp0} contours of the maximum growth rate $\sigma_{0}$ of (\ref{eq:WKBJ_2nd_Taylor_sigma0}) in the parameter space $(\Omega_{0},\theta)$ for two values of $Ri$: $Ri=0.1$ and $Ri=1$.
For both cases, we see that the growth rate is larger in the cyclonic regime $\Omega_{0}/S_{0}>0$ where the signs of the star's rotation $\Omega_{0}$ and the shear $S_{0}$ are the same. 
For $Ri=0.1$ in panel a, the maximum of $\sigma_{0}$ is found at the poles: $\theta=0^{\circ}$ and $180^{\circ}$ for slow rotation $0<\Omega_{0}/S_{0}<0.31$.
For fast rotation $\Omega_{0}/S_{0}>0.31$, $\sigma_{0}$ is maximal not at the poles but at colatitudes in the ranges: $0^{\circ}<\theta<40^{\circ}$ or $140^{\circ}<\theta<180^{\circ}$. 
For stronger stratification with $Ri=1$ in panel b, the maximum of $\sigma_{0}$ is found near the co-latitudes $\theta=40^{\circ}$ and $140^{\circ}$ for any cyclonic rotation $\Omega_{0}/S_{0}>0$. 
In the anti-cyclonic regime $\Omega_{0}/S_{0}<0$ where the signs of the star's rotation $\Omega_{0}$ and the shear $S_{0}$ are the opposite, the inertial instability is much more suppressed than the instability in the cyclonic regime. 
For the weakly stratified case $Ri=0.1$, the maximum of $\sigma_{0}$ is found near the poles in the ranges: $0^{\circ}<\theta<18^{\circ}$ and $162^{\circ}<\theta<180^{\circ}$ for slow anti-cyclonic rotation $-0.31<\Omega_{0}/S_{0}<0$ while the maximum is found at the poles $\theta=0^{\circ}$ and $180^{\circ}$ for fast anti-cyclonic rotation $\Omega_{0}/S_{0}<-0.31$. 
At the equator $\theta=90^{\circ}$, which corresponds to the case under the traditional $f$-plane approximation, we see that it is unstable only in the narrow range $-0.25<\Omega_{0}/S_{0}<0$ and stable elsewhere for $Ri=0.1$. 
For $Ri\geq1$, no instability is found in the anti-cyclonic regime $\Omega_{0}/S_{0}<0$ at any latitude. 

Figure \ref{Fig_sigma0_kp0} clearly shows that the cyclonic rotation can significantly promote the inertial instability at the poles or away from the poles depending on its magnitude $\Omega_{0}/S_{0}$ while the anticyclonic rotation with $\Omega_{0}/S_{0}<0$ promotes weak inertial instability near the poles or no instability if the stratification is strong enough. 
The complete rotation effect by the full Coriolis acceleration, which is applicable at any latitude, has not been thoroughly considered in previous studies on vertical shear instabilities in rotating fluids.
For instance, the traditional approximation of rotation \citep[TAR; see also][]{Mathis2019,Dhouib2021} neglecting the horizontal component of the rotation vector cannot predict the instability around the equator $\theta=90^{\circ}$.  
Our results bring attention to the importance of the full Coriolis acceleration on the inertial instability, as the instability of vertically-sheared flows is strong especially near the mid latitudes $\theta=45^{\circ}$ or $\theta=135^{\circ}$, which cannot be properly captured by the traditional approach.

\subsection{WKBJ analysis in the high-diffusivity limit $\kappa_{0}\rightarrow\infty$}
We now consider the opposite limit $\kappa_{0}\rightarrow\infty$ (i.e., $Pe\rightarrow0$) where the thermal diffusion is strong, a situation more relevant to the interior of stars.
For $k_{x}=0$ at which the inertial instability has the maximum growth, the 4th-order ODE (\ref{eq:lse_4thODE}) can be re-written as 
\begin{eqnarray}
\label{eq:lse_4thODE_k0}
\lefteqn{\frac{\mathrm{d}^{4}\hat{w}}{\mathrm{d}z^{4}}+\frac{\mathrm{i}k_{y}f_{\mathrm{v}}(U'+2f_{\mathrm{h}})}{f_{\mathrm{v}}^{2}+\sigma^{2}}\frac{\mathrm{d}^{3}\hat{w}}{\mathrm{d}z^{3}}}\nonumber\\
&&-\frac{k_{y}^{2}\left[2\sigma^{2}+f_{\mathrm{v}}^{2}+f_{\mathrm{h}}(U'+f_{\mathrm{h}})\right]-3\mathrm{i}k_{y}f_{\mathrm{v}}U'''}{f_{\mathrm{v}}^{2}+\sigma^{2}}\frac{\mathrm{d}^{2}\hat{w}}{\mathrm{d}z^{2}}\nonumber\\
&&-\frac{\mathrm{i}k_{y}^{3}f_{\mathrm{v}}(U'+2f_{\mathrm{h}})+2k_{y}^{2}f_{\mathrm{h}}U''}{f_{\mathrm{v}}^{2}+\sigma^{2}}\frac{\mathrm{d}\hat{w}}{\mathrm{d}z}\nonumber\\
&&+\frac{k_{y}^{4}\left[\sigma^{2}+f_{\mathrm{h}}(U'+f_{\mathrm{h}})\right]-\mathrm{i}k_{y}^{3}f_{\mathrm{v}}U''-k_{y}^{2}f_{\mathrm{h}}U'''+\mathrm{i}k_{y}f_{\mathrm{v}}U''''}{f_{\mathrm{v}}^{2}+\sigma^{2}}\hat{w}\nonumber\\
&&=O\left(\frac{1}{\kappa_{0}}\right).
\end{eqnarray}
We neglect the terms of the order of $O(\kappa_{0}^{-1})$ for large $\kappa_{0}$ and consider only the leading-order terms for large $k_{y}$ to apply the WKBJ approximation. 
The above ODE (\ref{eq:lse_4thODE_k0}) is then simplified as
\begin{eqnarray}
\label{eq:WKBJ_ODE4}
\lefteqn{\frac{\mathrm{d}^{4}\hat{w}}{\mathrm{d}z^{4}}+\frac{\mathrm{i}k_{y}f_{\mathrm{v}}(U'+2f_{\mathrm{h}})}{f_{\mathrm{v}}^{2}+\sigma^{2}}\frac{\mathrm{d}^{3}\hat{w}}{\mathrm{d}z^{3}}}\nonumber\\
&&-\frac{k_{y}^{2}\left[2\sigma^{2}+f_{\mathrm{v}}^{2}+f_{\mathrm{h}}(U'+f_{\mathrm{h}})\right]}{f_{\mathrm{v}}^{2}+\sigma^{2}}\frac{\mathrm{d}^{2}\hat{w}}{\mathrm{d}z^{2}}-\frac{\mathrm{i}k_{y}^{3}f_{\mathrm{v}}(U'+2f_{\mathrm{h}})}{f_{\mathrm{v}}^{2}+\sigma^{2}}\frac{\mathrm{d}\hat{w}}{\mathrm{d}z}\nonumber\\
&&+\frac{k_{y}^{4}\left[\sigma^{2}+f_{\mathrm{h}}(U'+f_{\mathrm{h}})\right]}{f_{\mathrm{v}}^{2}+\sigma^{2}}\hat{w}=0.
\end{eqnarray}
If we apply the same WKBJ approximation (\ref{eq:WKBJ}), we find that $\delta=k_{y}^{-1}$ and the leading-order function $K_{0}(z)$ satisfies the following 4th-order polynomial equation:
\begin{equation}
\label{eq:polynomial_K0}
K_{0}^{'4}+\mathcal{K}_{1} K_{0}^{'3}-(1+\mathcal{K}_{2})K_{0}^{'2}-\mathcal{K}_{1}K'_{0}+\mathcal{K}_{2}=0,
\end{equation}
where 
\begin{equation}
\mathcal{K}_{1}=\frac{\mathrm{i}f_{\mathrm{v}}(U'+2f_{\mathrm{h}})}{\sigma^{2}+f^{2}_{\mathrm{v}}},~~
\mathcal{K}_{2}=\frac{\sigma^{2}+f_{\mathrm{h}}(U'+f_{\mathrm{h}})}{\sigma^{2}+f^{2}_{\mathrm{v}}}.
\end{equation}
The quartic equation (\ref{eq:polynomial_K0}) has four solutions
\begin{equation}
K'_{0}=\pm1,~~
K'_{0}=\frac{-\mathcal{K}_{1}\pm\sqrt{\mathcal{K}_{1}^{2}+4\mathcal{K}_{2}}}{2}.
\end{equation}
The first two WKBJ solutions with $K'_{0}=\pm 1$ imply that $\hat{w}\simeq a_{+}\exp(k_{y}z)+a_{-}\exp(-k_{y}z)$ and thus we have $a_{\pm}=0$ after considering the boundary conditions that decay exponentially as $z\rightarrow\pm\infty$.
The remaining WKBJ solutions are
\begin{eqnarray}
\label{eq:WKBJ_ODE4_w}
\lefteqn{\hat{w}=\exp\left[-\frac{\mathrm{i}k_{y}f_{\mathrm{v}}(U+2f_{\mathrm{h}}z)}{2(\sigma^{2}+f^{2}_{\mathrm{v}})}\right]\times}\nonumber\\
&&\left[C_{+}\exp\left(k_{y}\int_{z}\sqrt{\Gamma}\mathrm{d}z\right)+C_{-}\exp\left(-k_{y}\int_{z}\sqrt{\Gamma}\mathrm{d}z\right)\right],
\end{eqnarray}
where
\begin{equation}
\Gamma=\frac{\sigma^{2}+f_{\mathrm{h}}(U'+f_{\mathrm{h}})}{\sigma^{2}+f^{2}_{\mathrm{v}}}-\frac{f^{2}_{\mathrm{v}}(U'+2f_{\mathrm{h}})^{2}}{4(\sigma^{2}+f^{2}_{\mathrm{v}})^{2}}.
\end{equation}
Similar to the WKBJ analysis in the previous subsection, we facilitate the turning point analysis by introducing a new variable $\hat{V}=\hat{w}/\mathcal{Q}$ where $\mathcal{Q}$ is the prefactor on the right-hand side term in (\ref{eq:WKBJ_ODE4_w})
\begin{equation}
\mathcal{Q}=\exp\left[-\frac{\mathrm{i}k_{y}f_{\mathrm{v}}(U+2f_{\mathrm{h}}z)}{2(\sigma^{2}+f^{2}_{\mathrm{v}})}\right].
\end{equation}
For the variable $\hat{V}$, we find the following 4th-order ODE at the leading order:
\begin{eqnarray}
\label{eq:WKBJ_ODE4_V}
\lefteqn{\frac{\mathrm{d}^{4}\hat{V}}{\mathrm{d}z^{4}}-\frac{\mathrm{i}k_{y}f_{\mathrm{v}}(U'+2f_{\mathrm{h}})}{f_{\mathrm{v}}^{2}+\sigma^{2}}\frac{\mathrm{d}^{3}\hat{V}}{\mathrm{d}z^{3}}}\nonumber\\
&&-\frac{k_{y}^{2}\left[2\sigma^{2}+f_{\mathrm{v}}^{2}+f_{\mathrm{h}}(U'+f_{\mathrm{h}})\right]}{f_{\mathrm{v}}^{2}+\sigma^{2}}\frac{\mathrm{d}^{2}\hat{V}}{\mathrm{d}z^{2}}+\frac{\mathrm{i}k^{3}_{y}f_{\mathrm{v}}(U'+2f_{\mathrm{h}})\Gamma}{\sigma^{2}+f^{2}_{\mathrm{v}}}\frac{\mathrm{d}\hat{V}}{\mathrm{d}z}\nonumber\\
&&+k_{y}^{4}\Gamma\left[1+\frac{f^{2}_{\mathrm{v}}\left(U'+2f_{\mathrm{h}}\right)^{2}}{4\left(\sigma^{2}+f^{2}_{\mathrm{v}}\right)^{2}}\right]\hat{V}=O\left(\frac{1}{k_{y}}\right).\nonumber\\
\end{eqnarray}
The WKBJ solution for $\hat{V}$ can be expressed as an evanescent solution if $\Gamma>0$:
\begin{eqnarray}
\label{eq:WKBJ_ODE4_V_exp}
\hat{V}=D_{+}\exp\left(k_{y}\int_{z}\sqrt{\Gamma}\mathrm{d}z\right)+D_{-}\exp\left(-k_{y}\int_{z}\sqrt{\Gamma}\mathrm{d}z\right),
\end{eqnarray}
or a wavelike solution if $\Gamma<0$:
\begin{eqnarray}
\label{eq:WKBJ_ODE4_V_wave}
\hat{V}=E_{+}\exp\left(\mathrm{i}k_{y}\int_{z}\sqrt{-\Gamma}\mathrm{d}z\right)+E_{-}\exp\left(-\mathrm{i}k_{y}\int_{z}\sqrt{-\Gamma}\mathrm{d}z\right),
\end{eqnarray}
where $D_{\pm}$ and $E_{\pm}$ are constants to be determined by the boundary conditions. 

\begin{figure}
   \centering
   \includegraphics[width=5cm]{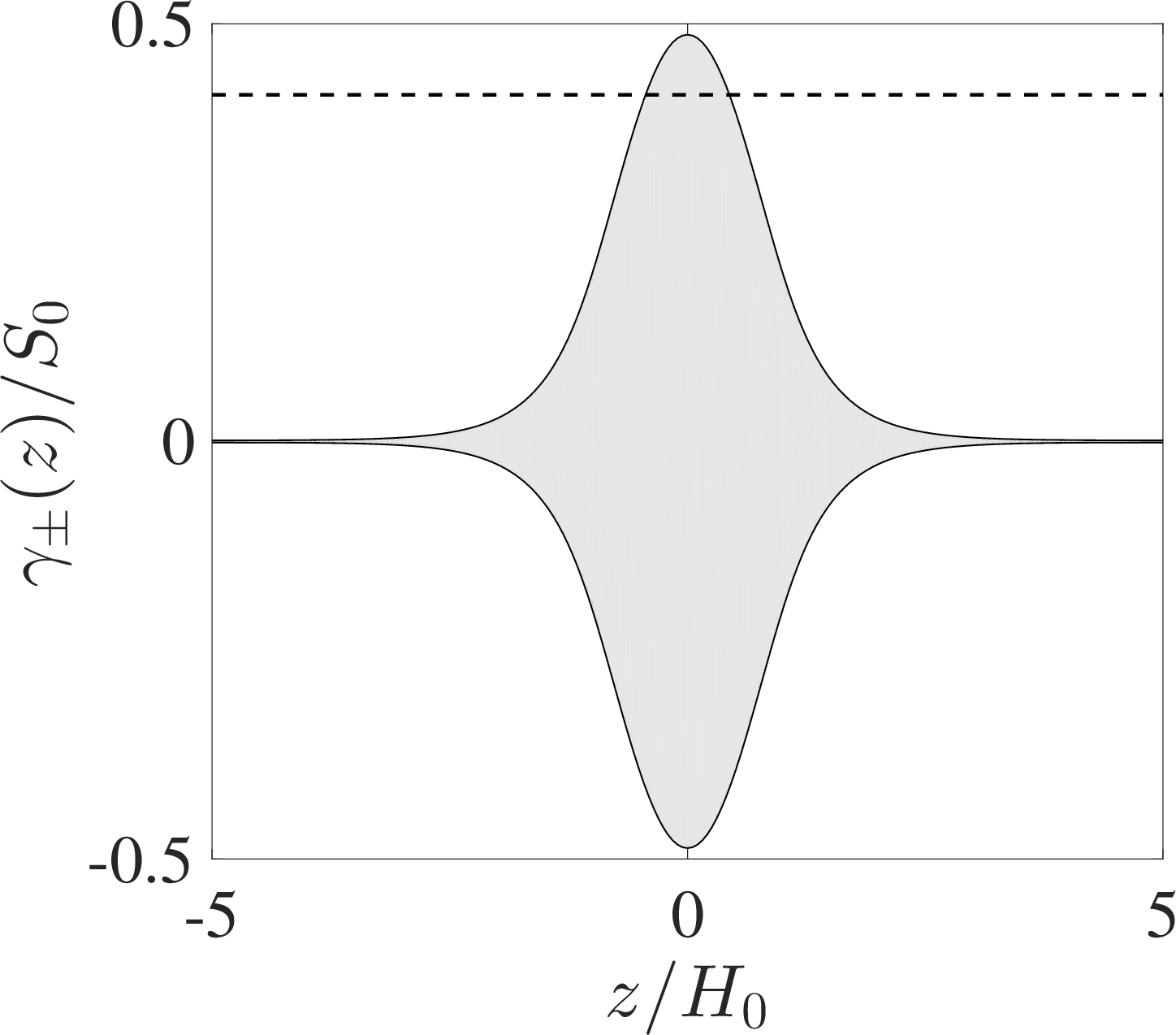}
              \caption{Epicyclic growth rates $\gamma_{\pm}$ (solid lines) for $\Omega_{0}/S_{0}=-0.25$ and $\theta=45^{\circ}$. 
              White and grey areas denote the regions where $\Gamma(z)$ is positive and negative, i.e. the WKBJ solution is evanescent-like and wavelike, respectively. 
              The dashed line indicates the growth rate $\sigma/S_{0}=0.415$ numerically obtained for $Pe=0.01$ at $(k_{x}H_{0},k_{y}H_{0})=(0,5)$. 
      }
         \label{Fig_epicyclic_kpinf}
\end{figure}
To find turning points $z_{t}$ where $\Gamma=0$, we define the epicyclic growth rates $\gamma_{\pm}$:
\begin{equation}
\gamma_{\pm}(z)=\pm\sqrt{\frac{-F_{1}+\sqrt{F_{1}^{2}+f_{\mathrm{v}}^{2}U^{'2}}}{2}}
\end{equation}
where $F_{1}(z)=f_{\mathrm{v}}^{2}+f_{\mathrm{h}}(U'(z)+f_{\mathrm{h}})$.
Figure \ref{Fig_epicyclic_kpinf} shows an example of the epicyclic growth rate $\gamma_{\pm}$ for the parameters $\Omega_{0}/S_{0}=-0.25$ and $\theta=45^{\circ}$. 
The profiles of $\gamma_{\pm}$ are similar to those of $\sigma_{\pm}(z)$ in Fig.~\ref{Fig_epicyclic} such that $\gamma_{+}$ and $\gamma_{-}$ reaches their maximum and minimum at $z=0$, respectively, due to their symmetry at $z=0$. 
The use of the epicyclic growth rate $\gamma_{\pm}$ is the same as that of $\sigma_{\pm}$: the solution is wavelike if the growth rate $\sigma$ is in the range $\gamma_{-}<\sigma<\gamma_{+}$ while it is evanescent if $\sigma>\gamma_{+}$ or $\sigma<\gamma_{-}$.
For the growth rate $\sigma/S_{0}=0.415$, which is computed numerically at $Pe=0.01$ and $(k_{x}H_{0},k_{y}H_{0})=(0,5)$, the turning points are at $z_{t\pm}=\pm0.43H_{0}$ and we can view from the WKBJ analysis that the eigenfunction is evanescent for $|z/H_{0}|>0.43$ while it is wavelike in the region $-0.43<z/H_{0}<0.43$.

The eigenfunction can be constructed as follows. 
We first consider an exponentially-decaying solution in the region $z>z_{t+}$:
\begin{equation}
\label{eq:WKBJ_ODE4_V_ztp}
\hat{V}=D_{+}\exp\left(-k_{y}\int_{z_{t+}}^{z}\sqrt{\Gamma(l)}\mathrm{d}l\right),
\end{equation}
where $D_{+}$ is a constant. 
The WKBJ solution (\ref{eq:WKBJ_ODE4_V_ztp}) for $z>z_{t+}$ can be connected with the WKBJ solution for $z<z_{t+}$ by considering the local equation around $z_{t+}$ with a new local variable $z_{\epsilon}=(z-z_{t+})/\epsilon$ with small $\epsilon$ and the Taylor expansion $\Gamma\simeq\Gamma'_{t+}\epsilon z_{\epsilon}$:
\begin{equation}
\label{eq:Airy_V_ztp}
\frac{\mathrm{d}^{2}\hat{V}}{\mathrm{d}z_{\epsilon}^{2}}-z_{\epsilon}\hat{V}=O\left(\frac{1}{k_{y}^{1/3}}\right),
\end{equation}
where we consider the small parameter $\epsilon=\left(k_{y}^{2}\Gamma'_{t+}\right)^{-2/3}$. 
Neglecting the terms of order $O\left({k_{y}^{-1/3}}\right)$ on the right-hand side, we recover the Airy equation for $\hat{V}$. 
By considering the asymptotic behaviors of the Airy functions $\hat{V}(z_{\epsilon})=c_{1}\mathrm{Ai}(z_{\epsilon})+d_{1}\mathrm{Bi}(z_{\epsilon})$ in the limit $z_{\epsilon}\rightarrow\pm\infty$, we find the following WKBJ solution in the region $z_{t-}<z<z_{t+}$:
\begin{equation}
\label{eq:WKBJ_ODE4_V_ztpztm}
\hat{W}=E_{+}\exp\left(\mathrm{i}k_{y}\int^{z_{t+}}_{z}\sqrt{-\Gamma(l)}\mathrm{d}l\right)+E_{-}\exp\left(-\mathrm{i}k_{y}\int^{z_{t+}}_{z}\sqrt{-\Gamma(l)}\mathrm{d}l\right),
\end{equation}
where constants $E_{\pm}$ satisfy
\begin{equation}
E_{+}=\exp\left(-\mathrm{i}\frac{\pi}{4}\right)D_{+},~
E_{-}=\exp\left(\mathrm{i}\frac{\pi}{4}\right)D_{+}.
\end{equation}
The similar analysis as in the previous subsection can be done to connect the WKBJ solution (\ref{eq:WKBJ_ODE4_V_ztpztm}) in the region $z_{t-}<z<z_{t+}$ with an exponentially-decaying solution in the region $z<z_{t-}$:
\begin{equation}
\label{eq:WKBJ_ODE4_V_ztm}
\hat{V}=D_{+}\exp\left(k_{y}\int_{z_{t-}}^{z}\sqrt{\Gamma(l)}\mathrm{d}l\right).
\end{equation}
The turning point analysis leads to the following quantization condition:
\begin{equation}
\label{eq:WKBJ_ODE4_quantization}
k_{y}\int_{z_{t-}}^{z_{t+}}\sqrt{-\Gamma(z)}\mathrm{d}z=\left(m-\frac{1}{2}\right)\pi,
\end{equation}
where $m$ is the mode number.
Same as in (\ref{eq:WKBJ_2nd_Taylor}), we can apply the Taylor expansion of $\Gamma$ around $z=0$ and express the growth rate $\sigma$ as
\begin{equation}
\label{eq:WKBJ_ODE4_sigma}
\sigma=\gamma_{0}-\frac{\gamma_{1}}{k_{y}}+O\left(\frac{1}{k_{y}^{2}}\right),
\end{equation}
where $\gamma_{0}$ is the leading-order term
\begin{equation}
\label{eq:WKBJ_ODE4_gamma0}
\gamma_{0}=\sqrt{\frac{-f_{\rm{v}}^{2}-f_{\rm{h}}(S_{0}+f_{\rm{h}})+\sqrt{\left[f_{\rm{v}}^{2}+(S_{0}+f_{\rm{h}})^{2}\right]\left(f_{\mathrm{v}}^{2}+f_{\mathrm{h}}^{2}\right)}}{2}},
\end{equation}
and $\gamma_{1}$ is the first-order term of the expansion (\ref{eq:WKBJ_ODE4_sigma})
\begin{equation}
\label{eq:WKBJ_ODE4_gamma1}
\gamma_{1}=\left(m-\frac{1}{2}\right)\frac{(\gamma_{0}^{2}+f^{2}_{\rm{v}})\sqrt{2S_{0}\left[f_{\rm{v}}^{2}S_{0}-2f_{\rm{h}}\gamma_{0}^{2}\right]}}{\gamma_{0}H_{0}\left[4\gamma_{0}^{2}+2f^{2}_{\rm{v}}+2f_{\rm{h}}(S_{0}+f_{\rm{h}})\right]}.
\end{equation}
The leading-order term $\gamma_{0}$ corresponds to the maximum growth rate obtained in the limit $k_{y}\rightarrow\infty$ since $\gamma_{1}>0$.
We see in Fig.~\ref{Fig_growth_rate_inertial} that the asymptotic growth rates from (\ref{eq:WKBJ_ODE4_sigma}) show very good agreement with numerical results for a highly diffusive case with $Pe=0.01$, especially when $k_{y}$ is sufficiently large. 

\begin{figure}
   \centering
      \includegraphics[width=7cm]{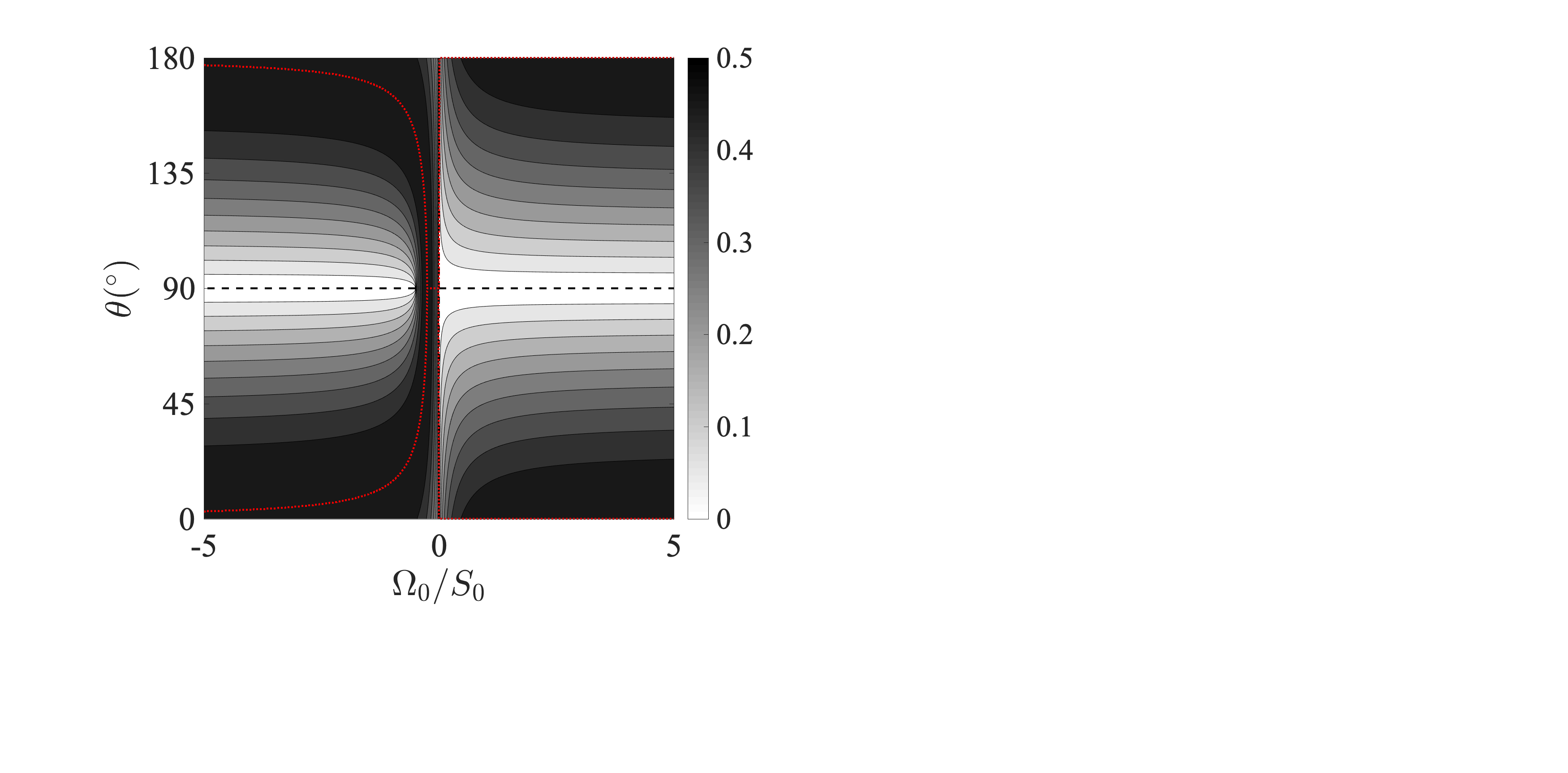}
              \caption{The maximum growth rate $\gamma_{0}/S_{0}$ (\ref{eq:WKBJ_ODE4_gamma0}) in the parameter space $(\Omega_{0},\theta)$. Black dashed lines indicate contour lines on which the growth rate becomes zero. 
              Red dotted lines indicate the co-latitude $\theta$ at which $\gamma_{0}$ is locally maximal.}
         \label{Fig_sigma0_kpinf}
\end{figure}
The advantage of using analytic expressions such as the growth rate (\ref{eq:WKBJ_ODE4_sigma}) or the maximum growth rate (\ref{eq:WKBJ_ODE4_gamma0}) is that we can easily undertake parametric investigations on the growth rate in the space $(\Omega_{0},\theta)$ as shown in Fig.~\ref{Fig_sigma0_kpinf}. 
For cyclonic rotation with $\Omega_{0}/S_{0}>0$, the maximum of $\gamma_{0}$ is attained at the poles $\theta=0^{\circ}$ and $180^{\circ}$.
At the equator $\theta=90^{\circ}$ (i.e. $f_{\rm{v}}=0$, $f_{\rm{h}}=2\Omega_{0}$), the maximum growth rate $\gamma_{0}$ is positive only in a narrow anti-cyclonic range of $-0.5<\Omega_{0}/S_{0}<0$ since $\gamma_{0}=\sqrt{-2\Omega_{0}(S_{0}+2\Omega_{0})}$ and the maximum of $\gamma_{0}$ is attained at $\Omega_{0}/S_{0}=-0.25$ as $\gamma_{0,\max}=S_{0}/2$. 
In the anti-cyclonic rotation regime $\Omega_{0}/S_{0}<0$, $\gamma_{0,\max}$ outside the equator is attained below $\Omega_{0}/S_{0}=-0.25$ and $\gamma_{0}$ is higher near the poles.  

We note that the contours of the maximum growth rate $\sigma_{0}$ for the non-diffusive case $\kappa_{0}=0$ shown in Fig.~\ref{Fig_sigma0_kp0} are quite different from those in the high-diffusivity limit $\kappa_{0}\rightarrow\infty$ shown in Fig.~\ref{Fig_sigma0_kpinf}, the latter which are more relevant to the stellar radiation zones with high thermal diffusive fluids.
The non-diffusive case has the maximum instability in the middle latitudes (i.e. co-latitudes around $\theta=45^{\circ}$ or $\theta=135^{\circ}$) while the highly-diffusive case has the maximum instability around the poles (i.e. $\theta=0^{\circ}$ or $\theta=180^{\circ}$). 
It is also noteworthy that in the high-diffusivity limit $\kappa_{0}\rightarrow\infty$, our finding on the inertial instability of vertical shear flow is different from that of horizontal shear flow where the growth rate is maximum near the equator $\theta=90^{\circ}$ \citep[][]{Park2021}.
This emphasizes the importance of understanding carefully the role of the full Coriolis acceleration in deriving the instability and turbulence in the radiation zones in stellar interiors.

Figures \ref{Fig_sigma0_kp0} and \ref{Fig_sigma0_kpinf} show the strong latitudinal dependence of the growth rate of the inertial instability. 
As in the case of the inflectional instability, we can therefore expect regions where the turbulent energy dissipation, momentum transport, and matter mixing will be localised. 
For instance, the inertial instability will trigger turbulent transport in the polar regions while it will vanish around the equator in the stellar regime with strong heat diffusion.

\subsection{Charateristic time scale of turbulent dissipation induced by inertial instability}
While inflectional instability is difficult to treat analytically, we derived in the previous subsections analytical expressions of the growth rate for the inertial instability.
Using these expressions, we can further deduce a scaling law for the charateristic time scale of turbulent dissipation induced by the inertial instability.  
We first consider the shellular rotation of stars varying with radius (i.e. $\Omega=\Omega(r)$). 
In the local frame rotating with $\Omega_{0}$ at $r$, we have the relative azimuthal velocity $U$:
\begin{equation}
U=r\sin\theta (\Omega-\Omega_{0}),
\end{equation}
and its radial gradient reads
\begin{equation}
\frac{\partial U}{\partial r}=\sin\theta\left(\Omega+r\frac{\partial\Omega}{\partial r}-\Omega_{0}\right).
\end{equation}
At given co-latitude $\theta$ and radius $r$, we rewrite the radial gradient in terms of the local vertical shear $S_{0}$:
\begin{equation}
S_{0}(\epsilon)=\epsilon2\Omega_{0}\sin\theta,
\end{equation}
where $\epsilon(r)$ is the nondimensional shear parameter defined as
\begin{equation}
\epsilon(r)=\frac{1}{2\Omega_{0}}\frac{\partial}{\partial r}\left[r\left(\Omega-\Omega_{0}\right)\right].
\end{equation}
By putting $S_{0}$ into the maximum growth rate (\ref{eq:WKBJ_ODE4_gamma0}) in the high-diffusivity limit $\kappa_{0}\rightarrow\infty$, we retain the following expression:
\begin{equation}
\label{eq:growth_max_epsilon}
\frac{\sigma^{2}}{(2\Omega_{0})^{2}}=\frac{\sqrt{(\epsilon^{2}+2\epsilon)\sin^{2}\theta+1}-1-\epsilon\sin^{2}\theta}{2}.
\end{equation}
The above expression (\ref{eq:growth_max_epsilon}) implies that the growth rate $\sigma$ is always positive except at the poles (i.e. $\sin\theta=0$) and equator (i.e. $\sin\theta=1$). 

\begin{figure}
   \centering
      \includegraphics[width=6cm]{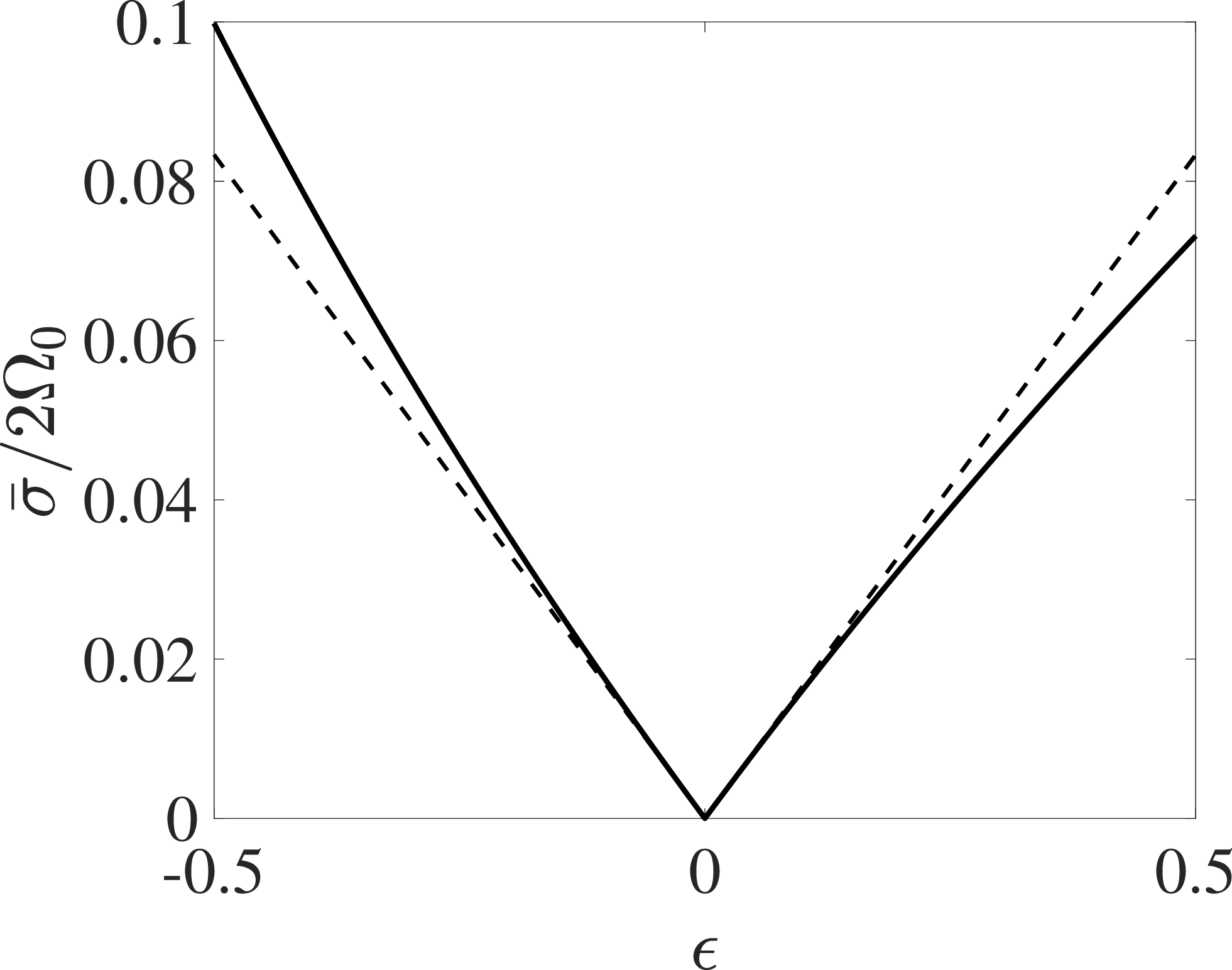}
              \caption{The average growth rate $\bar{\sigma}$ (\ref{eq:growth_avg}) as a function of $\epsilon$ (solid line) and its small-$
              \epsilon$ linear approximation (\ref{eq:growth_avg_linear}) (dashed line).}
         \label{Fig_growth_epsilon}
\end{figure}
We can consider a latitudinally-averaged growth rate $\bar{\sigma}$ defined as
\begin{equation}
\label{eq:growth_avg}
\bar{\sigma}=\frac{1}{2}\int_{0}^{\pi}\sigma\sin\theta\mathrm{d}\theta,
\end{equation}
\citep[see also,][]{Chaboyer1992}.
Figure \ref{Fig_growth_epsilon} shows how $\bar{\sigma}$ varies with $\epsilon$. 
The symmetry with respect to $\epsilon=0$ breaks down as $|\epsilon|$ increases.
However, in the limit $|\epsilon|\rightarrow0$, we can expand $\bar{\sigma}$ using the Taylor expansion and find a linear relation as
\begin{equation}
\label{eq:growth_avg_linear}
\frac{\bar{\sigma}}{2\Omega_{0}}\simeq \frac{1}{6}|\epsilon|,
\end{equation}
where the prefactor $1/6$ is obtained analytically by considering a small-$\epsilon$ limit.

If we assume that turbulence in the stellar interior is generated dominantly by the vertical shear instability, we can consider the viscosity ratio as
\begin{equation}
    \frac{\bar{\nu}_{\rm{v},\rm{v}}}{\bar{\nu}_{\rm{h},\rm{v}}}=\frac{l_{\parallel}^{2}}{l_{\perp}^2},
\end{equation}
where $l_{\parallel}$ and $l_{\perp}$ are the characteristic length scales of turbulence in the directions parallel and perpendicular to the stratification, respectively \citep[i.e. vertical and horizontal directions; see also,][]{Mathisetal2018}.
The dynamical time scale $\tau$ characterizing the turbulence induced by the vertical shear instability follows the scaling law as an inverse of the growth rate $\bar{\sigma}$:
\begin{equation}
    \tau=\frac{l_{\perp}^{2}}{\bar{\nu}_{\rm{h},\rm{v}}}=\frac{l_{\parallel}^{2}}{\bar{\nu}_{\rm{v},\rm{v}}}=\frac{1}{\bar{\sigma}}\simeq\frac{6}{2\Omega_{0}|\epsilon|}.
\end{equation}
A scaling law for the chracteristic time scale $\tau$ is similarly derived in the context of turbulence induced by the horizontal shear instability in \citet[][]{Park2021}. 

\section{Link between inertial and symmetric instabilities}
\label{sec:SI}
The inertial instability as well as the Goldreich-Schubert-Fricke (GSF) instability \citep[][]{Barker2020} are closely related to another type of instability called the symmetric instability, which has been studied mainly in the context of geophysics \citep[e.g., slantwise convection; see also][]{Holton}.
Recent studies by \cite{Tort2016} and \cite{Zeitlin2018} revealed that the full Coriolis acceleration can promote the symmetric instability, similar to what this paper claims for the inertial instability.
In this section, we revisit this symmetric instability problem without taking any assumption on thermal diffusivity (e.g., $Pr=\nu_{0}/\kappa_{0}=1$ in the case of \citet{Zeitlin2018}) and aim to demonstrate mathematically that these instabilities are essentially the same even if they are originally derived from different formulations. 

To explore the symmetric instability, we consider vertical shear flow with a linear profile as $U=\Lambda z$ where the vertical shear $\Lambda$ is a constant. 
We assume that there is no variation in the longitudinal direction $x$ (i.e. $\partial/\partial x=0$). 
The linearized perturbation equations are then expressed as
\begin{equation}
\label{eq:ptb_continuity_sym}
	\frac{\partial \tilde{v}}{\partial y}+\frac{\partial \tilde{w}}{\partial z}=0,
\end{equation}
\begin{equation}
\label{eq:ptb_mom_x_sym}
	\frac{\partial \tilde{u}}{\partial t}-f_{\rm{v}}\tilde{v}+\left(\Lambda+f_{\rm{h}}\right)\tilde{w}=\nu_{0}\nabla^{2}\tilde{u},
\end{equation}
\begin{equation}
\label{eq:ptb_mom_y_sym}
	\frac{\partial \tilde{v}}{\partial t}+f_{\rm{v}}\tilde{u}=-\frac{1}{\rho_{0}}\frac{\partial \tilde{p}}{\partial y}+\nu_{0}\nabla^{2}\tilde{v},
\end{equation}
\begin{equation}
\label{eq:ptb_mom_z_sym}
	\frac{\partial \tilde{w}}{\partial t}-f_{\rm{h}}\tilde{u}=-\frac{1}{\rho_{0}}\frac{\partial \tilde{p}}{\partial z}+\alpha_{0}g\tilde{T}+\nu_{0}\nabla^{2}\tilde{w},
\end{equation}
\begin{equation}
\label{eq:ptb_energy_sym}
	\frac{\partial \tilde{T}}{\partial t}-\frac{f_{\rm{v}}\Lambda}{\alpha_{0}g}\tilde{v}+\mathcal{T}_{\Theta}\tilde{w}=\kappa_{0}\nabla^{2}\tilde{T}.
\end{equation}
We note that, unlike the case of hyperbolic tangent shear flow, the wide-jet approximation is not required for a linear shear flow as the second derivative $U''$ is zero and the vertical temperature gradient $\partial\bar{T}/\partial z$ in Eq.~(\ref{eq:base_temperature_vertical}) becomes constant as $\partial\bar{T}/\partial z=\mathcal{T}_{\Theta}$. 
For the stability analysis, we consider the following normal mode expression:
\begin{equation}
\label{eq:ptb_normal_mode_sym}
\left(\tilde{\vec{u}},\tilde{p},\tilde{T}\right)=\Re\left[\left(\check{\vec{u}},\rho_{0}\check{p},\mathcal{T}_{\Theta}\check{T}\right)\exp\left(\mathrm{i}k_{y} y+\mathrm{i}k_{z} z++\sigma t\right)\right],
\end{equation}
where $\check{\vec{u}}=\left(\check{u},\check{v},\check{w}\right)$ denotes the perturbation velocity amplitude, $\tilde{p}$ and $\tilde{T}$ are normalized amplitudes of the perturbation pressure and temperature, respectively, and $k_{z}$ is the wavenumber in the vertical direction $z$. 
Applying the normal mode (\ref{eq:ptb_normal_mode_sym}) to the equations (\ref{eq:ptb_continuity_sym})-(\ref{eq:ptb_energy_sym}) leads to the following set of equations:
\begin{equation}
\label{eq:ptb_modal_continuity_sym}
\mathrm{i}(k_{y}\check{v}+k_{z}\check{w})=0,
\end{equation}
\begin{equation}
\label{eq:ptb_modal_mom_x_sym}
\sigma\check{u}-f_{\mathrm{v}}\check{v}+\left(\Lambda+f_{\mathrm{h}}\right)\check{w}=-\nu_{0}k^{2}\check{u}
\end{equation}
\begin{equation}
\label{eq:ptb_modal_mom_y_sym}
\sigma\check{v}+f_{\mathrm{v}}\check{u}=-\mathrm{i}k_{y}\check{p}-\nu_{0}k^{2}\check{v}
\end{equation}
\begin{equation}
\label{eq:ptb_modal_mom_z_sym}
	\sigma\check{w}-f_{\rm{h}}\check{u}=-\mathrm{i}k_{z}\check{p}+\alpha_{0}g\mathcal{T}_{\theta}\check{T}-\nu_{0}k^{2}\check{w},
\end{equation}
\begin{equation}
\label{eq:ptb_modal_energy_sym}
	\sigma\check{T}-\frac{f_{\rm{v}}\Lambda}{\alpha_{0}g\mathcal{T}_{\theta}}\check{v}+\check{w}=-\kappa_{0}k^{2}\check{T},
\end{equation}
where $k^{2}=k_{y}^{2}+k_{z}^{2}$.
From the energy equation (\ref{eq:ptb_modal_energy_sym}), we have
\begin{equation}
\check{T}=\left(\frac{f_{\rm{v}}\Lambda}{\alpha_{0}g\mathcal{T}_{\theta}}\check{v}-\check{w}\right)/(\sigma+\kappa_{0}k^{2}).
\end{equation}
From the two momentum equations (\ref{eq:ptb_modal_mom_y_sym})-(\ref{eq:ptb_modal_mom_z_sym}), we can eliminate the pressure $\check{p}$ and find the relation
\begin{eqnarray}
\label{eq:ptb_modal_no_pressure_sym}
&&\sigma\left(k_{z}\check{v}-k_{y}\check{w}\right)=-(k_{y}f_{\mathrm{h}}+k_{z}f_{\mathrm{v}})\check{u}-\nu_{0}k^{2}(k_{z}\check{v}-k_{y}\check{w})\nonumber\\
&&-k_{y}\alpha_{0}g\mathcal{T}_{\theta}\left(\frac{f_{\rm{v}}\Lambda}{\alpha_{0}g\mathcal{T}_{\theta}}\check{v}-\check{w}\right)/(\sigma+\kappa_{0}k^{2}).
\end{eqnarray}
Using the momentum equation (\ref{eq:ptb_modal_mom_x_sym}) and the continuity equation $\check{w}=-(k_{y}/k_{z})\check{v}$, we find the relation
\begin{equation}
(\sigma+\nu_{0}k^{2})\check{u}=\left(f_{\mathrm{v}}+\frac{k_{y}(\Lambda+f_{\mathrm{h}})}{k_{z}}\right)\check{v}.
\end{equation}
Putting these equations into Eq.~(\ref{eq:ptb_modal_no_pressure_sym}) leads to the following dispersion relation:
\begin{eqnarray}
\label{eq:dispersion_sym}
&&\left[\frac{\sigma k^{2}}{k_{z}}+\frac{\nu_{0}k^{4}}{k_{z}}+\frac{k_{y}(k_{z}f_{\mathrm{v}}\Lambda +k_{y}N^{2})}{k_{z}(\sigma+\kappa_{0}k^{2})}\right](\sigma+\nu_{0}k^{2})\nonumber\\
&&+(k_{y}f_{\mathrm{h}}+k_{z}f_{\mathrm{v}})\left(f_{\mathrm{v}}+\frac{k_{y}(\Lambda+f_{\mathrm{h}})}{k_{z}}\right)=0,
\end{eqnarray}
which can further be simplified into the following cubic equation
\begin{equation}
\label{eq:growth_cubic_sym}
\sigma^{3}+A_{2}\sigma^{2}+A_{1}\sigma+A_{0}=0,
\end{equation}
where 
\begin{eqnarray}
A_{2}&=&(2\nu_{0}+\kappa_{0})k^{2},\nonumber\\
A_{1}&=&\nu_{0}(\nu_{0}+2\kappa_{0})k^{4}\nonumber\\
&&+\frac{k_{y}\left(k_{z}f_{\mathrm{v}}\Lambda+k_{y}N^{2}\right)+(k_{y}f_{\mathrm{h}}+k_{z}f_{\mathrm{v}})\left[k_{z}f_{\mathrm{v}}+k_{y}(\Lambda+f_{\mathrm{h}})\right]}{k^{2}},\nonumber\\
A_{0}&=&\nu_{0}^{2}\kappa_{0}k^{6}+\nu_{0}k_{y}(k_{z}f_{\mathrm{v}}\Lambda+k_{y}N^{2})\nonumber\\
&&+\kappa_{0}(k_{z}f_{\mathrm{v}}+k_{y}f_{\mathrm{h}})\left[k_{z}f_{\mathrm{v}}+k_{y}(\Lambda+f_{\mathrm{h}})\right].
\end{eqnarray}
The growth rate $\sigma$ can be obtained by solving the cubic equation (\ref{eq:growth_cubic_sym}).
The advantage of the symmetric instability analysis is that both the diffusivity $\kappa_{0}$ and viscosity $\nu_{0}$ are included in the dispersion relation (\ref{eq:growth_cubic_sym}).
This allows us to examine explicitly without significant computations on how the viscous and thermal diffusions affect the stability of vertical shear flow in stably stratified-rotating fluids.

For the inviscid case with $\nu_{0}=0$, we can express the growth rate more explicitly. 
On the one hand, in the low-diffusivity (ld) limit $\kappa_{0}\rightarrow0$, the growth rate $\sigma$ can be expanded as
\begin{equation}
\sigma=\sigma_{\mathrm{ld}}^{(0)}-\kappa_{0}k^{2}\sigma_{\mathrm{ld}}^{(1)}+O(\kappa_{0}^{2}),
\end{equation}
where $\sigma_{\mathrm{ld}}^{(0)}$ is the leading-order growth rate
\begin{equation}
\label{eq:growth_nu0_kp0}
\sigma_{\mathrm{ld}}^{(0)}=\frac{1}{k}\sqrt{-k_{y}(k_{z}f_{\mathrm{v}}\Lambda +k_{y}N^{2})-(k_{y}f_{\mathrm{h}}+k_{z}f_{\mathrm{v}})\left[k_{z}f_{\mathrm{v}}+k_{y}(\Lambda+f_{\mathrm{h}})\right]}
\end{equation}
and $\sigma_{\mathrm{ld}}^{(1)}$ is the first-order term
\begin{equation}
\label{eq:growth_nu0_kp0_sigma1}
\sigma_{\mathrm{ld}}^{(1)}=\frac{k_{y}(k_{z}f_{\mathrm{v}}\Lambda+k_{y}N^{2})}{2\left[k_{y}(k_{z}f_{\mathrm{v}}\Lambda +k_{y}N^{2})+(k_{y}f_{\mathrm{h}}+k_{z}f_{\mathrm{v}})\left(k_{z}f_{\mathrm{v}}+k_{y}(\Lambda+f_{\mathrm{h}})\right)\right]}.
\end{equation}
On the other end, if we take the high-diffusivity (hd) limit $\kappa_{0}\rightarrow\infty$, the growth rate $\sigma$ can be expanded as
\begin{equation}
\sigma=\sigma_{\mathrm{hd}}^{(0)}-\frac{\sigma_{\mathrm{hd}}^{(1)}}{\kappa_{0}k^{2}}+O(\kappa_{0}^{-2}),
\end{equation}
where $\sigma_{\mathrm{hd}}^{(0)}$ is the leading-order growth rate in the limit $\kappa_{0}\rightarrow\infty$
\begin{equation}
\label{eq:growth_nu0_kpinf}
\sigma_{\mathrm{hd}}^{(0)}=\frac{1}{k}\sqrt{-(k_{y}f_{\mathrm{h}}+k_{z}f_{\mathrm{v}})\left[k_{z}f_{\mathrm{v}}+k_{y}(\Lambda+f_{\mathrm{h}})\right]}
\end{equation}
and $\sigma_{\mathrm{hd}}^{(1)}$ is the first-order term:
\begin{equation}
\label{eq:growth_nu0_kpinf_sigma1}
\sigma_{\mathrm{hd}}^{(1)}=\frac{k_{y}(k_{z}f_{\mathrm{v}}\Lambda+k_{y}N^{2})}{2k^{2}}.
\end{equation}
The leading-order terms $\sigma_{\mathrm{ld}}^{(0)}$ and $\sigma_{\mathrm{hd}}^{(0)}$ correspond to the growth rates in the limits $\kappa_{0}\rightarrow0$ and $\kappa_{0}\rightarrow\infty$, respectively. 
$\sigma_{\mathrm{ld}}^{(0)}$ and $\sigma_{\mathrm{hd}}^{(0)}$ vary in the wavenumber space $(k_{y},k_{z})$ and we can find their maxima: $\max\left(\sigma_{\mathrm{ld}}^{(0)}\right)$ and $\max\left(\sigma_{\mathrm{hd}}^{(0)}\right)$ and the corresponding wavenumbers at which the maxima are attained. 
To simplify the analysis, we introduce a phase angle $\phi$ between the two wavenumbers $k_{y}$ to $k_{z}$ as
\begin{equation}
\label{eq:wavenumber_angle}
\phi=\tan^{-1}\left(\frac{k_{z}}{k_{y}}\right),
\end{equation}
where $k_{y}=k\cos\phi$ and $k_{z}=k\sin\phi$.
By defining the phase angle $\phi$, we can express $\sigma_{\mathrm{ld}}^{(0)}$ and $\sigma_{\mathrm{hd}}^{(0)}$ in terms of $\phi$ as
\begin{eqnarray}
\label{eq:growth_nu0_kp0_phi}
\sigma_{\mathrm{ld}}^{(0)}&=&\left\{-\cos^{2}\phi\left[N^{2}+f_{\mathrm{h}}(\Lambda+f_{\mathrm{h}})\right]-\sin^{2}\phi f_{\mathrm{v}}^{2}\right.\nonumber\\
&&\left.-2\cos\phi\sin\phi f_{\mathrm{v}}\left(\Lambda+f_{\mathrm{h}}\right)\right\}^{\frac{1}{2}},
\end{eqnarray}
\begin{equation}
\label{eq:growth_nu0_kpinf_phi}
\sigma_{\mathrm{hd}}^{(0)}=\left[-\cos^{2}\phi f_{\rm{h}}(\Lambda+f_{\rm{h}})-\sin^{2}\phi f_{\rm{v}}^{2}-\cos\phi\sin\phi f_{\rm{v}}(\Lambda+2f_{\rm{h}})\right]^{\frac{1}{2}}.
\end{equation}
We see that the dependence on the wavenumber magnitude $k$ is eliminated in (\ref{eq:growth_nu0_kp0_phi})-(\ref{eq:growth_nu0_kpinf_phi}). 
\begin{figure}
   \centering
      \includegraphics[width=7cm]{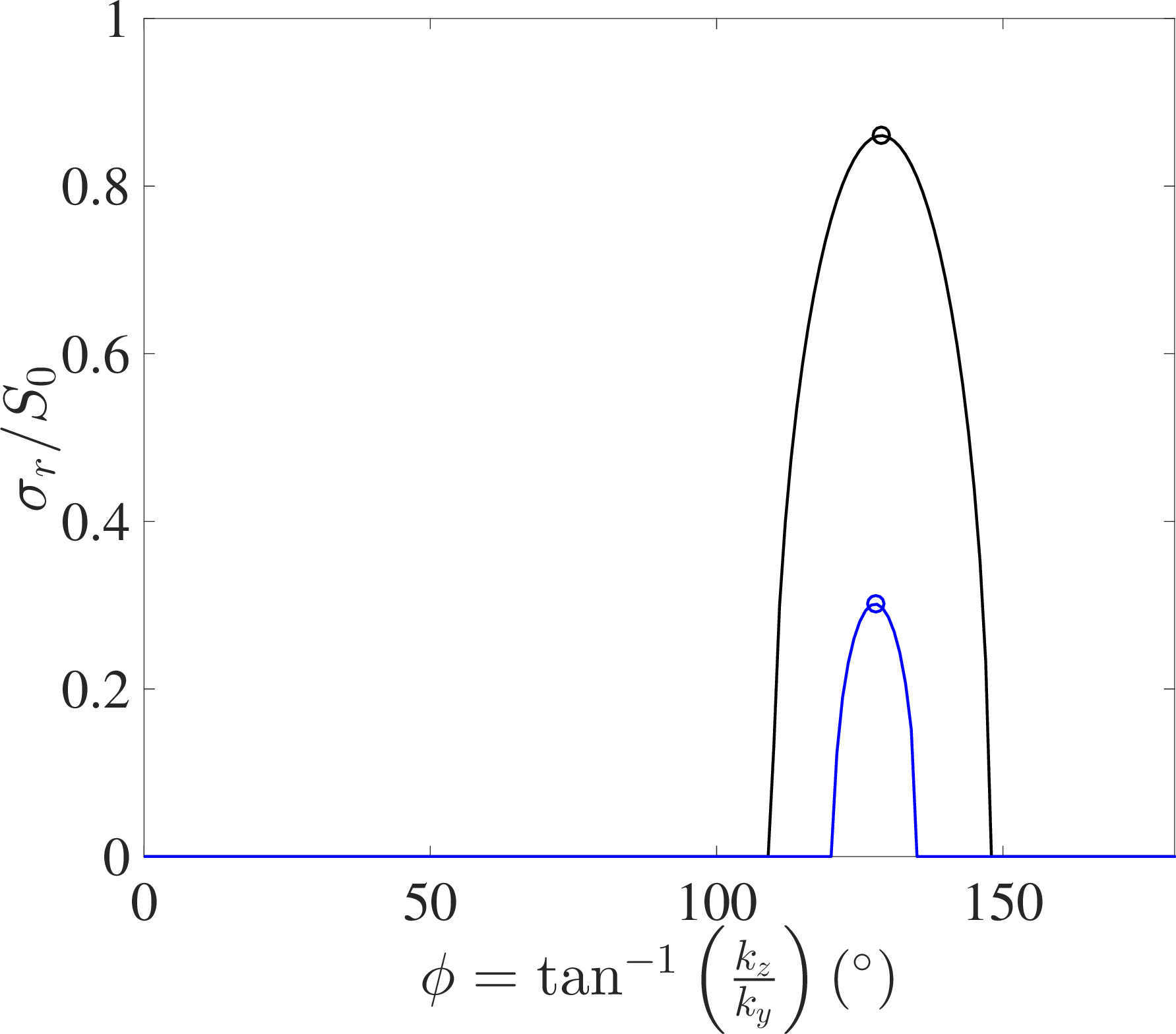}
              \caption{The growth rates $\sigma_{\mathrm{ld}}^{(0)}$ from (\ref{eq:growth_nu0_kp0_phi}) (black) and $\sigma_{\mathrm{hd}}^{(0)}$ from (\ref{eq:growth_nu0_kpinf_phi}) versus the phase angle $\phi$ at $\Omega_{0}/S_{0}=1$, $\theta=45^{\circ}$, and $Ri=0.1$.  
              Black and blue empty circles denote the maximums (\ref{eq:sigmax_nu0_kp0}) and (\ref{eq:sigmax_nu0_kpinf}) at the critical phase angles (\ref{eq:kykz_nu0_kp0}) and (\ref{eq:kykz_nu0_kpinf}), respectively.}
         \label{Fig_sigma_phi}
\end{figure}
In Fig.~\ref{Fig_sigma_phi}, we display these growth rates at the leading order as a function of the phase angle $\phi$ at a mid-latitude $\theta=45^{\circ}$ for $\Omega_{0}/S_{0}=1$ and $Ri=0.1$. 
For both cases, the growth rates are positive only in a finite range of $\phi$ and reach their maximums $\max(\sigma_{\mathrm{ld}}^{(0)})\simeq0.86$ and $\max(\sigma_{\mathrm{hd}}^{(0)})\simeq0.30$ around $\phi\simeq129^{\circ}$ and $\phi\simeq128^{\circ}$ (i.e. $k_{z}/k_{y}=1.1528$ and $k_{z}/k_{y}=1.1499$), respectivley. 
Such maximum values and critical phase angles can be easily obtained by taking the derivatives with respect to $\phi$ on the growth rates and finding their zeros (i.e. $\partial\sigma/\partial \phi=0$).
For $\sigma_{\mathrm{ld}}^{(0)}$ in the low-diffusivity limit $\kappa_{0}=0$, we have the critical phase angle $\phi_{c,\mathrm{ld}}$:
\begin{equation}
\label{eq:kykz_nu0_kp0}
\phi_{c,\mathrm{ld}}=\tan^{-1}\left[\frac{K_{0}-\sqrt{K_{0}^{2}+4f_{\rm{v}}^{2}(\Lambda+f_{\rm{h}})^{2}}}{2f_{\rm{v}}(\Lambda+f_{\rm{h}})}\right],
\end{equation}
where $K_{0}=f_{\mathrm{v}}^{2}-N^{2}-f_{\mathrm{h}}(\Lambda+f_{\rm{h}})$, at which the growth rate $\sigma_{\mathrm{ld}}^{(0)}$ reaches its maximum as
\begin{equation}
\label{eq:sigmax_nu0_kp0}
\max(\sigma_{\mathrm{ld}}^{(0)})=\sqrt{\frac{-K_{1}+\sqrt{K_{1}^{2}+4f_{\rm{v}}^{2}\left[(\Lambda+f_{\rm{h}})\Lambda-N^{2}\right]}}{2}},
\end{equation}
where $K_{1}=f_{\rm{v}}^{2}+N^{2}+f_{\rm{h}}(\Lambda+f_{\rm{h}})$.
We note that this maximum growth rate of the symmetric instability is equivalent to the maximum growth rate (\ref{eq:WKBJ_2nd_Taylor_sigma0}) for inertial instability obtained by the WKBJ analysis for the non-diffusive case $\kappa_{0}=0$ in the limit $k_{y}\rightarrow\infty$.
Similarly, we can find the critical phase angle and the corresponding maximum of $\sigma_{\mathrm{hd}}^{(0)}$ in the high-diffusivity limit $\kappa_{0}\rightarrow\infty$ as
\begin{equation}
\label{eq:kykz_nu0_kpinf}
\phi_{c,\mathrm{hd}}=\tan^{-1}\left[\frac{f_{\mathrm{v}}^{2}-(\Lambda+f_{\mathrm{h}})f_{\mathrm{h}}-\sqrt{\left[f_{\mathrm{v}}^{2}+\left(\Lambda+f_{\mathrm{h}}\right)^{2}\right]\left(f_{\mathrm{v}}^{2}+f_{\mathrm{h}}^{2}\right)}}{f_{\mathrm{v}}\left(\Lambda+2f_{\mathrm{h}}\right)}\right],
\end{equation}
\begin{equation}
\label{eq:sigmax_nu0_kpinf}
\max(\sigma_{\mathrm{hd}}^{(0)})=\sqrt{\frac{-f_{\mathrm{v}}^{2}-f_{\mathrm{h}}(\Lambda+f_{\mathrm{h}})+\sqrt{\left[f_{\mathrm{v}}^{2}+\left(\Lambda+f_{\mathrm{h}}\right)^{2}\right]\left(f_{\mathrm{v}}^{2}+f_{\mathrm{h}}^{2}\right)}}{2}}.
\end{equation}
This maximum growth rate is also equivalent to the maximum growth rate (\ref{eq:WKBJ_ODE4_gamma0}) of the inertial instability obtained by the WKBJ analysis in the high-diffusivity limit $\kappa_{0}\rightarrow\infty$. 

 \begin{figure*}
   \centering
   \includegraphics[height=4.8cm]{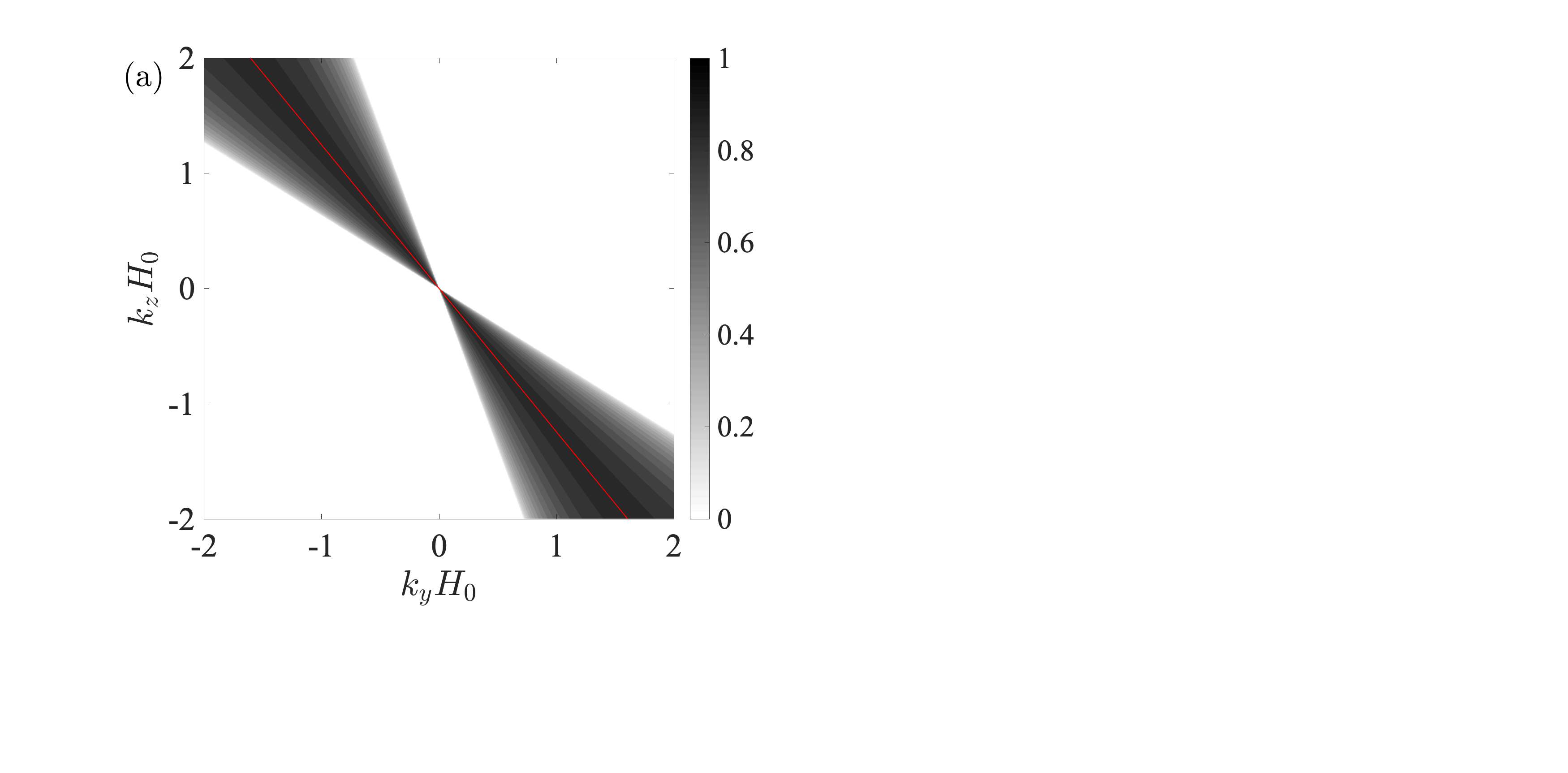}
   \includegraphics[height=4.8cm]{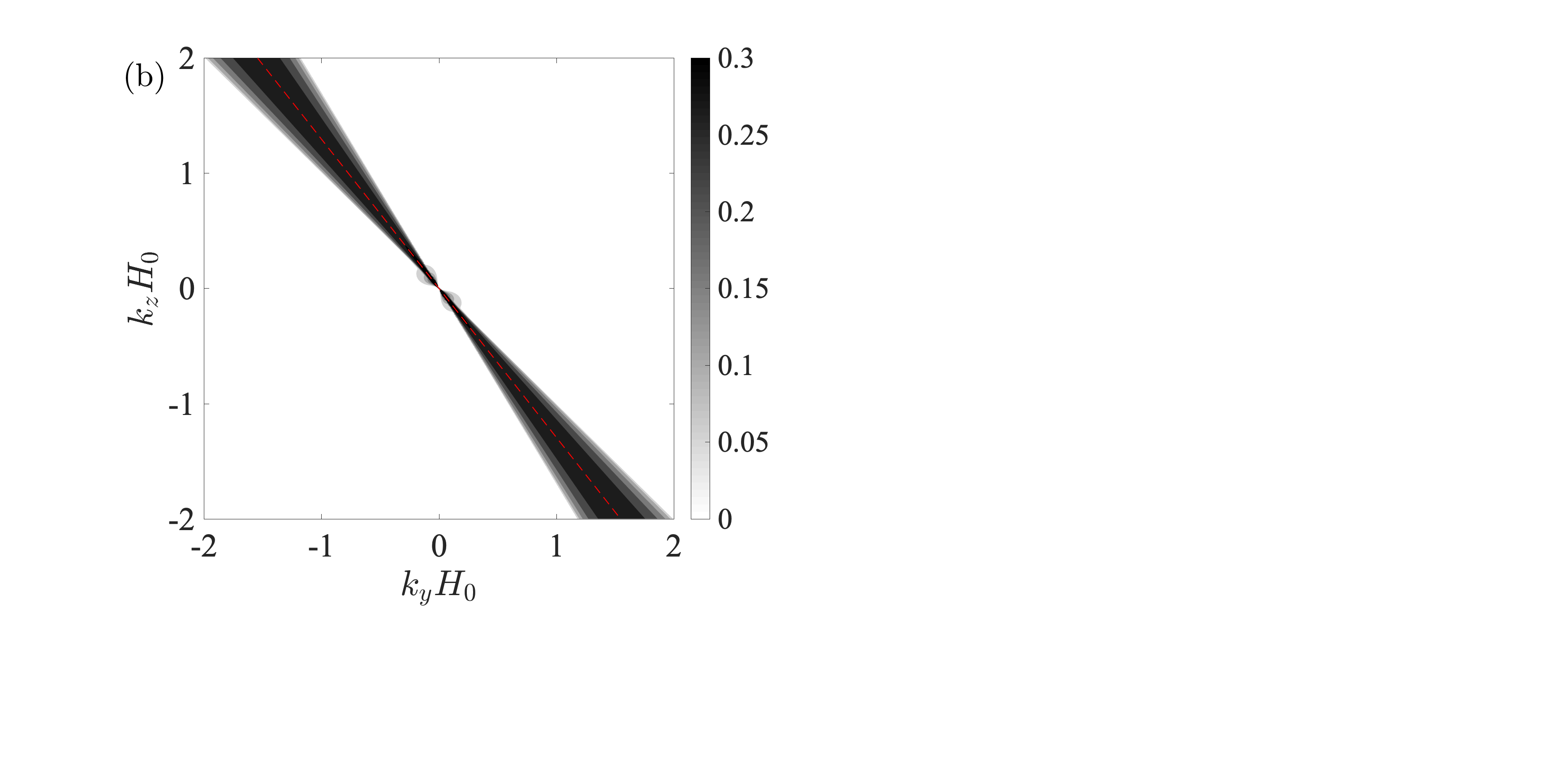}
   \includegraphics[height=4.8cm]{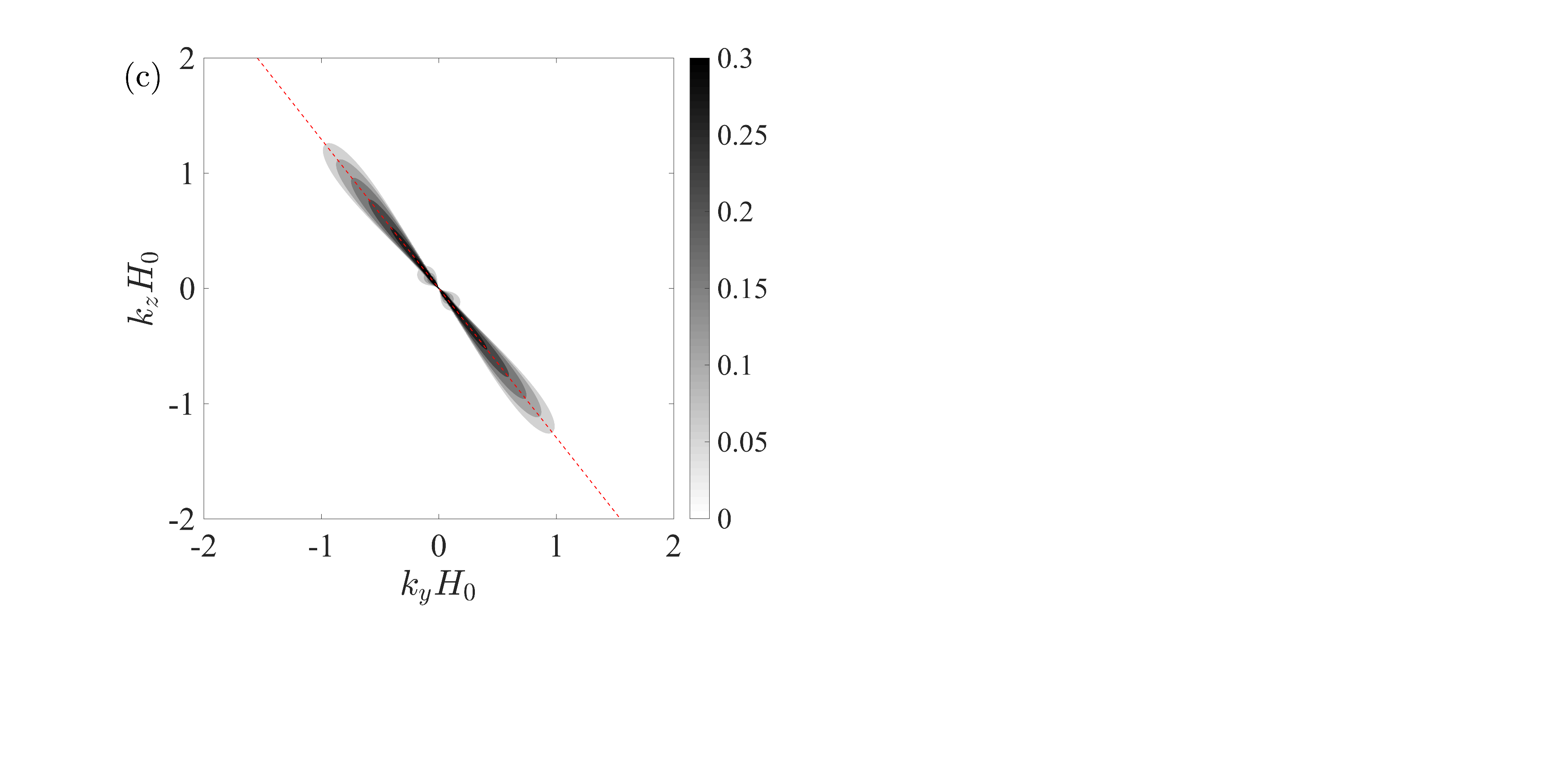}
      \caption{Contours of the growth rate $\sigma/S_{0}$ in the parameter space of wavenumbers $(k_{y},k_{z})$ for symmetric instability at $\Omega_{0}/S_{0}=1$, $\theta=45^{\circ}$, and $Ri=0.1$ for (a) $Re=Pe=\infty$, (b) $Re=\infty$ and $Pe=0.01$, and (c) $Re=10$ and $Pe=0.01$. 
      Red solid line in (a) indicates the wavenumber ratio (\ref{eq:kykz_nu0_kp0}) and red dashed lines in (b,c) indicate (\ref{eq:kykz_nu0_kpinf}).
      }
         \label{Fig_growth_contours_sym}
   \end{figure*}
Figure \ref{Fig_growth_contours_sym} shows contours of the growth rate $\sigma$ in the wavenumber space $(k_{y},k_{z})$ calculated from the cubic equation (\ref{eq:growth_cubic_sym}) for various $Re$ and $Pe$ at $\Omega_{0}/S_{0}=1$, $\theta=45^{\circ}$ and $Ri=0.1$.
For these parameters, $\sigma$ is positive in the second and fourth quadrants. 
For the inviscid and non-diffusive case shown in panel a, we see that the growth rate remains constant at a fixed phase angle $\phi=\tan^{-1}(k_{z}/k_{y})$ and reaches its maximum at the critical phase angle $\phi_{c,\mathrm{ld}}$.
For a highly-diffusive case at $Pe=0.01$ shown in panel b, we see that $\sigma$ is positive in a narrower range of $\phi$ in the second and fourth quadrants and it remains constant at a fixed phase angle $\phi$ only for a sufficiently large $k$.  
For a small $k<<1$, it is unstable in leaf-shaped regions in the second/fourth quadrants. 
For a viscous case in panel c, the symmetric instability is suppressed and the growth rate is only positive for finite wavenumbers $k_{y}$ and $k_{z}$, which implies that the growth rate now depends on the wavenumber amplitude $k$. 
For a fixed $k$, we verified that the maximum growth rate for viscous cases still occurs at the critical phase angle $\phi_{c}$, which is close to the one $\phi_{c,\mathrm{hd}}$ obtained for the inviscid case in the high-diffusivity limit $\kappa_{0}\rightarrow\infty$. 

We verified above that the maximum growth rates of the symmetric instability are equal to those of the inertial instability for both low and high-diffusivity limits, if we consider the same shear $\Lambda=S_{0}$. 
A question to raise is whether we can compare not only the maximum growth rates of the two instabilities but also their growth rates at arbitrary wavenumbers $k_{y}$ and $k_{z}$.  
The problem in the inertial instability of the hyperbolic tangent shear flow is that the velocity varies along the $z$-direction so the vertical wavenumber $k_{z}$ of inertial instability modes also changes locally as verified in the WKBJ solutions (\ref{eq:WKBJ_2nd_wavelike}) and (\ref{eq:WKBJ_ODE4_w}).
For instance, for the inviscid and non-diffusive case ($\nu_{0}=\kappa_{0}=0$), we can find the following local vertical wavenumber $k_{z}(z)$ by taking the derivative with respect to $z$ in the exponent of the WKBJ solution (\ref{eq:WKBJ_2nd_wavelike}):
\begin{equation}
k_{z}(z)=-\frac{k_{y}f_{\mathrm{v}}(U'(z)+f_{\mathrm{h}})}{\sigma^{2}+f_{\mathrm{v}}^{2}}\pm k_{y}\sqrt{-\Delta(z)},
\end{equation}
where $\pm$ signs appear due to the two wavelike solutions with non-zero amplitudes $B_{\pm}$ in the WKBJ solution (\ref{eq:WKBJ_2nd_wavelike}).
As we equate the shear $\Lambda$ of the symmetric instability with the inertial instability shear $S_{0}$ evaluated at $z=0$, we also consider the local wavenumber at $z=0$ as 
\begin{equation}
\label{eq:WKBJ_kz0}
k_{z,0}\equiv k_{z}(0)=-\frac{k_{y}f_{\mathrm{v}}(S_{0}+f_{\mathrm{h}})}{\sigma^{2}+f_{\mathrm{v}}^{2}}\pm k_{y}\sqrt{-\Delta(0)},
\end{equation}
where
\begin{equation}
\Delta(0)=\frac{\sigma^{2}+N^{2}+f_{\mathrm{h}}(S_{0}+f_{\mathrm{h}})}{\sigma^{2}+f^{2}_{\mathrm{v}}}-\frac{f_{\mathrm{v}}^{2}\left(S_{0}+f_{\mathrm{h}}\right)^{2}}{(\sigma^{2}+f_{\mathrm{v}}^{2})^{2}}.
\end{equation}
We can formulate the expression (\ref{eq:WKBJ_kz0}) as
\begin{equation}
\left[\frac{k_{z,0}}{k_{y}}+\frac{f_{\mathrm{v}}(S_{0}+f_{\mathrm{h}})}{\sigma^{2}+f_{\mathrm{v}}^{2}}\right]^{2}=\left[\frac{f_{\mathrm{v}}^{2}\left(S_{0}+f_{\mathrm{h}}\right)^{2}}{(\sigma^{2}+f_{\mathrm{v}}^{2})^{2}}-\frac{\sigma^{2}+N^{2}+f_{\mathrm{h}}(S_{0}+f_{\mathrm{h}})}{\sigma^{2}+f^{2}_{\mathrm{v}}}\right],
\end{equation}
and simplify it as
\begin{equation}
k_{z,0}^{2}+2\frac{k_{z,0}k_{y}f_{\mathrm{v}}(S_{0}+f_{\mathrm{h}})}{\sigma^{2}+f_{\mathrm{v}}^{2}}
+k_{y}^{2}\frac{\sigma^{2}+N^{2}+f_{\mathrm{h}}(S_{0}+f_{\mathrm{h}})}{\sigma^{2}+f^{2}_{\mathrm{v}}}=0.
\end{equation}
This leads to the following growth rate
\begin{align}
\label{eq:WKBJ_growth_sym}
   \sigma &=\nonumber\\
       &\!\!\sqrt{\frac{-k_{y}(k_{z,0}f_{\mathrm{v}}S_{0} +k_{y}N^{2})-(k_{y}f_{\mathrm{h}}+k_{z,0}f_{\mathrm{v}})\left[k_{z,0}f_{\mathrm{v}}+k_{y}(S_{0}+f_{\mathrm{h}})\right]}{k_{z,0}^{2}+k_{y}^{2}}}.
\end{align}
We see that this growth rate (\ref{eq:WKBJ_growth_sym}) is equal to that of the symmetric instability (\ref{eq:growth_nu0_kp0}) when $S_{0}=\Lambda$.

Similarly, for the high-diffusivity case $\kappa\rightarrow\infty$, we find the following local wavenumber $k_{z}(z)$ from the WKBJ solution (\ref{eq:WKBJ_ODE4_w}):
\begin{equation}
k_{z}(z)=-\frac{k_{y}f_{\mathrm{v}}(U'+2f_{\mathrm{h}})}{2(\sigma^{2}+f_{\mathrm{v}}^{2})}\pm k_{y}\sqrt{-\Gamma(z)}.
\end{equation}
By taking the local wavenumber at $z=0$, we find a relation
\begin{equation}
\left[k_{z,0}+\frac{k_{y}f_{\mathrm{v}}(S_{0}+2f_{\mathrm{h}})}{2(\sigma^{2}+f_{\mathrm{v}}^{2})}\right]^{2}=k_{y}^{2}\left[\frac{f^{2}_{\mathrm{v}}(S_{0}+2f_{\mathrm{h}})^{2}}{4(\sigma^{2}+f^{2}_{\mathrm{v}})^{2}}-\frac{\sigma^{2}+f_{\mathrm{h}}(S_{0}+f_{\mathrm{h}})}{\sigma^{2}+f^{2}_{\mathrm{v}}}\right],
\end{equation}
which leads to the following growth rate:
\begin{equation}
\label{eq:WKBJ_growth_sym_Pe0}
\sigma=\sqrt{\frac{-(k_{y}f_{\mathrm{h}}+k_{z,0}f_{\mathrm{v}})\left[k_{z,0}f_{\mathrm{v}}+k_{y}(S_{0}+f_{\mathrm{h}})\right]}{k_{z,0}^{2}+k_{y}^{2}}},
\end{equation}
which is the same as the growth rate of the symmetric instability (\ref{eq:growth_nu0_kpinf}) when $S_{0}=\Lambda$. 

As shown by the analysis above, symmetric instability of a linear shear flow $U=\Lambda z$ is closely linked to the inertial instability of a hyperbolic-tangent shear flow.
This is an important mathematical result on the inertial instability that shows that the growth rate does not depend on the profile of the shear but it rather depends only on the strength of the shear. 
We note that this argument is not applicable to the inflectional instability which arises due to an inflection point, a unique feature that depends strongly on the profile of the shear flow. 
Nevertheless, an advantage of using the symmetric instability is that we have analytic expressions of the growth rate even for finite viscosity $\nu$ and diffusivity $\kappa$.
This is practical when we use the analytical expressions of the growth rate to deduce turbulent viscosity models to be implemented in stellar evolution simulations. 
In the next section, we demonstrate how our results on vertical shear instabilities can further be used to deduce the turbulent effective viscosities and describe turbulent dissipation in stellar radiation zones. 

\section{Turbulent transport induced by vertical shear instabilities}
\label{sec:Turbulent}
\subsection{Evolution of perturbation energy and turbulent effective viscosities}
Stability results such as the growth rate, eigenfunction or wavenumbers have been used to construct turbulent viscosity models, which can be derived analytically from a simple eddy viscosity argument with basic scaling laws \citep[][]{Spruit2002,Fulleretal2019,Park2020,Park2021}.
Numerical simulations of turbulence in a shearing box can also be used to formulate and compute turbulent effective viscosities.
For instance, nonlinear simulations of the GSF instability using a local Cartesian model at either the equator \citep[][]{Barker2019} or at a general latitude \citep[][]{Barker2020,Dymott2023} can provide the characteristics of turbulence driven by the GSF instability. 
From these turbulence simulations, an effective viscosity $\nu_{E}$ is suggested to follow a scaling as $\nu_{E}=\nu_{0}\Omega_{0}^{2}S_{0}^{-1}N^{-1}Pr^{-1/2}<u_{x}u_{y}>$ where we use here our notations such as $S_{0}$ or $\Omega_{0}$ to facilitate the comparison and $<u_{x}u_{y}>$ is the averaged Reynolds stress of the order $O(10)-(100)$ obtained from their simulations.

In this section, we will use linear stability analysis results to examine turbulent dissipation induced by vertical shear instabilities and deduce turbulent viscosity models. 
The proposed viscosity models are similar to those from previous studies but we will derive them by considering more specifically nonlinear saturation of shear instabilities based on the perturbation energy argument.
We will focus on how turbulent dissipation depends on the two different instabilities (i.e. the inertial and inflectional ones) and parameters such as the P\'eclet number $Pe$, the Prandtl number $Pr$ or the Richardson number $Ri$.
To do so, we first define the total energy of perturbation per unit density as
\begin{equation}
\label{eq:perturbation_energy_total}
\tilde{E}=\frac{1}{2}\left<\tilde{u}^{2}+\tilde{v}^{2}+\tilde{w}^{2}+N^{2}\tilde{\mathcal{T}}^{2}\right>,
\end{equation}
\citep[see also,][]{Park2017} where $\tilde{\mathcal{T}}=\tilde{T}/\mathcal{T}_{\Theta}$ is the normalized temperature and $<\cdot>$ denotes the volume integral defined as
\begin{equation}
\label{eq:norm}
\left<\tilde{X}\right>=\int_{-l_{x}}^{l_{x}}\int_{-l_{y}}^{l_{y}}\int_{-l_{z}}^{l_{z}}\tilde{X}\mathrm{d}x\mathrm{d}y\mathrm{d}z,
\end{equation}
where $(l_{x},l_{y},l_{z})$ are the domain lengths in the Cartesian coordinate defined as $l_{x}=\pi/k_{x}$, $l_{y}=\pi/k_{y}$, $l_{z}=\pi/k_{z}$ or $l_{z}\rightarrow\infty$ if the boundary conditions are non-periodic and open in the $z$-direction.
The total energy per unit density $\tilde{E}$ is comprised of the kinetic energy per unit density $\tilde{E}_{\mathrm{K}}$ and potential energy per unit density $\tilde{E}_{\mathrm{P}}$ as $\tilde{E}=\tilde{E}_{\mathrm{K}}+\tilde{E}_{\mathrm{P}}$ where
\begin{equation}
\label{eq:perturbation_energy}
\tilde{E}_{\mathrm{K}}=\frac{1}{2}\left<\tilde{u}^{2}+\tilde{v}^{2}+\tilde{w}^{2}\right>,~~
\tilde{E}_{\mathrm{P}}=\frac{1}{2}\left<N^{2}\tilde{\mathcal{T}}^{2}\right>.
\end{equation}
After some manipulations applied to the perturbation equations (\ref{eq:ptb_x_mom})-(\ref{eq:ptb_diffusion}), we can express the time evolution of the total perturbation energy as
\begin{eqnarray}
\label{eq:energy_time}
\lefteqn{\frac{\partial\tilde{E}}{\partial t}=-\left<U'\tilde{u}\tilde{w}\right>+\left<U'f_{\mathrm{v}}\tilde{v}\tilde{\mathcal{T}}\right>-\nu_{0}\left<\left(\frac{\partial\tilde{u}}{\partial x}\right)^{2}+\left(\frac{\partial\tilde{{u}}}{\partial y}\right)^{2}+\left(\frac{\partial\tilde{u}}{\partial z}\right)^{2}\right.}\nonumber\\
&&\left.+\left(\frac{\partial\tilde{v}}{\partial x}\right)^{2}+\left(\frac{\partial\tilde{{v}}}{\partial y}\right)^{2}+\left(\frac{\partial\tilde{v}}{\partial z}\right)^{2}+\left(\frac{\partial\tilde{w}}{\partial x}\right)^{2}+\left(\frac{\partial\tilde{{w}}}{\partial y}\right)^{2}+\left(\frac{\partial\tilde{w}}{\partial z}\right)^{2}\right>\nonumber\\
&&-N^{2}\kappa_{0}\left<\left(\frac{\partial\tilde{\mathcal{T}}}{\partial x}\right)^{2}+\left(\frac{\partial\tilde{\mathcal{T}}}{\partial y}\right)^{2}+\left(\frac{\partial\tilde{\mathcal{T}}}{\partial z}\right)^{2}\right>.
\end{eqnarray}
We note that there is no contribution from pressure and nonlinear terms after the continuity equation (\ref{eq:ptb_continuity}) is taken into account \citep[see also,][]{Schmid2001}. 
The viscous and thermal dissipations act a stabilizing role on the perturbation energy with positive $\nu_{0}$ and $\kappa_{0}$ while the shear $U'$ can either destabilize or stabilize depending on the signs of $\tilde{u}$, $\tilde{w}$, $\tilde{v}$ and $\tilde{\mathcal{T}}$. 
The first term on the right-hand side of (\ref{eq:energy_time}) represents the growth by the Orr mechanism when the shear $U'$ is negatively correlated with $\tilde{u}\tilde{w}$ \citep[][]{Orr1907}. 
The second term denotes the energy growth occurring when the vertical Coriolis parameter $f_{\mathrm{v}}$ is non-zero and positive (negative) and the shear $U'$ is positively (negatively) correlated with $\tilde{v}\tilde{\mathcal{T}}$.
If we apply the normal mode (\ref{eq:ptb_normal_mode}) to the equation (\ref{eq:energy_time}), we obtain the following equation
\begin{eqnarray}
\label{eq:energy_time_mode_shape}
\lefteqn{2\sigma_{r}\hat{E}=\int_{-l_{z}}^{l_{z}}U'\left[\frac{f_{\mathrm{v}}\left(\hat{v}\hat{\mathcal{T}}^{*}+\hat{v}^{*}\hat{\mathcal{T}}\right)-\left(\hat{u}\hat{w}^{*}+\hat{u}^{*}\hat{w}\right)}{2}\right]\mathrm{d}z}\nonumber\\
&&-\nu_{0}\int_{-l_{z}}^{l_{z}}\left[k^{2}\left(|\hat{u}|^{2}+|\hat{v}|^{2}+|\hat{w}|^{2}\right)+\left|\frac{\partial\hat{u}}{\partial z}\right|^{2}+\left|\frac{\partial\hat{v}}{\partial z}\right|^{2}+\left|\frac{\partial\hat{w}}{\partial z}\right|^{2}\right]\mathrm{d}z\nonumber\\
&&-N^{2}\kappa_{0}\int_{-l_{z}}^{l_{z}}\left[k^{2}|\hat{T}|^{2}+\left|\frac{\partial\hat{T}}{\partial z}\right|^{2}\right]\mathrm{d}z,
\end{eqnarray}
where $\hat{E}$ is the modal energy per unit density defined as
\begin{equation}
\label{eq:ptb_energy_modal}
\hat{E}=\int_{-l_{z}}^{l_{z}}\left(|\hat{u}|^{2}+|\hat{v}|^{2}+|\hat{w}|^{2}+N^{2}|\hat{{T}}|^{2}\right)\mathrm{d}z.
\end{equation}

Let us now consider a nonlinear saturation process where perturbations grow due to vertical shear instabilities and reach an equilibrium state. 
At this nonlinearly-saturated equilibrium state, there is no growth (i.e., $\sigma_{r}=0$) and the base flow is distorted as $U+\delta U$, balancing with the modified perturbations $\hat{\textbf{q}}+\delta\hat{\textbf{q}}$ where $\hat{\textbf{q}}=(\hat{u},\hat{v},\hat{w},\hat{p},\hat{T})$. 
These base and perturbation states at nonlinear saturation with $\sigma_{r}=0$ lead to the following equation:
\begin{eqnarray}
&&0=\int_{-l_{z}}^{l_{z}}\left(U'+\delta U'\right)\left[\frac{f_{\mathrm{v}}\left(\hat{v}\hat{\mathcal{T}}^{*}+\hat{v}^{*}\hat{\mathcal{T}}\right)-\left(\hat{u}\hat{w}^{*}+\hat{u}^{*}\hat{w}\right)}{2}\right]\mathrm{d}z\nonumber\\
&&-\nu_{0}\int_{-l_{z}}^{l_{z}}\left[k^{2}\left(|\hat{u}|^{2}+|\hat{v}|^{2}+|\hat{w}|^{2}\right)+\left|\frac{\partial\hat{u}}{\partial z}\right|^{2}+\left|\frac{\partial\hat{v}}{\partial z}\right|^{2}+\left|\frac{\partial\hat{w}}{\partial z}\right|^{2}\right]\mathrm{d}z\nonumber\\
&&-N^{2}\kappa_{0}\int_{-l_{z}}^{l_{z}}\left[k^{2}|\hat{T}|^{2}+\left|\frac{\partial\hat{T}}{\partial z}\right|^{2}\right]\mathrm{d}z+O(\delta\hat{\textbf{q}}),
\end{eqnarray}
where $O(\delta\hat{\textbf{q}})$ denotes the term of order $\delta$ due to the distortion in perturbation mode shape $\delta\hat{\textbf{q}}$.
In the process of nonlinear saturation, both distortions $\delta U$ and $\delta\hat{\textbf{q}}$ are important and they interact each other to generate turbulent transport. 
However, if we assume that the changes in perturbation is negligible \citep[i.e. $\delta\hat{\textbf{q}}=0$; see also,][]{Stuart1958} and if we introduce a turbulent effective viscosity $\nu_{\mathrm{v}}$ that presumably maintains the balance between the base flow distortion and turbulent dissipation as follows:
\begin{eqnarray}
&&\delta U'\left[\frac{f_{\mathrm{v}}\left(\hat{v}\hat{\mathcal{T}}^{*}+\hat{v}^{*}\hat{\mathcal{T}}\right)-\left(\hat{u}\hat{w}^{*}+\hat{u}^{*}\hat{w}\right)}{2}\right]=\nonumber\\
&&-\nu_{\mathrm{v}}\left[k^{2}\left(|\hat{u}|^{2}+|\hat{v}|^{2}+|\hat{w}|^{2}+N^{2}|\hat{T}|^{2}\right)\right.\nonumber\\
&&\left.+\left|\frac{\partial\hat{u}}{\partial z}\right|^{2}+\left|\frac{\partial\hat{v}}{\partial z}\right|^{2}+\left|\frac{\partial\hat{w}}{\partial z}\right|^{2}+N^{2}\left|\frac{\partial\hat{T}}{\partial z}\right|^{2}\right],
\end{eqnarray}
which implies the turbulent energy growth induced by the shear distortion $\delta U'$ balances with turbulent dissipation with $\nu_{\mathrm{v}}$ as a turbulent viscosity coefficient, we can find the following relation for $\nu_{\mathrm{v}}$ in terms of the energy $\hat{E}$ and the growth rate $\sigma_{r}$:
\begin{eqnarray}
\label{eq:turbulent_viscosity}
&&\nu_{\mathrm{v}}\int_{-l_{z}}^{l_{z}}\left[k^{2}\left(|\hat{u}|^{2}+|\hat{v}|^{2}+|\hat{w}|^{2}+N^{2}|\hat{T}|^{2}\right)+\left|\frac{\partial\hat{u}}{\partial z}\right|^{2}+\left|\frac{\partial\hat{v}}{\partial z}\right|^{2}\right.\nonumber\\
&&\left.+\left|\frac{\partial\hat{w}}{\partial z}\right|^{2}+N^{2}\left|\frac{\partial\hat{T}}{\partial z}\right|^{2}\right]\mathrm{d}z=2\sigma_{r}\hat{E}.
\end{eqnarray}
Although the neglection of the perturbation distortion $\delta\hat{\textbf{q}}$ may not reflect precise turbulent processes, the equation (\ref{eq:turbulent_viscosity}) is still crucial as it shows a clear link between the turbulent effective viscosity and the growth rate of the vertical shear instabilities. 
%We also note that this semi-analytical expression is practical to be adopted in stellar evolution simulations. 
If $k_{x}=0$ is considered like in the case of symmetric instability, we obtain the following simplified equation 
\begin{equation}
\label{eq:turbulent_viscosity_sym_1}
\nu_{\mathrm{v}}(k_{y}^{2}+k_{z}^{2})2\check{E}=2\sigma_{r}\check{E},
\end{equation}
where $\check{E}=|\check{u}|^{2}+|\check{v}|^{2}+|\check{w}|^{2}+N^{2}|\check{T}|^{2}$.
This implies
%The turbulent viscosity can be simply expressed as
\begin{equation}
\label{eq:turbulent_viscosity_sym_2}
\nu_{\mathrm{v}}=\frac{\sigma_{r}}{k_{y}^{2}+k_{z}^{2}}.
\end{equation}
The turbulent viscosity $\nu_{\rm{v}}$ expressed in terms of the growth rate $\sigma_{r}$ is equivalent to the one proposed by \citet{Spruit2002,Fulleretal2019} where the turbulent viscosity is proportional to the growth rate.
We note that, while they propose the turbulent viscosity based on a simple scaling law, our model (\ref{eq:turbulent_viscosity_sym_2}) is derived using the perturbation energy argument and nonlinear saturation of the instability. 
%If we adopt descriptions of the turbulent viscosities $\nu_{\mathrm{h},\mathrm{v}}$ and $\nu_{\mathrm{v},\mathrm{v}}$ proposed by \citet{Mathisetal2018}, which describes the horizontal and vertical turbulent dissipation, respectively, we express the viscosities as follows:
%\begin{equation}
%\label{eq:turbulent_viscosities_hv}
%\nu_{\mathrm{h},\mathrm{v}}=\frac{\sigma_{r}}{k_{y}^{2}},~
%\nu_{\mathrm{v},\mathrm{v}}=\frac{\sigma_{r}}{k_{z}^{2}}.
%\end{equation}
%These viscosities satisfy the following relation
%\begin{equation}
%\frac{1}{\nu_{\mathrm{h},\mathrm{v}}}+\frac{1}{\nu_{\mathrm{v},\mathrm{v}}}=\frac{1}{\nu_{\mathrm{v}}}.
%\end{equation}

\subsection{Turbulent viscosities for inertial and inflectional instabilities}
\label{sec:turbulent_viscosity}

 \begin{figure*}
   \centering
   \includegraphics[height=4.6cm]{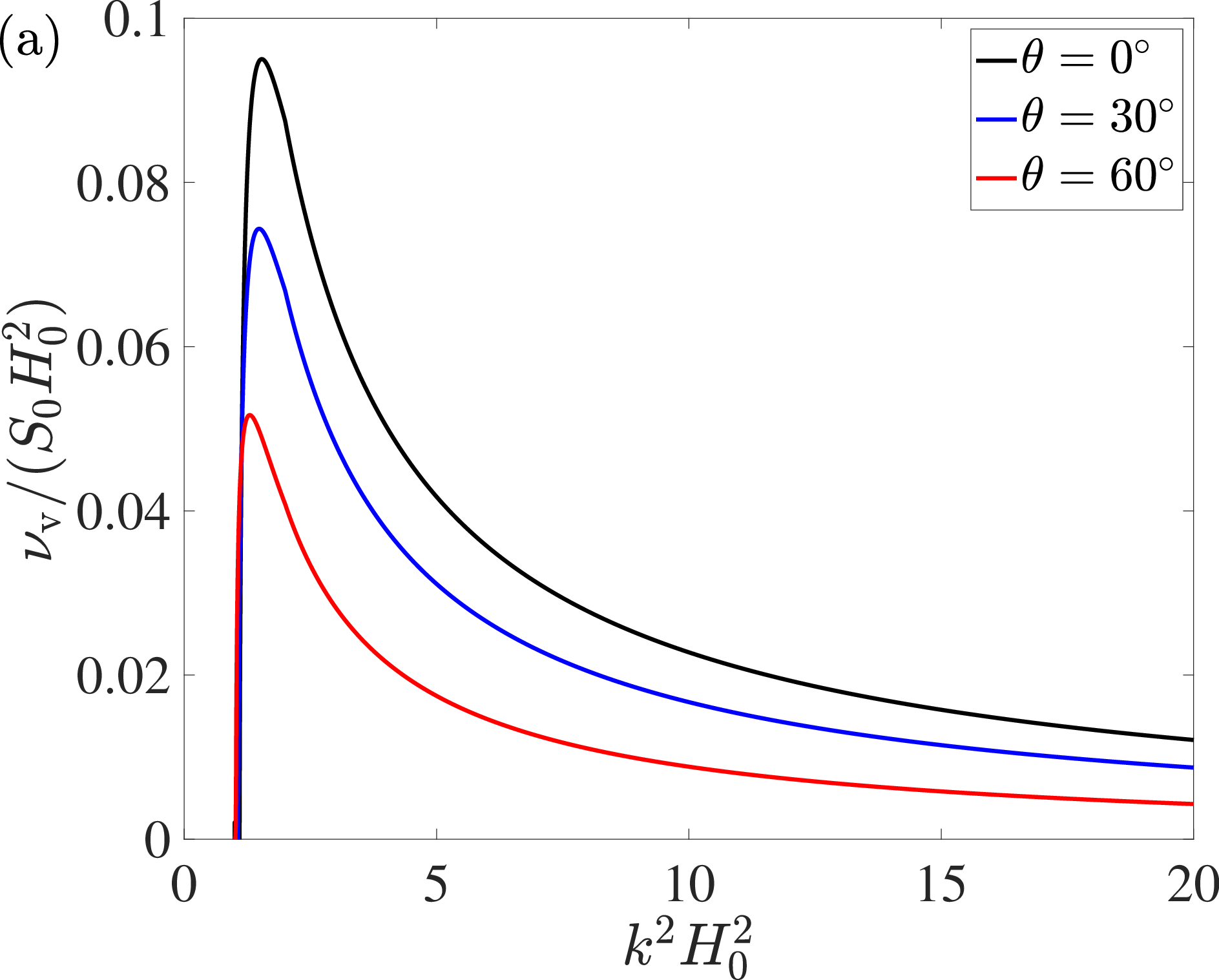}
   \includegraphics[height=4.6cm]{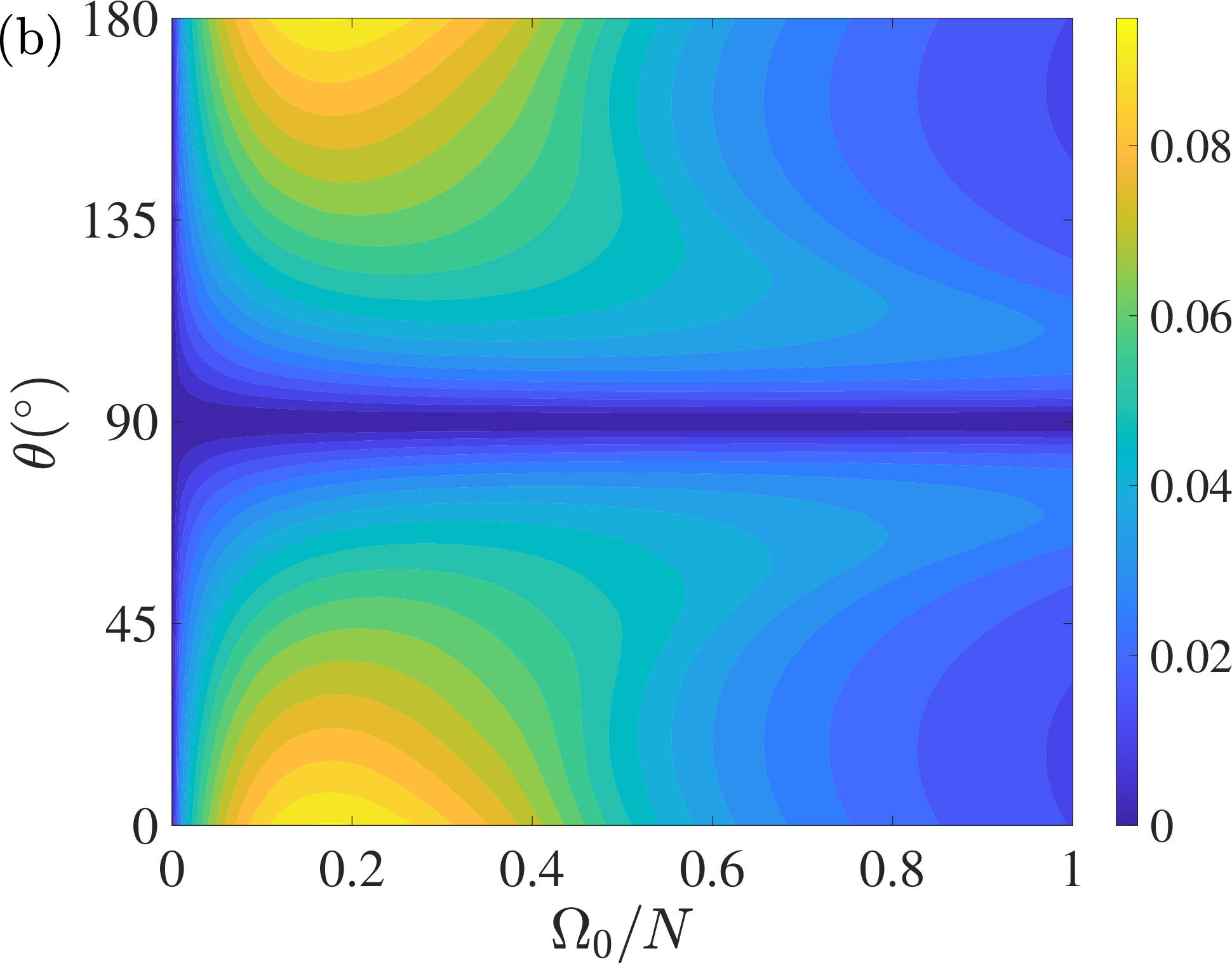}
   \includegraphics[height=4.6cm]{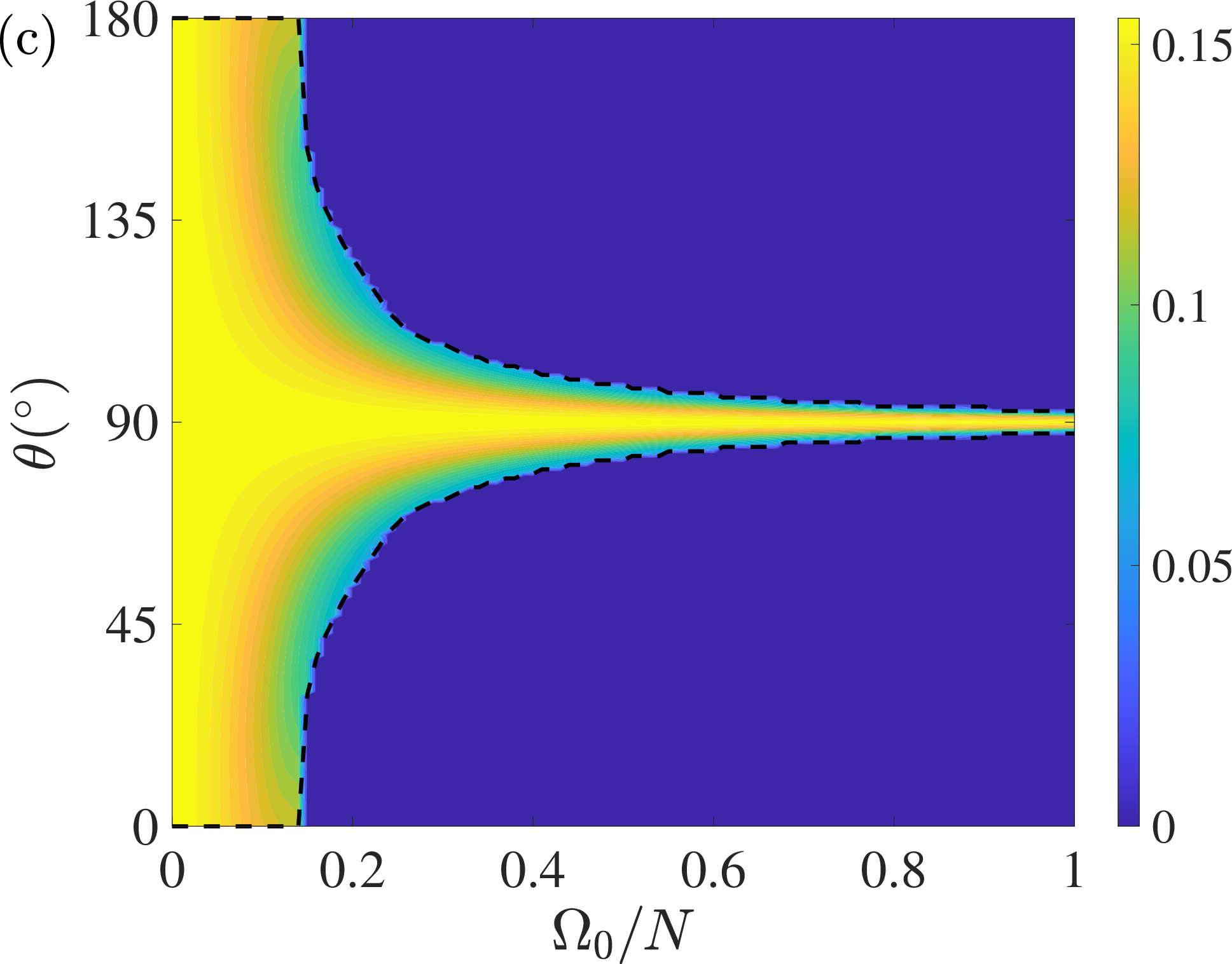}
      \caption{(a) Turbulent viscosity $\nu_{\rm{v}}$ for the inertial instability computed at $Pe=0.01$, $Pr=10^{-6}$, $Ri=0.25$ and $\Omega_{0}/N=0.2$ for different colatitudes $\theta$. (b,c) Contours of the maximum turbulent viscosity $\nu_{\rm{v},\max}$ at $Pe=0.01$, $Pr=10^{-6}$ and $Ri=0.25$ for (b) the inertial instability and (c) the inflectional instability. 
      The black dashed line in panel c denotes the cut-off colatitude $\theta_{c}$ at which the inertial instability dominates the inflectional instability.  
      }
         \label{Fig_nu_t}
   \end{figure*}

From the above analysis, we propose a general turbulent viscosity model using the growth rate and wavenumbers as
\begin{equation}
\label{eq:nu_v}
    \nu_{\rm{v}}=\frac{\sigma}{k^{2}},
\end{equation}
where $k^{2}=k_{x}^{2}+k_{y}^{2}+k_{z}^{2}$. 
%This proposition is coherent with the models proposed by \citet{Spruit2002,Fulleretal2019} and is applicable for both inflectional and inertial instabilities. 
%From the linear stability analysis, we can also propose scaling laws for the turbulent viscosity based on turbulent dissipation by inflectional and inertial instabilities.
%For both cases, we follow the proposition for the turbulent viscosity $\nu_{\rm{v}}$ as
Due to the inhomogeneity in the vertical $z$-direction for the hyperbolic tangent shear flow (\ref{eq:base_shear}), it is not straightforward to quantify the wavenumber $k_{z}$. 
For the inflectional instability that has the maximum growth rate in the range $|k_{x}H_{0}|<1$ at $|k_{y}|=|k_{y,\min}|$, we can approximate $k_{z}H_{0}\simeq1$ based on the observation that eigenfunctions are localised around the sheared region (see e.g. Fig.~\ref{Fig_modes}a). 
The situation is more complicated with the inertial instability that has the maximum growth rate for large $k_{y}$ at $k_{x}=0$.
On the one hand, for small $|k_{y}H_{0}|<1$, we expect a similar scaling law as $k_{z}H_{0}\simeq1$. 
On the other hand, for large $|k_{y}H_{0}|>1$, the eigenfunction is localised around the centre $z=0$ and has a small vertical length scale as $1/|k_{y}|$ (i.e. $|k_{z}|\simeq |k_{y}|$), which can also be presumed from the quantization conditions such as Eq.~(\ref{eq:WKBJ_2nd_quantization}). 
Therefore, for the inertial instability, we assume $k_{z}H_{0}=\max(1,k_{y}H_{0})$.

Panel a of Fig. \ref{Fig_nu_t} shows examples of the turbulent viscosity $
\nu_{\rm{v}}$ versus $k^{2}$ for the inertial instability at $Pe=0.01$, $Pr=10^{-6}$, $Ri=0.25$ and $\Omega_{0}/N=0.2$ for different latitudes $\theta=0,30,60^{\circ}$.
%While varying $k_{y}$, the case with $k_{x}=0$ is considered since the growth rate of the inertial instability is maximal. 
The growth rate of the inertial instability becomes zero at $k_{y}=0$, thus the viscosity $\nu_{\rm{v}}$ becomes zero at $k^{2}H_{0}^{2}=1$ (i.e. $k_{x}=k_{y}=0$, $k_{z}H_{0}$ is considered to be 1).
$\nu_{\rm{v}}$ increases sharply as $k^{2}$ increases from the unity and reaches its peak before it decays exponentially. 
It is not shown here but this feature is similarly observed for $\nu_{\rm{v}}$ of the inflectional instability computed at $|k_{y}|=|k_{y,\min}|$ while varying $k_{x}$. 
By picking up the maximum $\nu_{\rm{v}}$ over the range of $k^{2}$ at a given ratio $\Omega_{0}/N$ and a colatitude $\theta$, we compute contours of the maximum viscosity $\nu_{\rm{v},\max}$ for the inertial and inflectional instabilities at $Pe=0.01$, $Pr=10^{-6}$ and $Ri=0.25$ as shown in Fig.~\ref{Fig_nu_t} panels b and c.
For the inertial case in panel b, we see that $\nu_{\rm{v},\max}$ is maximal at the poles $\theta=0, 180^{\circ}$ around $\Omega_{0}/N=0.2$.
As it is inertially stable near the equator, we see that $\nu_{\rm{v},\max}$ is zero around the equator.
The inertial instability is driven by the rotation $\Omega_{0}$, thus $\nu_{\rm{v},\max}$ is also zero at $\Omega_{0}/N=0$. 
This parametric dependence of $\nu_{\rm{v},\max}$ in the parameter space $(\Omega_{0}/N,\theta)$ is different that of the inflectional instability as shown in panel c.
Without rotation (i.e. $\Omega_{0}=0$), the inflectional instability is independent of the colatitude $\theta$ so as the viscosity $\nu_{\rm{v},\max}$.
As $\Omega_{0}/N$ increases, $\nu_{\rm{v},\max}$ is maximal at the equator and decreases as the latitude is away from the equator, the feature similar to the growth rate contours in Fig.~\ref{Fig_growth_rates}c.
Same as the growth rate, the viscosity $\nu_{\rm{v}}$ is constant and independent of the ratio $\Omega_{0}/N$ at the equator. 
We also found that, at a certain colatitude $\theta_{c}$ (namely the cut-off colatitude hearafter), the growth rate of the inflectional instability decreases sharply and becomes smaller than the growth rate of the inertial instability whose maximum is attained at $k_{x}=0$ and $|k_{y}|=|k_{y,\min}|$. 
This happens for $\Omega_{0}/N>0.12$, the regime in which the inertial instability co-exists and dominates the inflectional instability. 
In panel c, we distinguish the turbulent dissipation induced by the inflectional instability from that by the inertial instability by filtering the viscosity outside the cut-off colatitude $\theta_{c}$. 
For the same parameters $Pe=0.01$, $Pr=10^{-6}$ and $Ri=0.25$ considered here, we see in panels b and c that the maximum of $\nu_{\rm{v},\max}$ for the inflectional instability is higher while contours of $\nu_{\rm{v},\max}$ for the inertial instability span wider in the parameter space $(\Omega_{0}/N,\theta)$. 
From these contours in panels b and c, we see that the turbulent dissipation induced by the inertial instability is strong near the poles while the dissipation induced by the inflectional instability is strong near the equator.\\ 

\begin{figure}
   \centering
      \includegraphics[width=7cm]{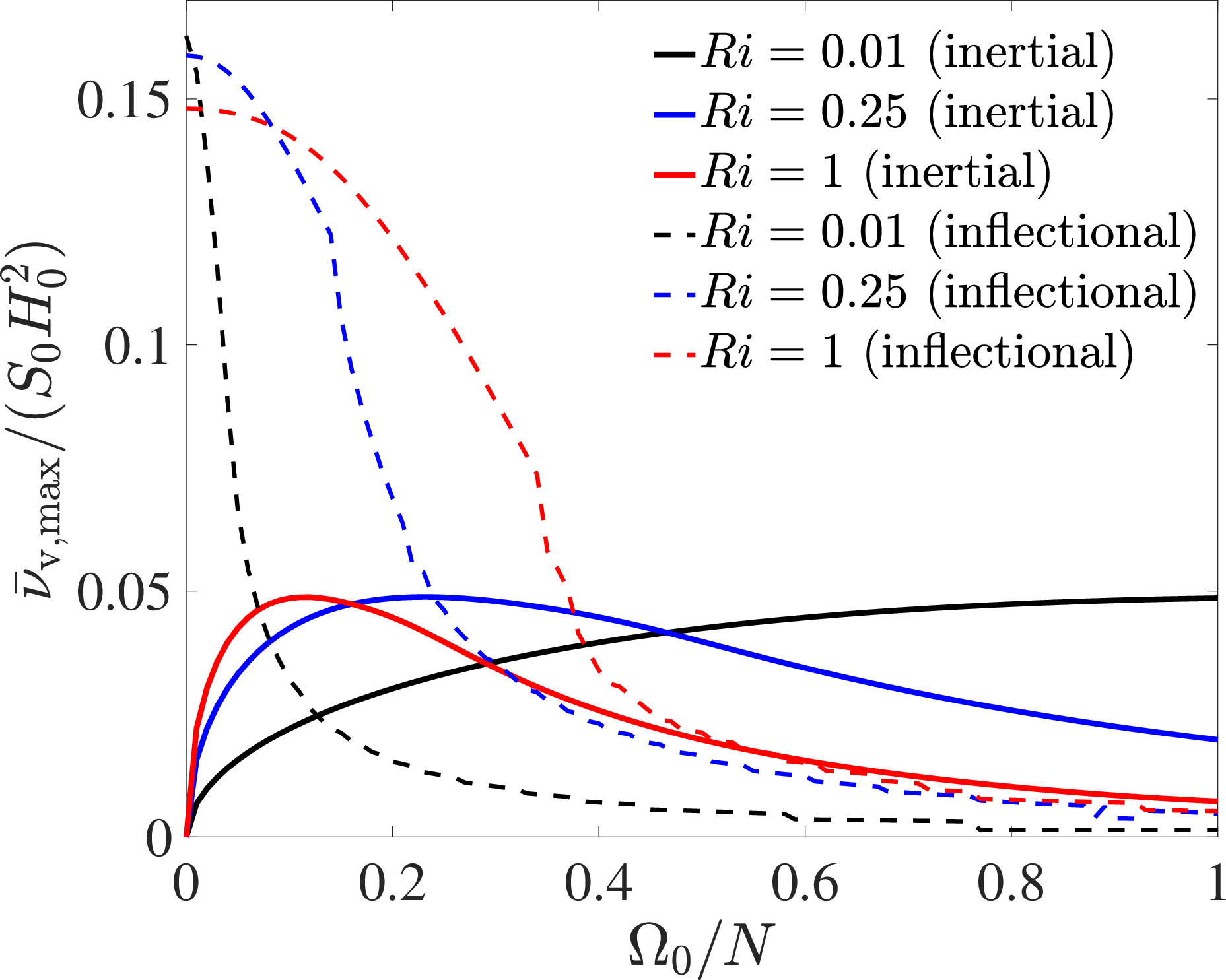}
              \caption{Latitudinally-averaged turbulent viscosity $\bar{\nu}_{\rm{v},\max}$ versus the ratio $\Omega_{0}/N$ for the inertial (solid lines) and inflectional (dashed lines) instabilities for $Ri=0.01$ (black), $Ri=0.25$ (blue) and $Ri=1$ (red) at $Pe=0.01$ and $Pr=10^{-6}$. 
      }
         \label{Fig_nut_averaged}
\end{figure}

\begin{table*}[!t]
\caption{Summary table for vertical shear instabilities and their variation with the full Coriolis acceleration and thermal diffusion}
\label{table:summary}
\centering
\begin{tabular}{l | c | c | c}
\hline
\hline
Instability type & $\kappa_{0}\uparrow$ ($Pe\downarrow$) & $\Omega_{0}/N\uparrow$ & Maximum turbulent transport at high $\kappa_{0}$\\
\hline
Inflectional & \textcolor{red}{$\uparrow$} & \textcolor{blue}{$\downarrow$}(except at $\theta=90^{\circ}$) & Near the equator ($\theta=90^{\circ}$) \\
Inertial  &  \textcolor{red}{$\uparrow$} & \textcolor{red}{$\uparrow$} (small $\Omega_{0}/N$) \textcolor{blue}{$\downarrow$} (large $\Omega_{0}/N$) & Near the poles ($\theta=0^{\circ},~180^{\circ}$)\\
\hline
\end{tabular}
\end{table*}

In Fig.~\ref{Fig_nut_averaged}, we plot the latitudinally-averaged turbulent viscosity $\bar{\nu}_{\rm{v},\max}\left({\Omega_{0}}/{N}\right)$ defined as
\begin{equation}
\label{eq:nu_v_averaged}
\bar{\nu}_{\rm{v},\max}=\frac{1}{2}\int_{0}^{\pi}\nu_{\rm{v},\max}(\Omega_{0}/N,\theta)\sin\theta\mathrm{d}\theta,
\end{equation}
for different values of $Ri$ at $Pe=0.01$ and $Pr=10^{-6}$.
For the inflectional instability, we see that the viscosity $\bar{\nu}_{\rm{v},\max}$ is maximal for the non-rotating case $\Omega_{0}=0$ and decreases as the ratio $\Omega_{0}/N$ increases.
While the $\bar{\nu}_{\rm{v},\max}$ curves descend as $Ri$ increases when $\Omega_{0}/N$ is close to zero, as predicted by \cite{Zahn1992} in which the Coriolis acceleration is neglected (i.e. $\Omega_{0}/N=0$), the viscosity $\bar{\nu}_{\rm{v},\max}$ increases and the overall curves ascend with increasing $Ri$ when $\Omega_{0}/N$ is large. 
We found that the turbulent viscosity for the inflectional instability is large and positive only for strong thermal diffusion with low $Pe$ as $Pe\ll1$.
For weak thermal diffusion with $Pe\geq1$, the growth rate $\sigma$ decays sharply as the latitude is away from the equator and thus the averaged viscosity $\bar{\nu}_{\rm{v},\max}$ becomes zero. 

For the inertial instability, $\bar{\nu}_{\rm{v},\max}$ is zero at $\Omega_{0}/N=0$ and increases as $\Omega_{0}/N$ increases. 
After the viscosity reaches its peak, it decays monotonically with the ratio $\Omega_{0}/N$. 
For the same $Pe=0.01$ and $Pr=10^{-6}$, the averaged turbulent viscosity $\bar{\nu}_{\rm{v},\max}$ of the inertial instability becomes larger than that of the inflectional instability for large $\Omega_{0}/N$. 
It is also shown that, as the the stratification increases (i.e. $Ri$ increases), the curves for the averaged turbulent viscosity $\bar{\nu}_{\rm{v},\max}$ shrink with their peaks moving towards $\Omega_{0}/N=0$.
The trend with varying $Ri$ is opposite for the inflectional instability as the overall curves move away from $\Omega_{0}/N=0$ as $Ri$ increases. 
This implies that the inflectional instability plays an important role in turbulent transport for strongly-stratified slowly-rotating stars (i.e. high $Ri$ and low $\Omega_{0}/N$) while the inertial instability becomes more important for weakly-stratified fast-rotating stars (i.e. low $Ri$ and high $\Omega_{0}/N$). 

Our proposition of a new turbulent viscosity model such as (\ref{eq:nu_v}) or the latitudinally-averaged one (\ref{eq:nu_v_averaged}) advances our understanding of turbulent transport (and mixing) in stellar radiation zones as the model (i) includes the effect of the full Coriolis acceleration to account for turbulent dissipation at a general latitude and (ii) distinguishes the contribution to turbulence induced by the inflectional instability from that by the inertial instability. 

Our model is, however, deduced from a linear framework and is required to be compared and validated against results from nonlinear numerical simulations. 
Further analytical and numerical investigations are, therefore, crucial and should be conducted to propose a more complete modelling of turbulence to be implemented in 1-D and 2-D stellar structure and evolution codes, as previously studied in \citet{Mathisetal2018} in the case of 1-D codes. 

\section{Conclusion}
\label{sec:Conclusion}
In this paper, we study vertical shear instabilities induced by differential rotation along the radial direction in stellar radiation zones where fluid is stably stratified and the seat of a strong thermal diffusion. 
The novelty in this study is that we consider the effects of thermal diffusion and the full Coriolis acceleration, the latter considering both the vertical and horizontal rotation components, on vertical shear instabilities. 
This configuration allows us to explore shear instabilities at any latitude where the rotation vector is not perpendicular but is inclined to the local plane of stellar radiation zones where the thermal diffusion is strong. 
To explore vertical shear instabilities, we consider a canonical shear flow with a hyperbolic-tangent profile and analyse its stability in stably stratified, thermally-diffusive, and rotating fluids with the full Coriolis acceleration. 
Two types of instabilities are identified: the inflectional instability due to an inflection point of the shear flow and the inertial instability occurring due to an imbalance between the centrifugal acceleration and pressure gradient.
The inflectional instability is found to be destabilized as either thermal diffusion becomes stronger or stratification becomes weaker.
The regime of inflectional instability exists in a narrow wavenumber range $k_{x}<1$ and the instability is maximal at $k_{y}=0$ or $k_{y}=k_{y,\min}$ if the wide-jet approximation is considered for cases outside the equator.  
At the equator with $f_{\rm{v}}=0$, the maximum growth rate of the inflectional instability is independent of the horizontal rotation component $f_{\rm{h}}$.
At a general latitude with non-zero $f_{\rm{v}}$, the inflectional instability is suppressed as the ratio $\Omega_{0}/N$ increases at a co-latitude $\theta$ outside the equator. 
While parametric dependence of the inertial instability on stratification and thermal diffusion is similar to that of the inflectional instability, the effect of the full Coriolis acceleration on the inertial instability is different. 
In non-diffusive fluids, the inertial instability is found to be maximal in the mid latitudes for the cyclonic rotation with $\Omega_{0}/S_{0}>0$. 
For highly-diffusive fluids, a case more relevant to stellar radiation zones, the inertial instability and its turbulent dissipation are maximal at the poles.  
These findings are confirmed by both numerical linear stability analysis and the WKBJ analysis with explicit expressions of the dispersion relation for inertial instability. 
Furthermore, it is revealed that symmetric instability, which is analytically more approachable as a linear shear flow is considered in the analysis, is found to be analogous not only qualitatively but also quantitatively to the inertial instability of a hyperbolic-tangent shear flow when the same value of the vertical shear at $z=0$ is considered for both cases. 
As a consequence of these results, we foresee that the inflectional instability plays a key role for turbulent transport and mixing in the equatorial region of strongly-stratified radiation zones of slowly-rotating stars while the inertial instability will promote turbulent transport in the polar regions of weakly-stratified radiation zones in fast-rotating stars.
The summary of our findings on vertical shear instabilities is provided in Table \ref{table:summary}.\\

In this work, we have thus provided a first complete linear analysis of vertical shear instabilities with taking simultaneously the full Coriolis acceleration and potentially strong heat diffusion into account. 
The prescriptions we have computed for the vertical eddy viscosity to model the turbulent transport of momentum and chemicals are ready to be implemented in 2D stellar structure and evolution numerical models. 
They can also be implemented in 1D models by making an average over latitudes. The next steps will be to undertake the nonlinear study of the inflectional and inertial instabilities of a vertical shear using direct numerical simulations, to study the instabilities of shear varying both with the radius and the co-latitude \citep[e.g.][]{Garaudetal2024}, and to evaluate the effects of the presence of a potential magnetic field \citep[e.g.][]{Lecoanetetal2013} by taking the full Coriolis acceleration and heat diffusion into account. 
This would contribute to build step by step a complete picture of shear instabilities in stellar radiation zones, which can be used in forthcoming generations of stellar structure and evolution models.

\section*{Acknowledgements}
      J. Park acknowledges support from the Royal Astronomical Society and Office of Astronomy for Development through the RAS-OAD astro4dev grant and from the Engineering and Physical Sciences Research Council (EPSRC) through the EPSRC mathematical sciences small grant (EP/W019558/1). The authors acknowledge support from the European Research Council through ERC grant SPIRE 647383. S. Mathis acknowledges support from the European Research Council (ERC) under the Horizon Europe programme (Synergy Grant agreement 101071505: 4D-STAR), from the CNES SOHO-GOLF and PLATO grants at CEA-DAp, and from PNPS (CNRS/INSU). While partially funded by the European Union, views and opinions expressed are however those of the author only and do not necessarily reflect those of the European Union or the European Research Council. Neither the European Union nor the granting authority can be held responsible for them. 

%%%%%%%%%%%%%%%%%%%%%%%%%%%%%%%%%%%%%%%%%%%%%%%%%%
\section*{Data Availability}
The data underlying this article will be shared on reasonable request to the corresponding author. 
%The inclusion of a Data Availability Statement is a requirement for articles published in MNRAS. Data Availability Statements provide a standardised format for readers to understand the availability of data underlying the research results described in the article. The statement may refer to original data generated in the course of the study or to third-party data analysed in the article. The statement should describe and provide means of access, where possible, by linking to the data or providing the required accession numbers for the relevant databases or DOIs.



\bibliographystyle{mnras}
\bibliography{mnras_PM2025} % if your bibtex file is called example.bib

%\begin{thebibliography}{99}
%\end{thebibliography}

%%%%%%%%%%%%%%%%%%%%%%%%%%%%%%%%%%%%%%%%%%%%%%%%%%

%%%%%%%%%%%%%%%%% APPENDICES %%%%%%%%%%%%%%%%%%%%%

%\appendix



%%%%%%%%%%%%%%%%%%%%%%%%%%%%%%%%%%%%%%%%%%%%%%%%%%

% Don't change these lines
\bsp	% typesetting comment
\label{lastpage}
\end{document}